\documentclass[iop]{emulateapj}

\usepackage{amsmath}
\usepackage{mathrsfs}
\usepackage{xspace}
\usepackage{color}

\shorttitle{Influence of stellar flares on exoplanets atmospheres}
\shortauthors{Venot, Rocchetto, Carl, Hashim \& Decin}

\begin{document}

\title{Influence of stellar flares on the chemical composition of exoplanets and spectra}

\author{Olivia Venot $\dagger$\altaffilmark{1}, Marco Rocchetto \altaffilmark{2}, Shaun Carl \altaffilmark{3}, Aysha Roshni Hashim \altaffilmark{3}, Leen Decin \altaffilmark{1}}
\affil{$^1$ Instituut voor Sterrenkunde, Katholieke Universiteit Leuven, Celestijnenlaan 200D, 3001 Leuven, Belgium \\
$^2$ University College London, Department of Physics and Astronomy, Gower Street, London WC1E 6BT, UK\\
$^3$ Department of Quantum Chemistry and Physical Chemistry, Katholieke Universiteit Leuven, Celestijnenlaan 200F, 3001 Leuven, Belgium}
\email{$\dagger$ olivia.venot@kuleuven.be}
  

\begin{abstract}
More than three thousand exoplanets have been detected so far, and more and more spectroscopic observations of exoplanets are performed. Future instruments (JWST, E-ELT, PLATO, Ariel,\,\dots) are eagerly awaited as they will be able to provide spectroscopic data with a greater accuracy and sensitivity than what is currently available. This will allow more accurate conclusions to be drawn on the chemistry and dynamics of the exoplanet atmospheres, on condition that the observational data are processed carefully. An important aspect to consider is temporal stellar atmospheric disturbances that can influence the planetary composition, and hence spectra, and potentially can lead to incorrect assumptions about the steady-state atmospheric composition of the planet.

In this paper, we focus on perturbations that come from the host star in the form of flare events that significantly increase the photon flux impingement on the exoplanetÕs atmosphere. In some cases, and particularly for M stars, this sudden increase may last for several hours. We aim at answering the question to what extent a stellar flare is able to modify the chemical composition of the planetary atmosphere and, therefore influence the resulting spectra.

We use a one-dimensional thermo-photochemical model to study the neutral atmospheric composition of two hypothetic planets located around the star AD Leo. We place the two planets at different distances from the star, which results in effective atmospheric temperatures of 412~K and 1303~K. 

AD Leo is an active star that has already been observed during a flare. We therefore use the spectroscopic data from this flare event to simulate the evolution of the chemical composition of the atmospheres of the two hypothetic planets. We compute synthetic spectra to evaluate the implications for observations.

The increase of the incoming photon flux affects the chemical abundances of some important species (such as H and NH$_3$) down to altitudes associated with an atmospheric pressure of 1 bar, that can lead to variations in planetary spectra (up to 150 ppm) if performed during transit. We find that each exoplanet has a post-flare steady-state composition that is significantly different from the pre-flare steady-state. We predict that these variations could be detectable with both current and future spectroscopic instruments if sufficiently high signal-to-noise spectra are obtained.

\end{abstract}

 
\section{Introduction}

M stars are the most common stars in our Galaxy, and thus are likely to harbour most of the planetary systems. A particularity of M stars is their high activity. This most active class of stars undergo great stellar variability, such as star spots, granulation and flares. The latter are violent and unpredictable outbursts of the photosphere that are caused by sudden changes in magnetic flux profiles. During such events, the emitted radiative energy can increase by several orders of magnitude over the complete wavelength range, including the prominent H$_{\alpha}$ emission.\\
A few studies on the effect of flares on exoplanets have been conducted to date.  \cite{segura2010effect} focused on the composition of Earth-like planets. They simulated an Earth-like planet located in the habitable zone of the active M dwarf AD Leo and tracked the evolution of the abundances of water and ozone in the planetary atmosphere during a flare event. They found that the enhanced UV radiation emitted during a flare would not affect the habitability of the planet. They did not address the effect of the flare on the planetary spectra. Complementary to this latter paper, \cite{tofflemire2012} studied the consequences of M dwarf stellar flares on the observation of exoplanets. They found that if a flare with an energy of ~10$^{31}$-10$^{32}$ erg, i.e. a standard flare occuring in M-type dwarfs, occurs during a transit, it would have no impact on the infrared (IR) detection and on the characterisation of exoplanets above a level of $\sim$5 -- 11 mmag. The most energetic flares (E $>$ 10$^{34}$ erg) would be easily detected in IR observations, as they would produce an increase in magnitude significantly larger than the photometric noise obtained with most observational instruments. Here, again, the authors did not consider the effect of flares on the atmospheric composition of exoplanets.
Recently, \cite{Rugheimer2015} studied the atmospheres and the spectra of Earth-like planets orbiting different kinds of M dwarfs (both active and inactive) leading to a very useful spectral database. They did not, however, vary the stellar flux during their simulations in order to mimic stellar activity.\\
Far from astrobiological consideration, in this paper, we simulate two exoplanets (sub-Neptune/super-Earth) orbiting the star AD Leo and study the consequences of a stellar flare on both the chemical composition of the atmospheres and the synthetic spectra. We model two cases: a warm and a hot planet with effective temperatures of 412 K and 1303 K, respectively. In the context of the future space missions (James Webb Space Telescope (JWST), PLAnetary Transits and Oscillations of stars (PLATO), etc.), the aim is to determine whether or not planetary spectra acquired during a stellar flare can be differentiated from ``quiescent" spectra, and whether or not false conclusions can be drawn, i.e., we aim to study to what extent a stellar flare can influence the chemical composition of the atmosphere and the resulting synthetic spectra.
In Sect.~\ref{sec:model} we describe the  one-dimensional (1D) kinetic model and the radiative transfer model together with the adopted thermal profiles and stellar fluxes. We present our results in Sect.~\ref{sec:results} and discuss them in Sect.~\ref{sec:disc}. Our conclusions are given in Sect.~\ref{sec:concl}.

\section{Models}\label{sec:model}

\subsection{1D neutral chemical model}

We used our thermo-photochemical model that has been developed especially for warm atmospheres \citep{ven2012}. This code has been used to study exoplanets \citep{ven2012, venot2013high, ven2014, ven2015, agu2014} but also the deep atmosphere of giant planets of the Solar System \citep{Cavalie2014, Mousis2014}. 
We use the numerical solver DLSODES\footnote{D- meaning the Double precision version of LSODE (Livermore Solver for Ordinary Differential Equations), and -S standing for Sparse matrix}. It can be found in ODEPACK: a collection of powerful numerical solvers for ordinary differential equation systems. To resolve stiff problems (i.e. those containing processes having a great disparity in rates), as systems of time-dependent chemical rate equations, DLSODES uses the BDF (backward differentiation formula) method. This multiple steps method has been implemented for the first time by C. W. Gear \citep{gear1971numerical} and is known to yield very accurate solutions in such cases. The superiority of LSODES to resolve astrochemical kinetics systems, over other ODE solvers, has been shown by \cite{nejad2005}. Also, the advantage of using DLSODES instead of the Cranck-Nicholson method (a common method used in photochemical models) is presented in \cite{dob2010}. More details about this solver can be found in the work of \citet{hindmarsh1983odepack} and \citet{radhakrishnan1993description}.\\ In our study, the time step for the solver starts at 10$^{-10}$s and increases continuously as chemistry evolves.\\

One of the strengths of our model is its robust chemical scheme that has been validated against experiments over a large range of pressures (0.01 -- 100~bar) and temperatures (300 -- 2500~K). This reaction network has been constructed thanks to a close collaboration with specialists in combustion (see details in \citealt{ven2012} and references therein). This C$_0$-C$_2$ scheme describes the kinetics of species with up to two carbon atoms. The network contains 105 neutral species involved in 960 reversible reactions. To model the effect of UV irradiation, we added to this network a set of 55 photodissociations. References to the associated cross-sections and quantum yields used can be found in \cite{ven2012} and \cite{dob2014}. For the cross-sections of CO$_2$ \citep{venot2013high} and NH$_3$ \citep{venot_NH3}, we used our recent measurements at 500~K.\\
We recently developed an extended chemical scheme (C$_0$-C$_6$) that is able to reproduce the kinetics of species with up to 6 carbon atoms \citep{ven2015}. This chemical network is very useful to study atmospheres rich in carbon atoms. As we have shown in \cite{ven2015}, for atmospheres with solar C/O ratios, the C$_0$-C$_2$ scheme is sufficient. It yields the same results  and is significantly computationally more efficient. Here, we have not not explored C/O ratios different than solar, so our C$_0$-C$_2$ scheme is sufficient.
In order to parametrise the vertical mixing, we choose a constant eddy diffusion coefficient $K_{zz}=10^8$ cm$^2$s$^{-1}$, which corresponds to the lower value employed by \cite{Fortney2013} (from which we take our pressure-temperature profiles, see Sect.~\ref{sec:PTprofiles}) and which is a commonly used value in studies of exoplanet's atmospheres \citep[e.g.,][]{lew2010, moses2011disequilibrium, line2011thermochemical, venot2013high, ven2015}. The planets have a radius $R_p$ of 0.238 $R_{Jup}$ (2.61 R$_{Earth}$) and a mass $M$ of 0.02 $M_{Jup}$ (6.35 M$_{Earth}$).

\subsection{Stellar fluxes}

\begin{figure}[!htb]
\centering
\includegraphics[angle=0,width=\columnwidth]{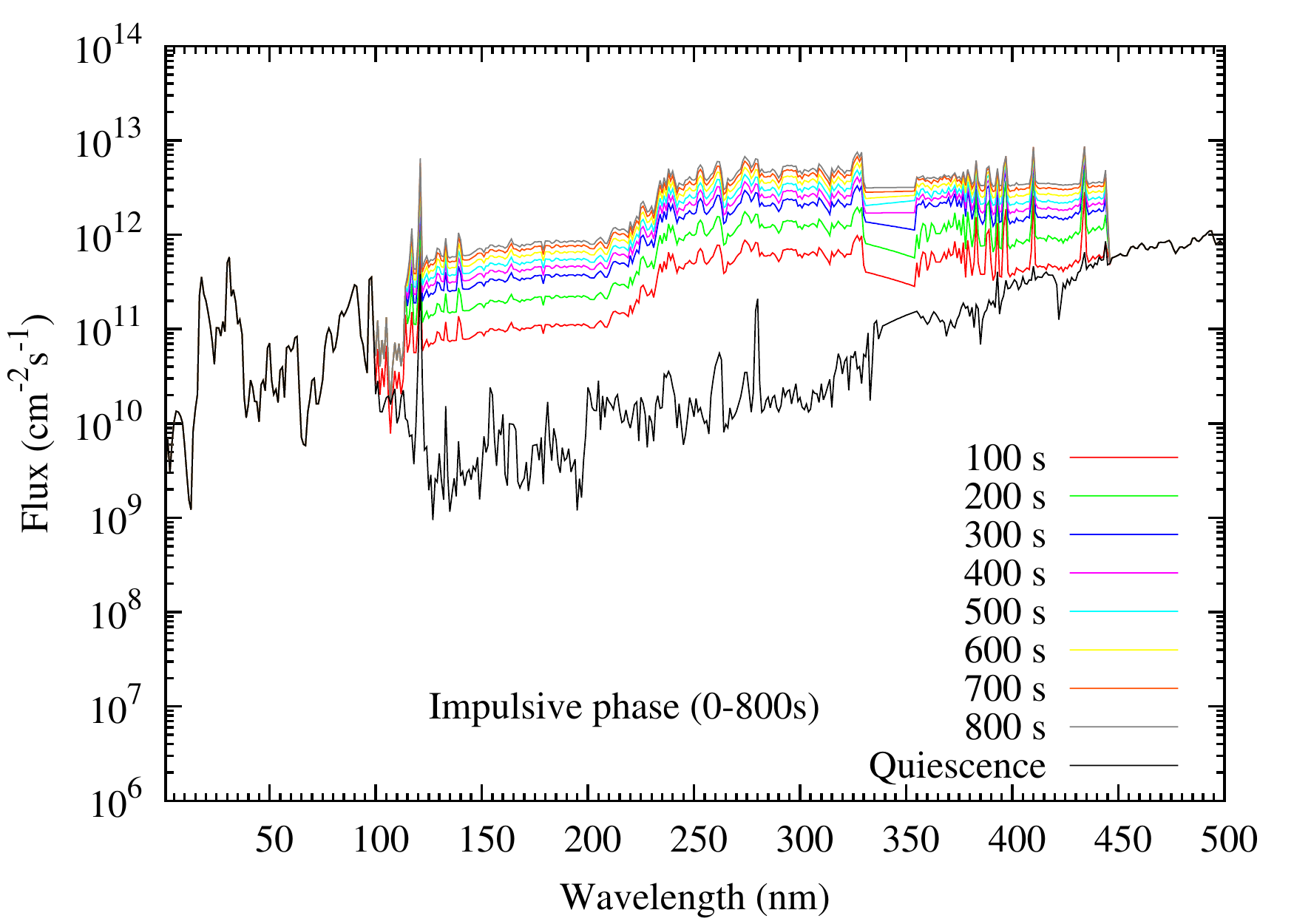}
\includegraphics[angle=0,width=\columnwidth]{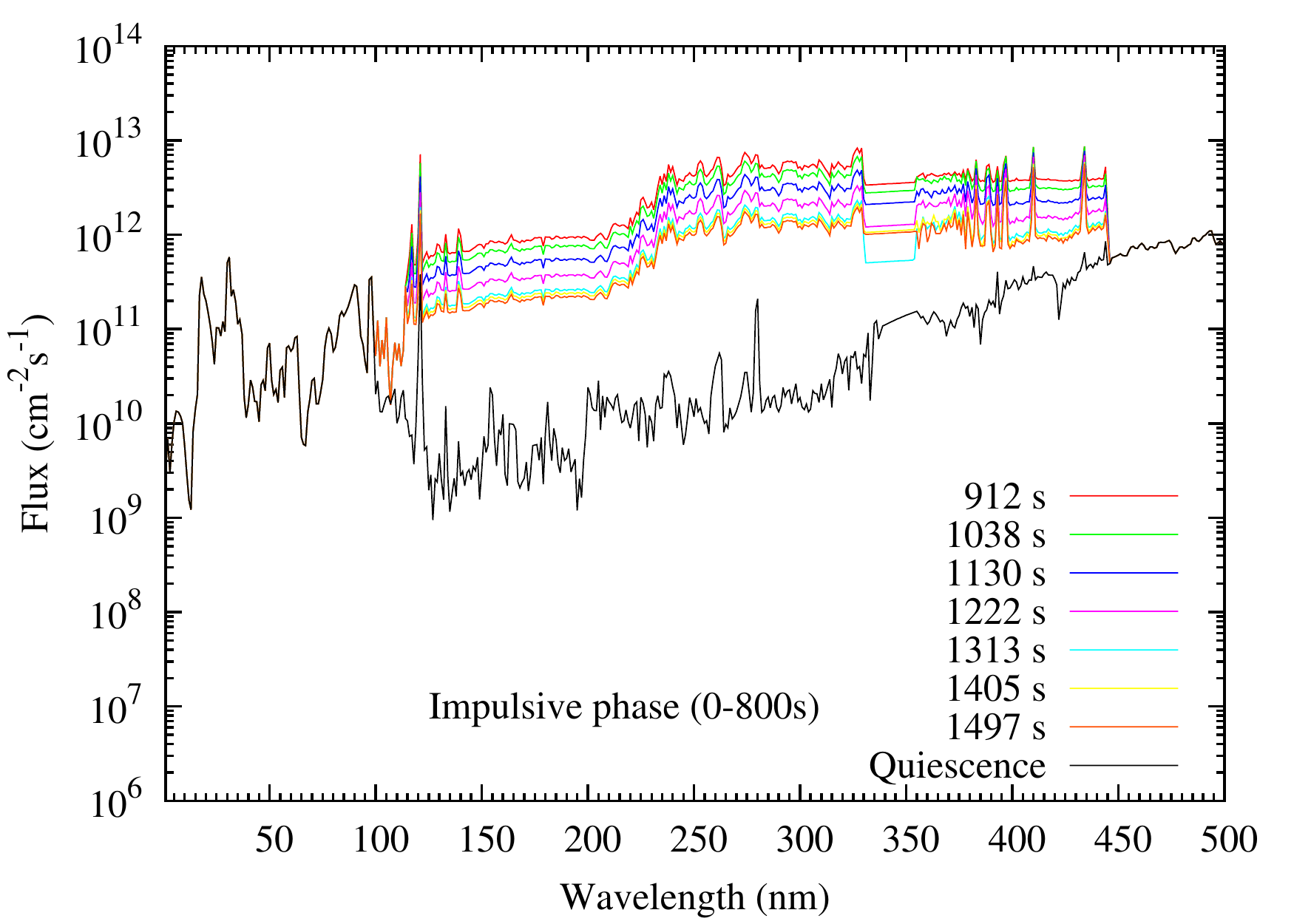}
\includegraphics[angle=0,width=\columnwidth]{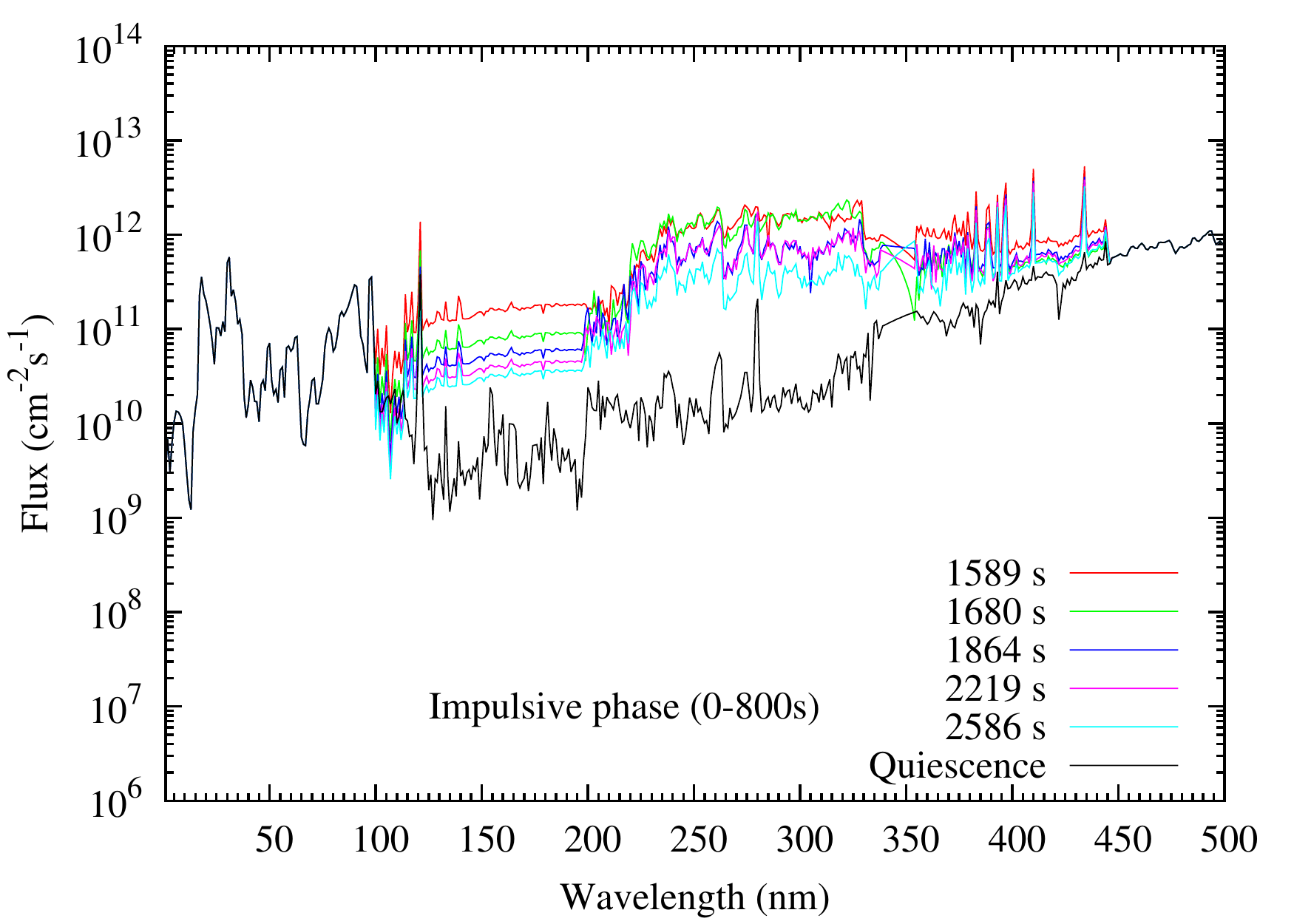}
\caption{Stellar spectra of AD Leo at a distance of 1 AU from the star. In the [1 -- 99] nm and [445 -- 900] nm range, all spectra are identical. Spectra at different time steps during the flare are represented by different colours, as labelled on the plots, and are used as input for the photochemical model at the corresponding time step. The quiescent spectra is represented in each figure in black. The three phases of the flare are represented: the first impulsive (\textit{top}), the second impulsive (\textit{middle}) and the gradual (\textit{bottom}).}
\label{fig:flux}
\end{figure}

AD Leo (also called Gl 388) is a M3.5V star \citep{shkolnik2009} located at 4.9 pc from the Sun \citep{rojas2012}. Its luminosity is 0.024 L$_{\sun}$ \citep{pettersen1981} and its effective temperature is 3\,390 $\pm$19 K \citep{rojas2012}. It is one of the most active flare stars known.
The great flare affecting AD Leo in 1985 was observed by \cite{hawley1991} in the visible and UV spectral region. Unfortunately, the time distribution of this acquisition was not well distributed (especially in UV spectral region). From these original data, \cite{segura2010effect} constructed a time-dependent sequence of spectra that can be used in photochemical models. A complete description of the method used can be found in their paper.\\
To model the flare event, we used these data between 100 nm and 444 nm. For wavelengths less than 100 nm, we use the mean of the Sun spectra at maximum and minimum activity \citep{gueymard2004}, multiplied by a factor of 100 in order to reproduce the high X-ray luminosity of AD Leo (L$_X\simeq4\times10^{28}$ erg.s$^{-1}$\citep{Favata2000} vs. L$_{X\sun}\simeq10^{26}-10^{27}$ erg.s$^{-1}$ for the Sun \citep{Judge2003}). For wavelengths larger than 444 nm, we used the quiescent spectrum of AD Leo from \cite{segura2005biosignatures}. To calculate the steady states before and after the flare event, we used the same spectrum except that the quiescent spectrum of AD Leo \citep{segura2005biosignatures} was used down to 100 nm. We note that for AD Leo, log$[L_X/L_{bol}] = -3.3$, which is close to the saturation level ($\sim-3$) that very active stars reach \citep[e.g.,][]{vilhu1987, stauffer1994}.
The quiescent and flares stellar fluxes are shown on Fig.~\ref{fig:flux}. They are separated in three phases: the first impulsive phase from 0 to 800 s, during which the stellar flux increases rapidly by one order of magnitude; the second impulsive phase, from 912 to 1497 s, during which the stellar flux, after reaching the maximum at 912 s, decreases by less than one order of magnitude over approximatively the same timescale as the first phase; and finally the gradual phase during which the stellar flux continues to decrease, but more slowly.

\subsection{Thermal profile}\label{sec:PTprofiles}

\begin{figure}[!htb]
\centering
\includegraphics[angle=0,width=\columnwidth]{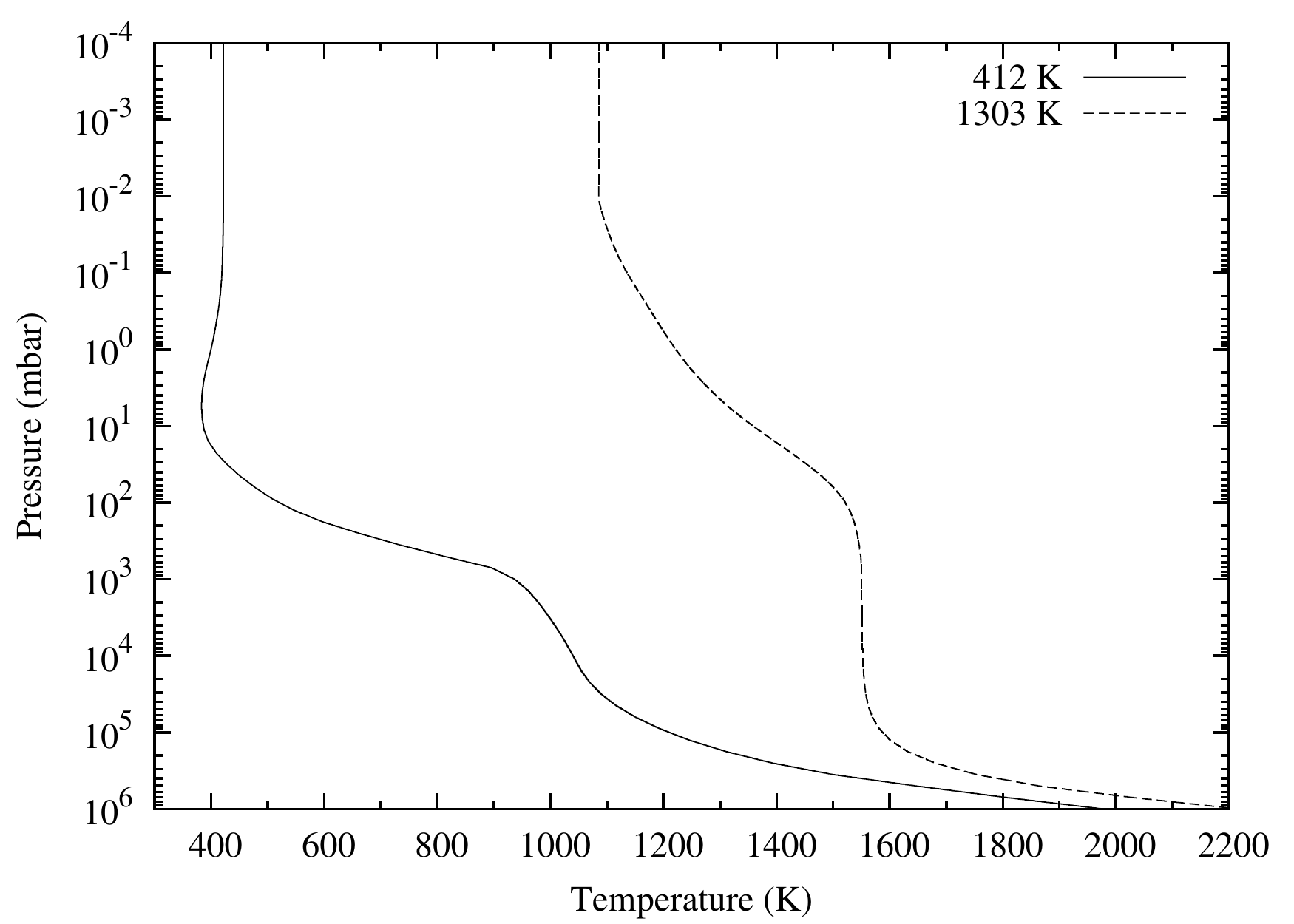}
\caption{Thermal profiles of the hypothetic AD Leo b planets.}\label{fig:figure_PT}
\end{figure}

To model our hypothetic planet's atmosphere, we use the thermal profiles from \cite{Fortney2013}. Their fully non-gray atmosphere code has been adapted to exoplanet atmospheres \citep[e.g.,][]{Fortney2005, Fortney2008, Morley2013}. The code considers both the incident radiation from the star and the thermal radiation from the planet's atmosphere, which are resolved using the algorithm developed by \cite{toon89}. Opacities are tabulated using the correlated-k approximation. Using this code, they computed several temperature profiles for GJ1214b-like planets, corresponding to different intensities of irradiation of the planet. All atmospheric models consider an enrichment of 50 times solar metallicity, an internal temperature of 60~K (and not 75~K as it is misprinted in the legend of their Fig. 3, J. Fortney, private communication), and a Bond albedo of $\sim$0.05. The albedo is here very low due to the absence of clouds in the model. However, clouds, which have been detected in GJ1214b \citep{Kreidberg2014}, can significantly increase albedo (to 0.4-0.6 according to \citealt{charnay2014}). More details about the data used to calculate these profiles can be found in \cite{Fortney2013}, and references therein.\\
We consider that using thermal profiles of GJ1214b as analogs to model our hypothetic ``AD Leo b" planets is appropriate considering that AD Leo and GJ 1214 are both M stars, with similar effective temperatures (respectively, 3$\,$390 $\pm$ 19 K and 3$\,$245$\pm$31 K, \citealt{rojas2012}). We used the profiles corresponding to the effective temperatures of 412 K and 1303 K (see Fig.~\ref{fig:figure_PT}). These profiles originally end at 10$^{-3}$ mbar, with a temperature increasing at low pressures, which is not a physical effect but an artefact due to boundary conditions of the model. In order to model completely the photochemical processes, we extrapolated the profiles to lower pressures, using an isothermal profile from 10$^{-2}$ to 10$^{-4}$ mbar to avoid this artefact. The distances between the star and the planets necessary to be consistent with these thermal profiles are 0.069 AU and 0.0069 AU, for 412 K and 1303 K, respectively. The hottest profile can raise the doubt about the possible existence of a sub-Neptune/super-Earth so close to its star. Planets observed at a distance less than 0.01 AU are usually smaller than our super-Earth models (i.e. R$\sim$ 0.07 R$_J$, \citealt{charpinet2011, muirhead2012}). However, \cite{Sanchis-Ojeda2015} reported very recently the discovery of an exoplanet with similar properties to our model (R = 0.22 R$_J$ at a distance of 0.008 AU) and they suggested that this planet was a disintegrating rocky planet with a cometary head and tail. This kind of object can be seen as a final step in the evolution of a planet being destroyed by tidal forces (S. Raymond, private communication). 

\subsection{Spectra model}

Synthetic transmission spectra in the wavelength range $0.8-20$\,$\mu$m for the different atmospheric compositions were obtained using the forward model described in \cite{Waldmann2015}, based on the Tau code by \cite{Hollis2013} and modified to allow variable pressure-dependent temperature profiles. The 1D radiative transfer model is based on a  code that calculates the optical path through the planetary atmosphere, resulting in a transmission spectrum of transit depth as a function of wavelength. The wavelength and temperature-dependent absorption cross sections for the absorbing molecules were computed using the line lists from ExoMol \citep{Tennyson2012}, HITRAN \citep{Rothman2009,Rothman2013} and HITEMP \citep{Rothman2010}. We only considered the following molecules: H$_2$O, CO$_2$, OH, HCN , NH$_3$, CH$_4$ and CO. Although these molecules represent only a fraction of the 105 molecules considered in the thermo-photochemical model, they are the most abundant and therefore dominate the spectral features. We also included collisional-induced cross sections for H$_2$-H$_2$ and  H$_2$-He \citep{Borysow2001,Borysow2002}. The atmosphere is assumed to be cloud-free. 

\section{Results}\label{sec:results}

\subsection{Chemical composition}

\begin{figure*}[!htb]
\centering
\includegraphics[angle=0,width=\columnwidth]{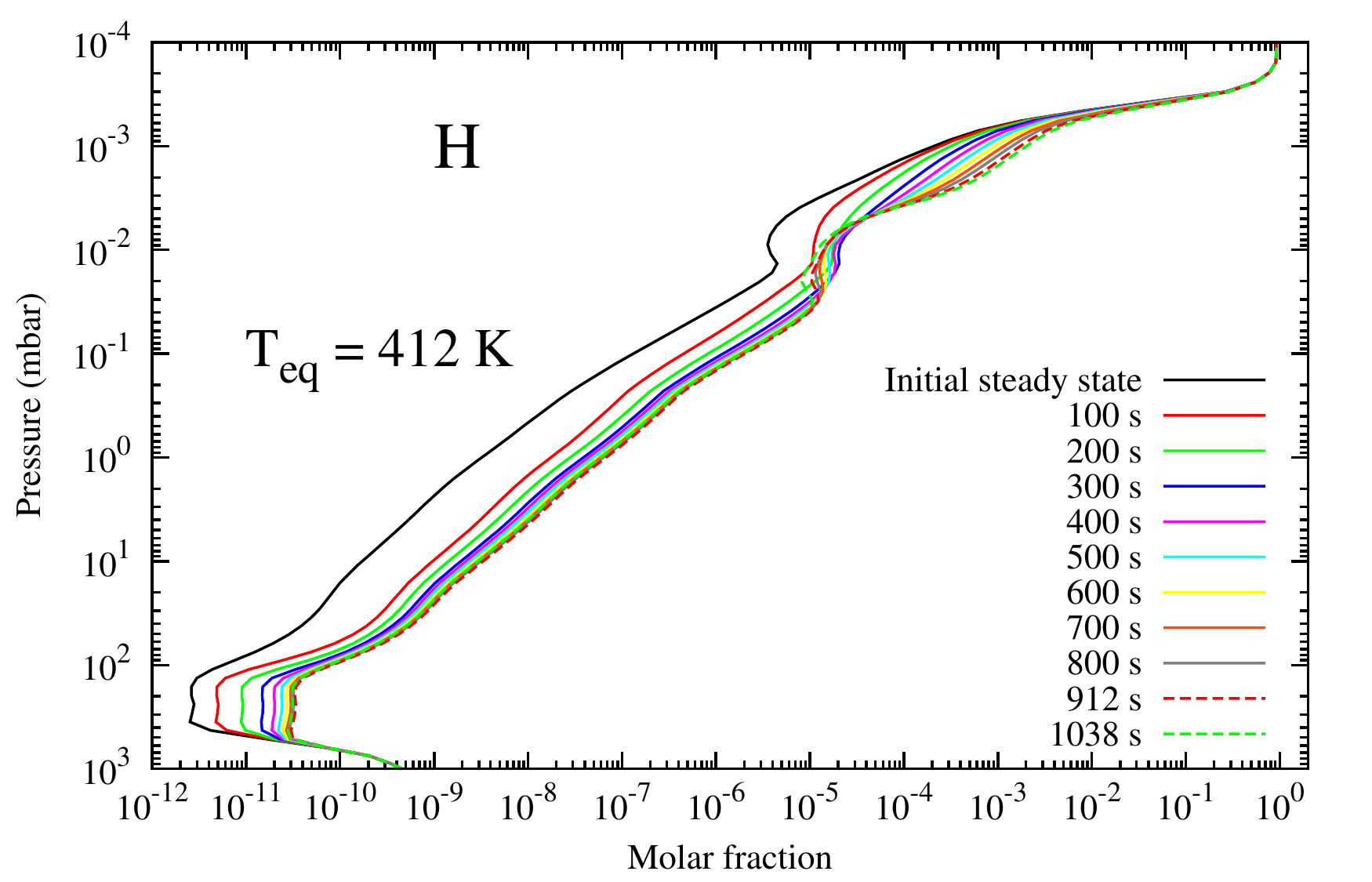}
\includegraphics[angle=0,width=\columnwidth]{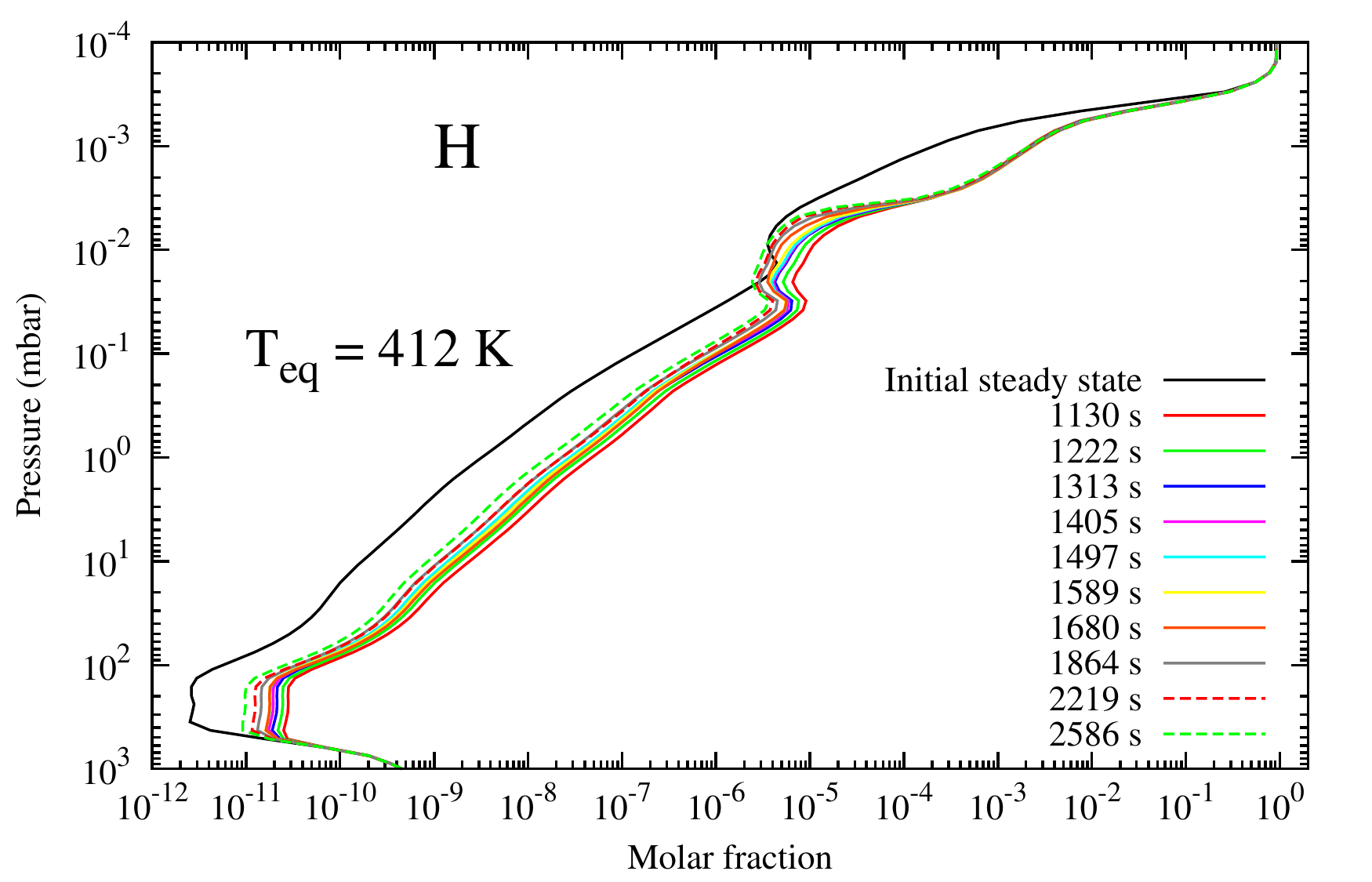}\\
\includegraphics[angle=0,width=\columnwidth]{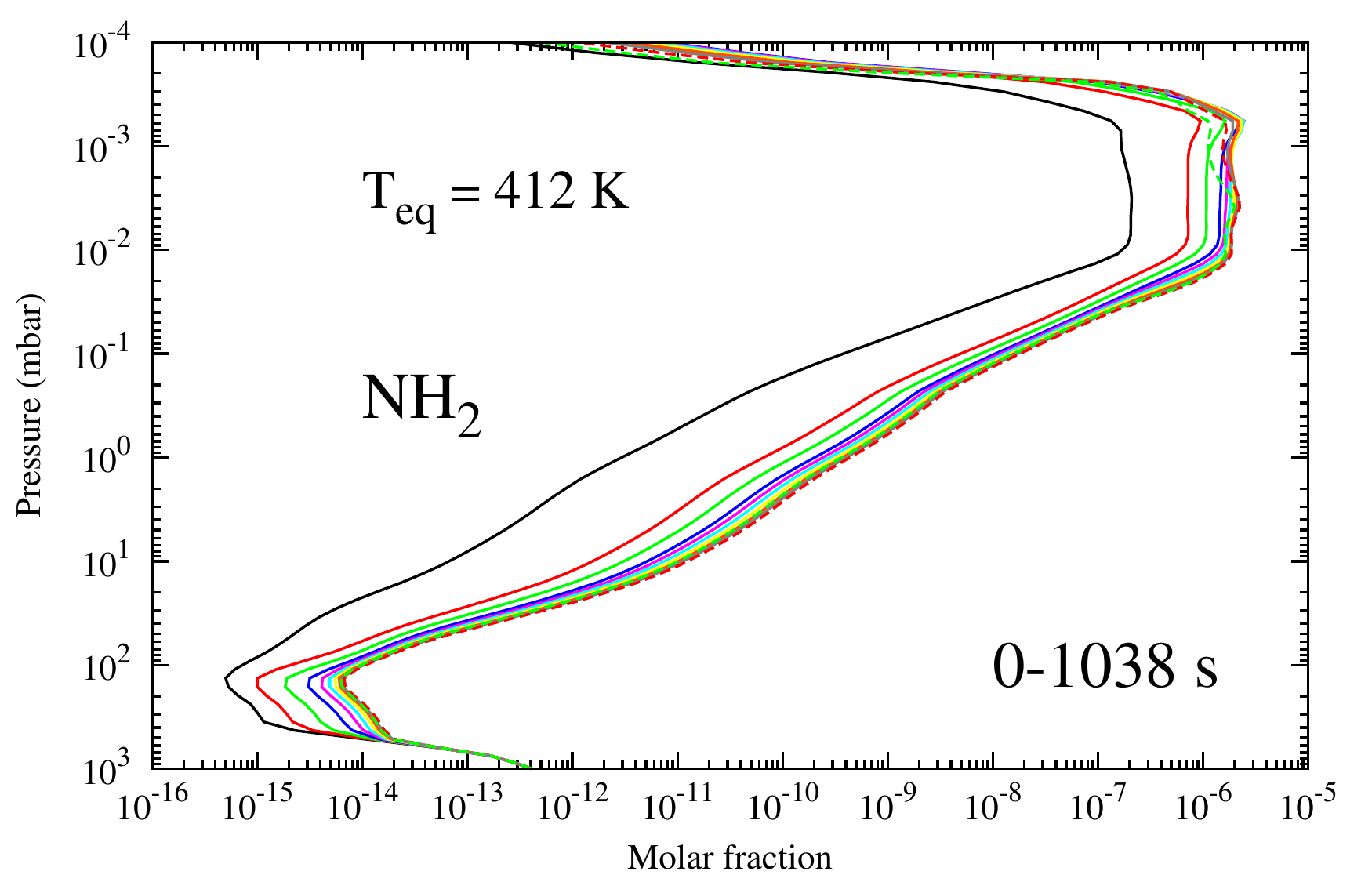}
\includegraphics[angle=0,width=\columnwidth]{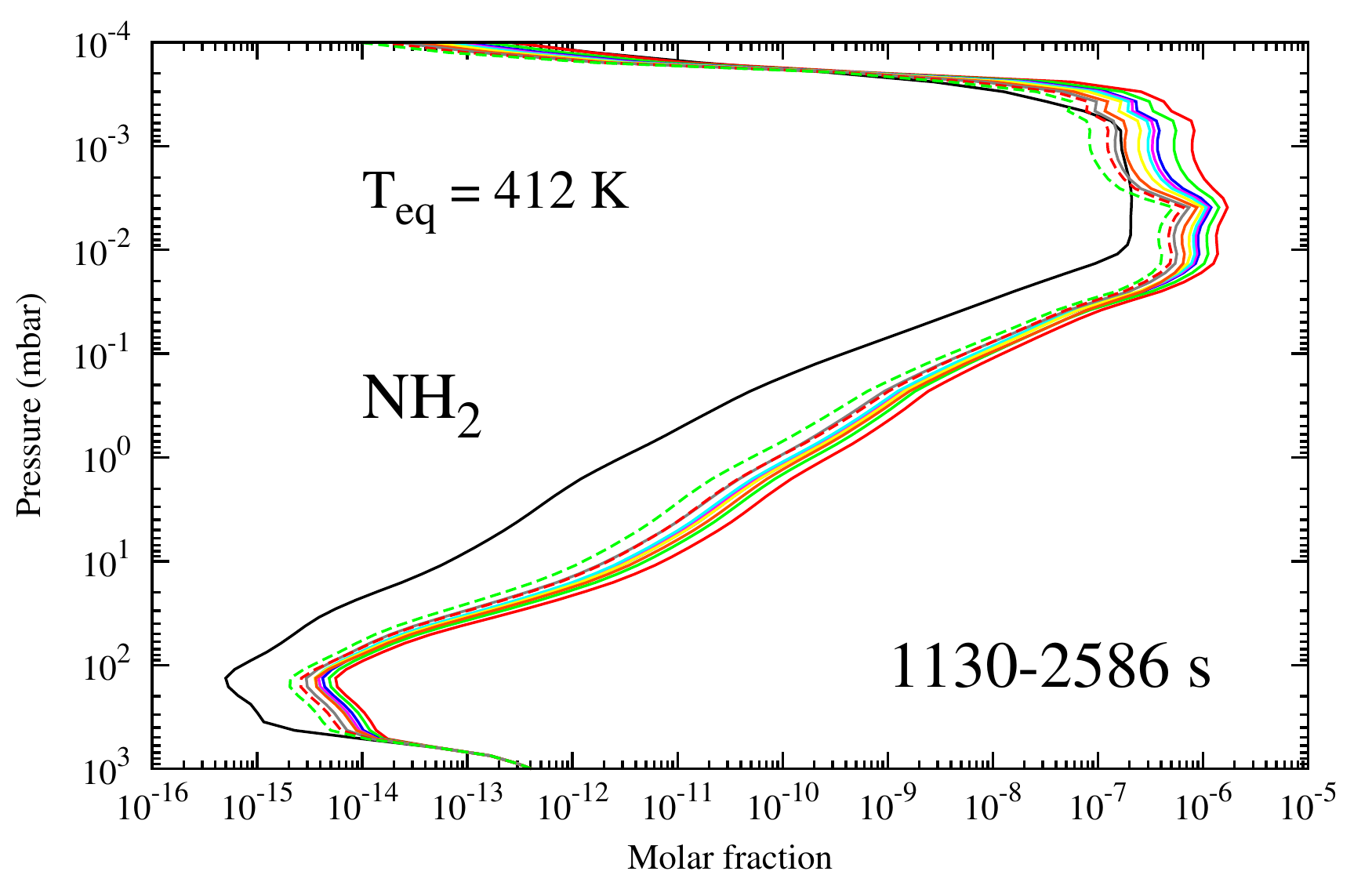}\\
\includegraphics[angle=0,width=\columnwidth]{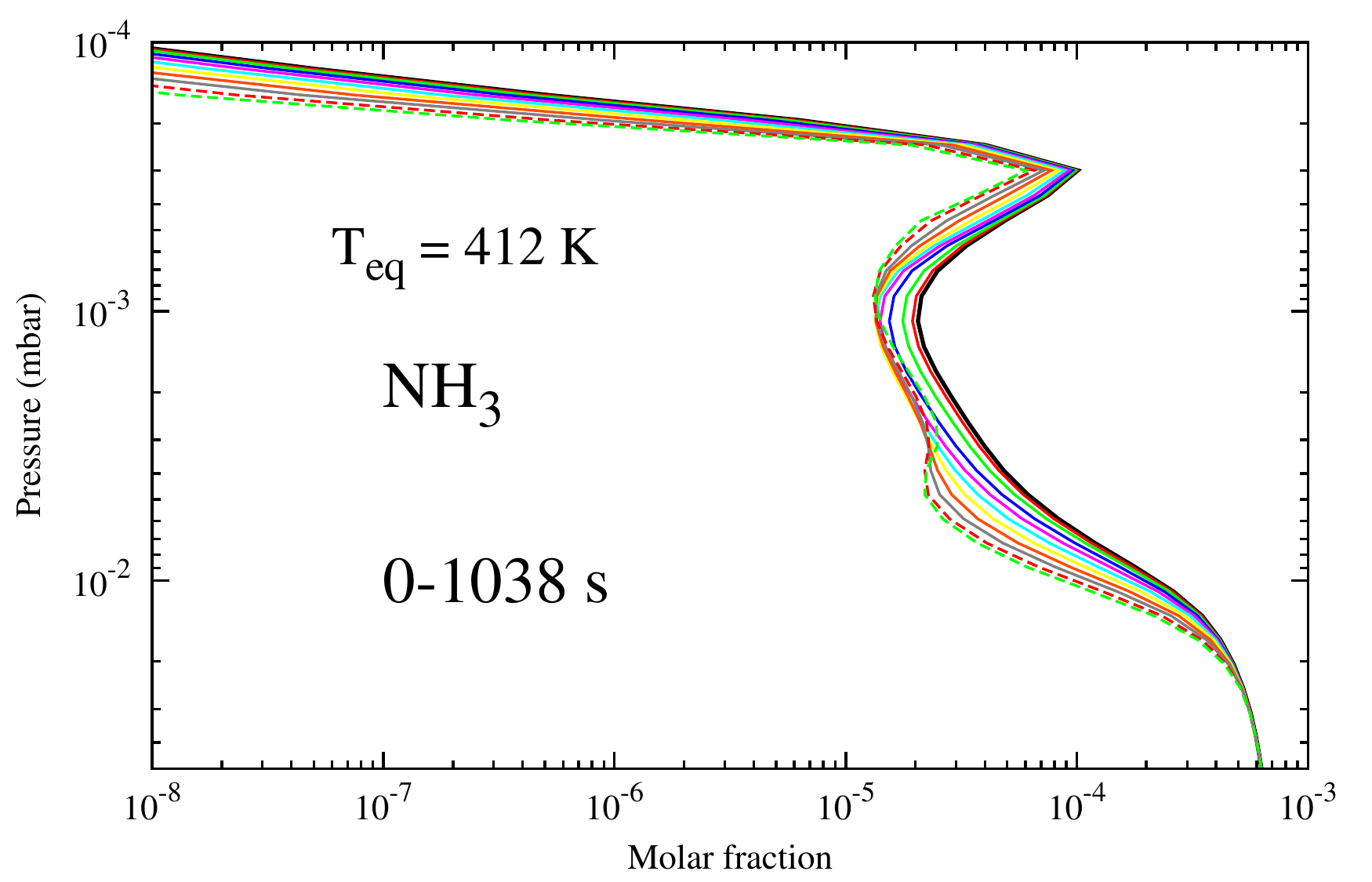}
\includegraphics[angle=0,width=\columnwidth]{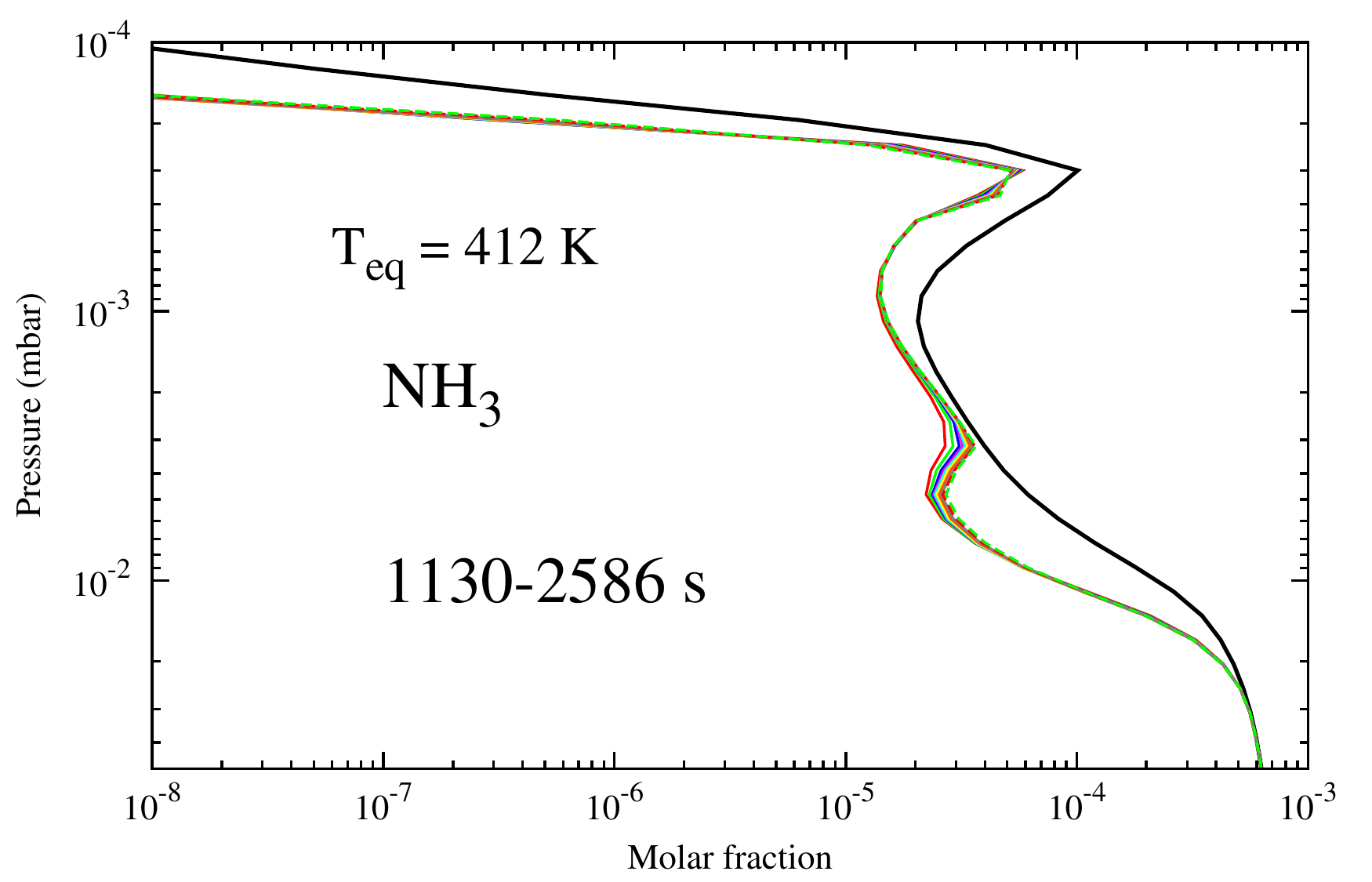}\\
\caption{Evolution of H, NH$_2$ and NH$_3$ mixing ratios during the flare event with the thermal profile corresponding to $T_{eq}$ = 412K. The legend for all figures is in the upper panels.} \label{fig:412K_NH3_H}

\end{figure*}

\begin{figure*}[!htb]
\centering
\includegraphics[angle=0,width=\columnwidth]{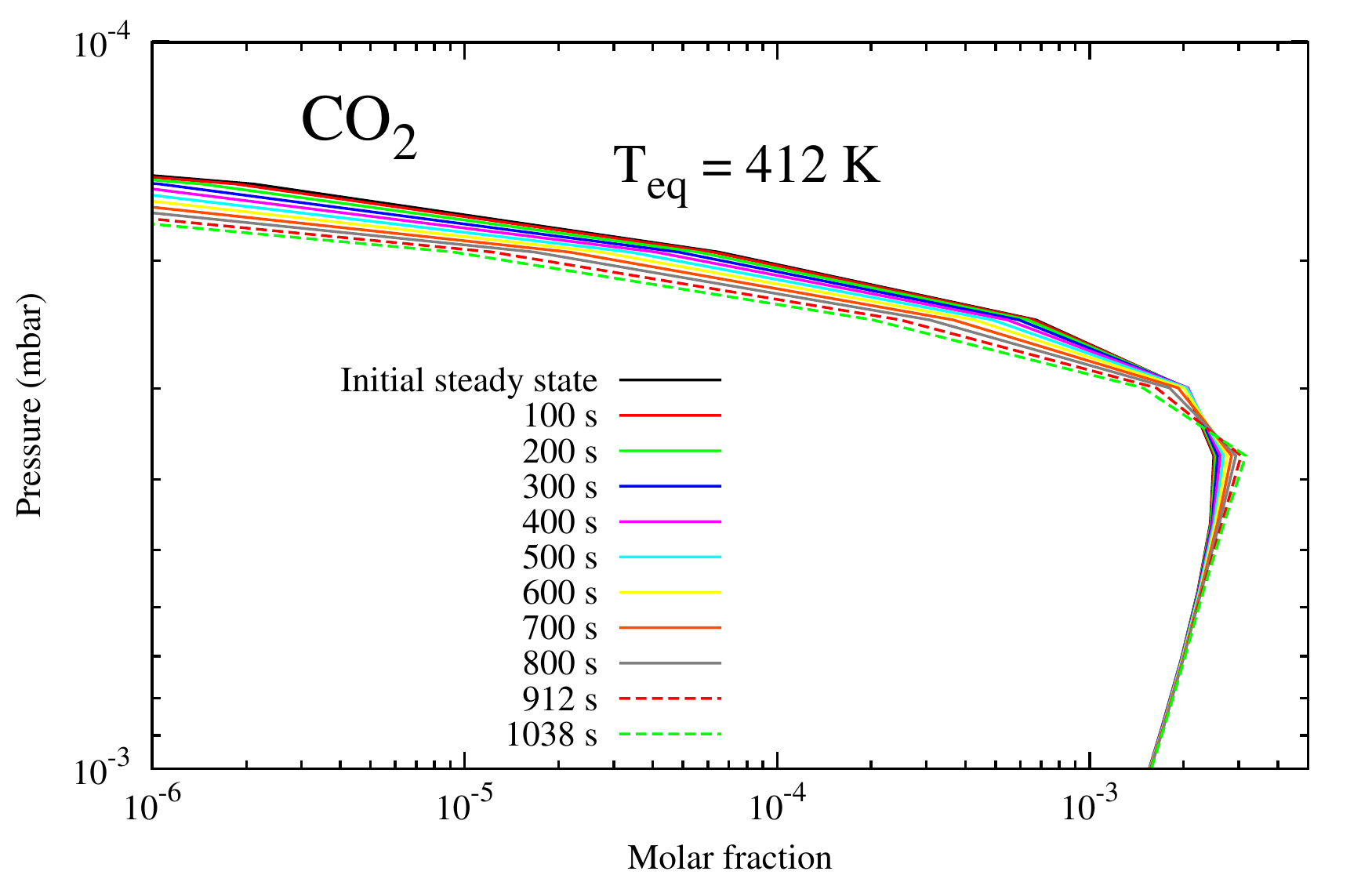}
\includegraphics[angle=0,width=\columnwidth]{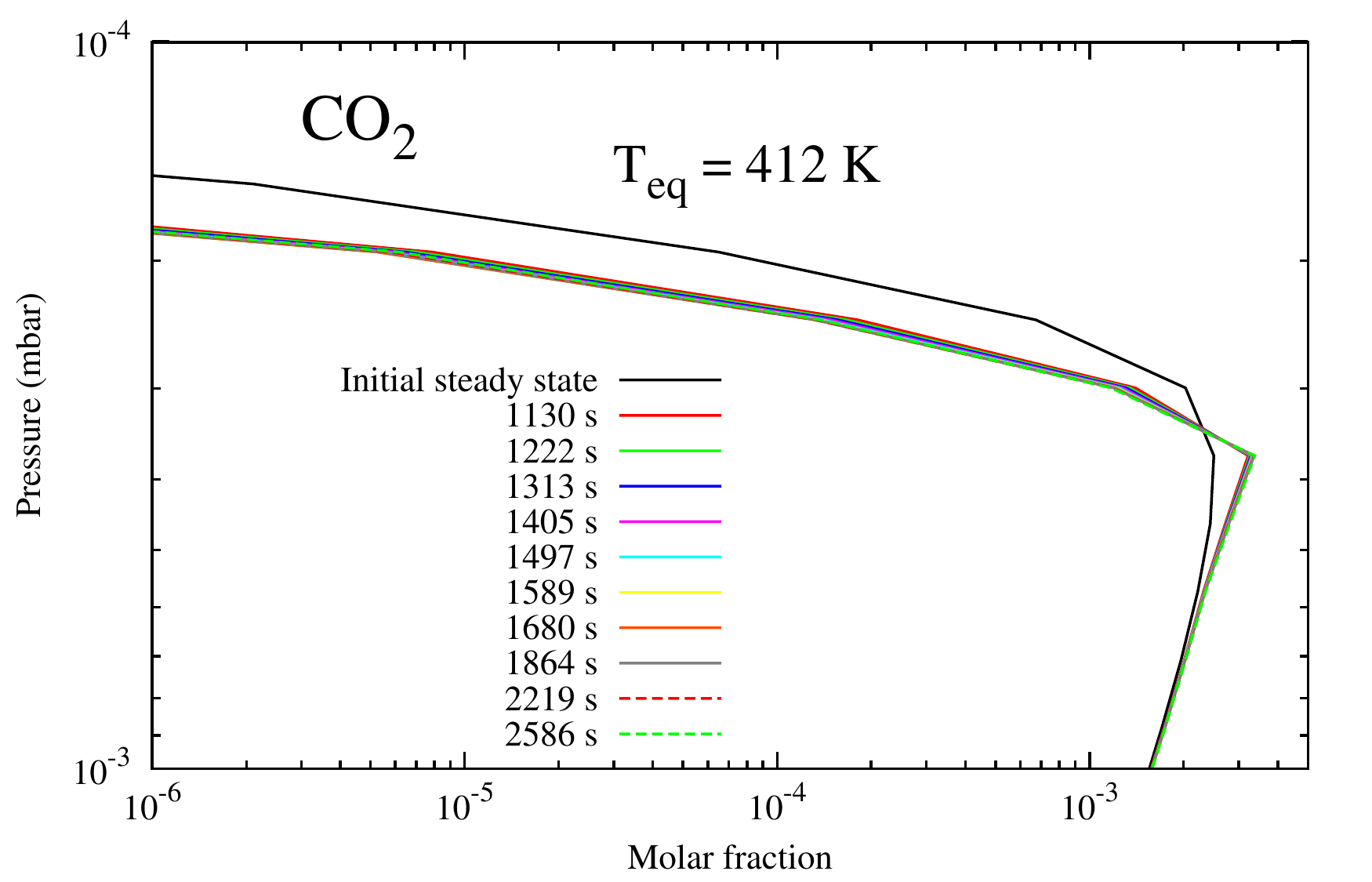}\\
\includegraphics[angle=0,width=\columnwidth]{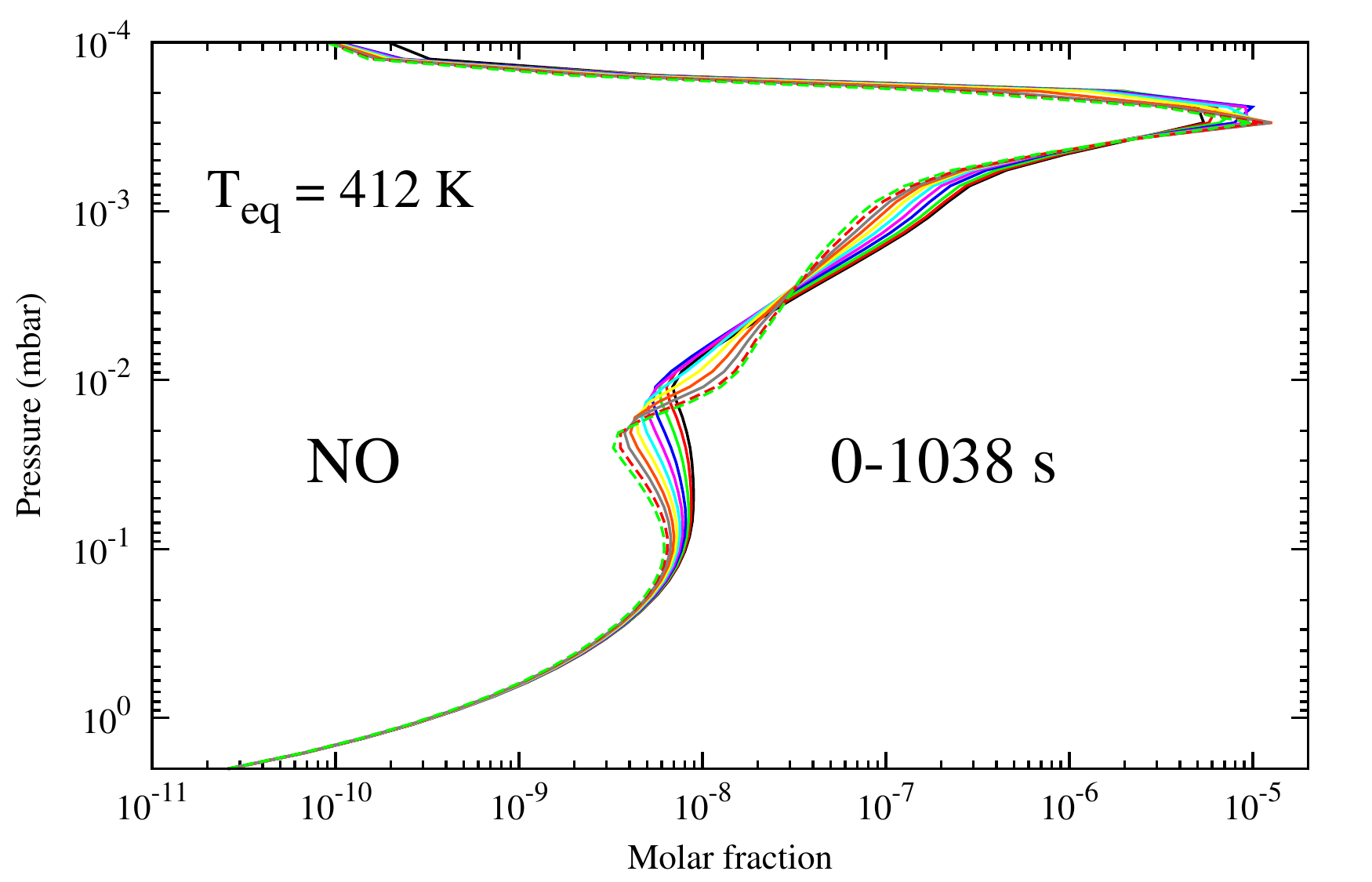}
\includegraphics[angle=0,width=\columnwidth]{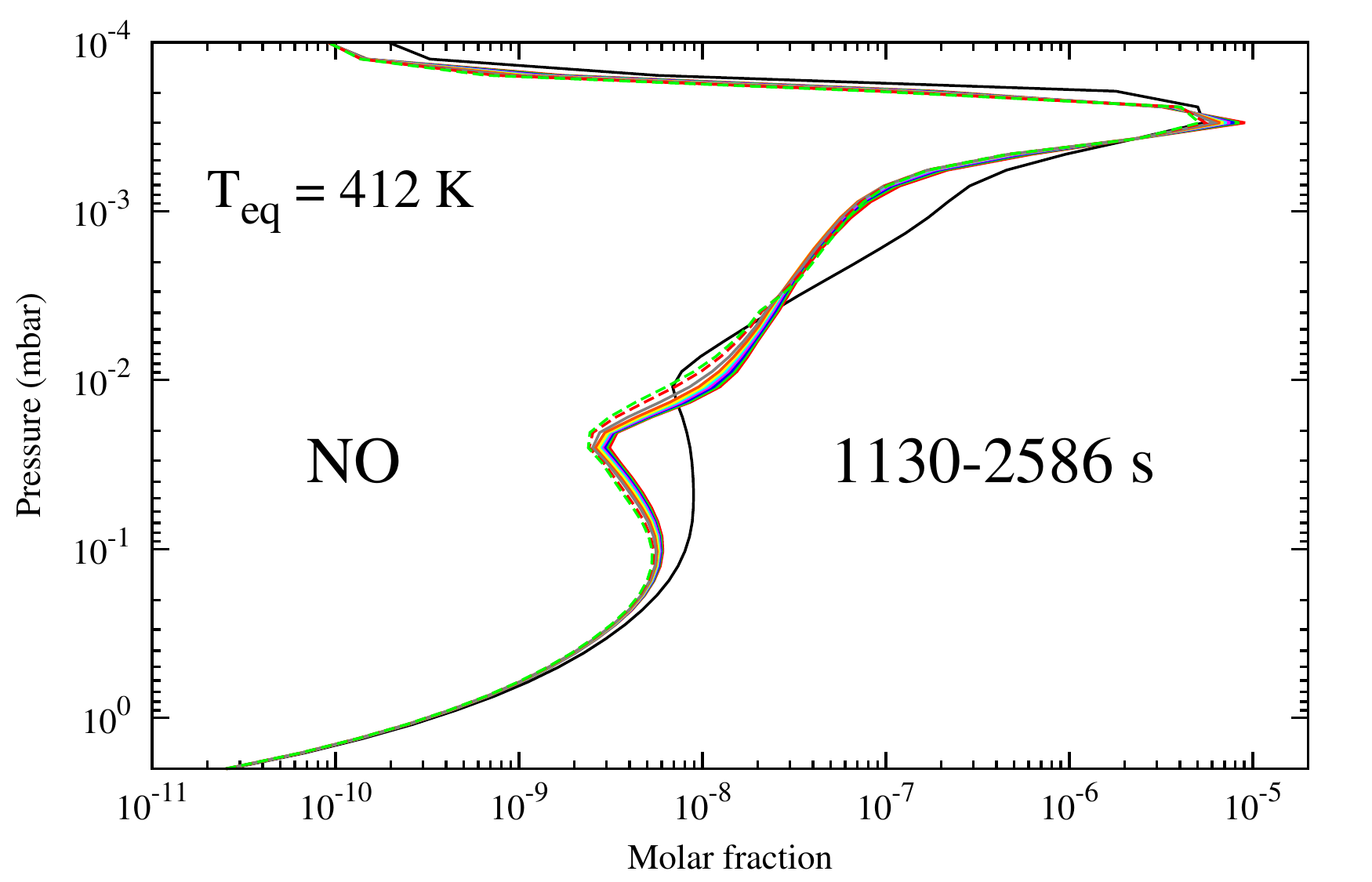}\\
\includegraphics[angle=0,width=\columnwidth]{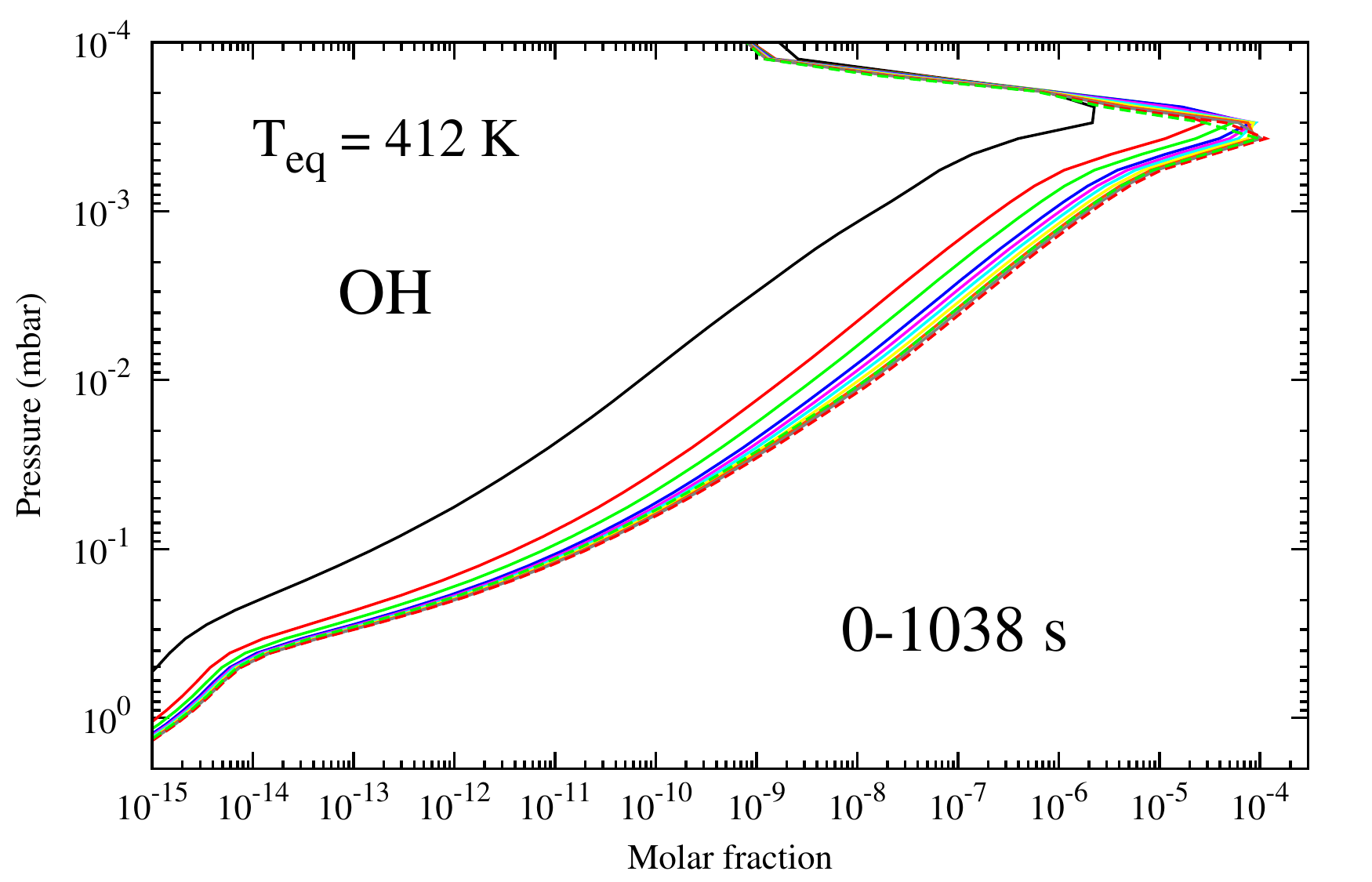}
\includegraphics[angle=0,width=\columnwidth]{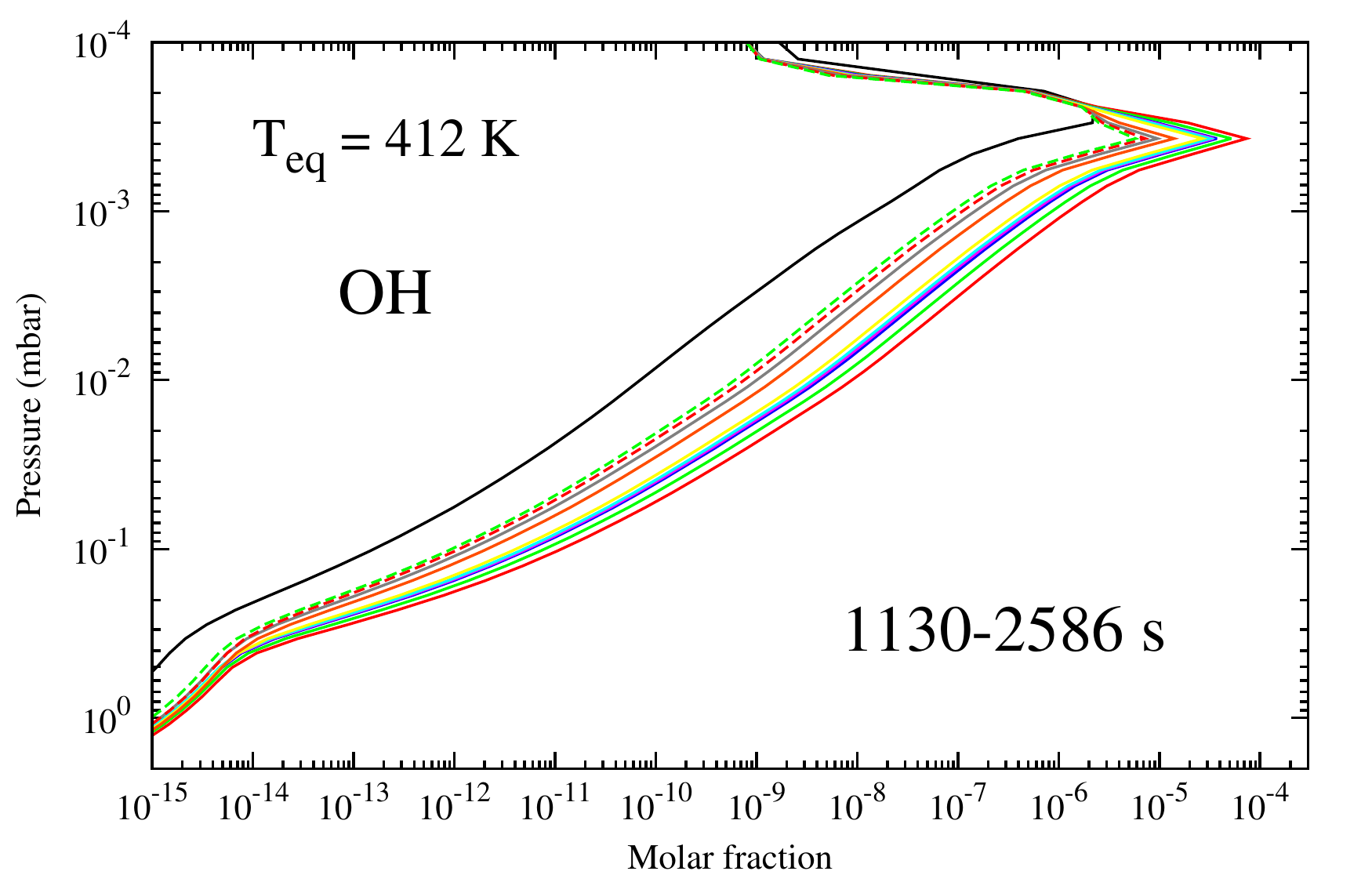}\\
\caption{Evolution of CO$_2$, NO and OH mixing ratios during the flare event with the thermal profile corresponding to $T_{eq}$ = 412K. The legend for all figures is in the upper panels.} \label{fig:412K_NO_OH}
\end{figure*}

\begin{figure*}[!htb]
\centering
\includegraphics[angle=0,width=\columnwidth]{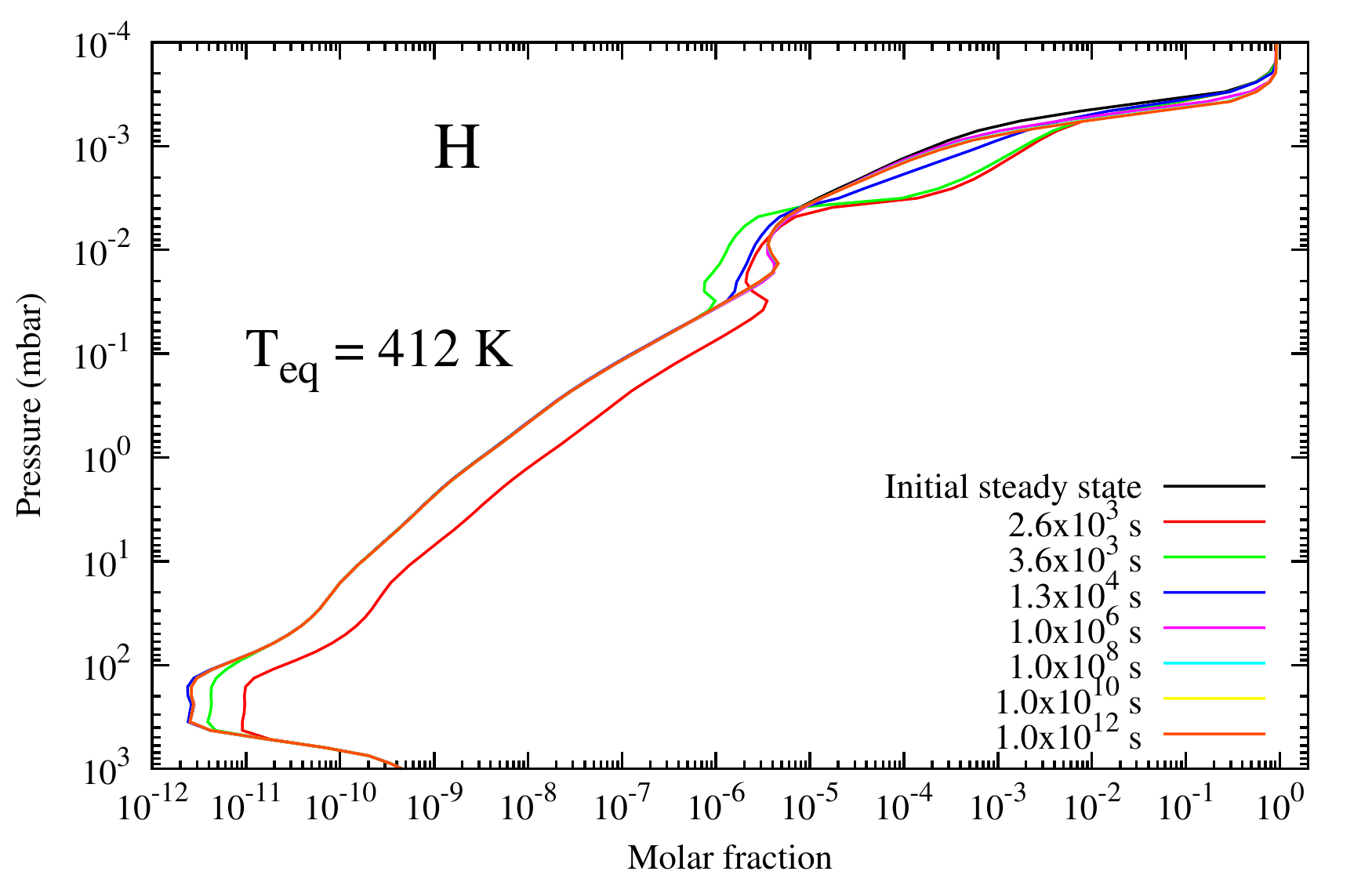}
\includegraphics[angle=0,width=\columnwidth]{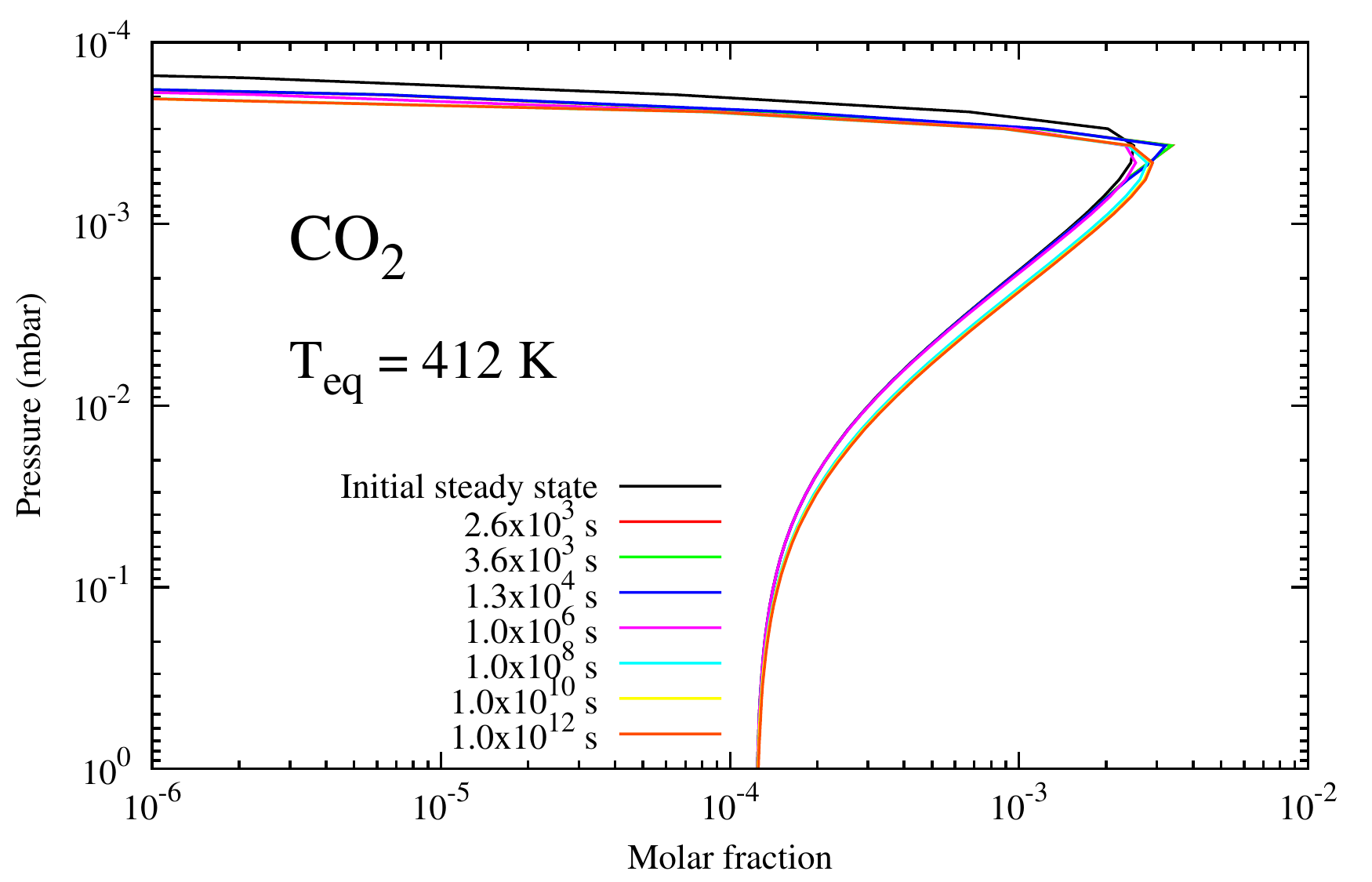}\\
\includegraphics[angle=0,width=\columnwidth]{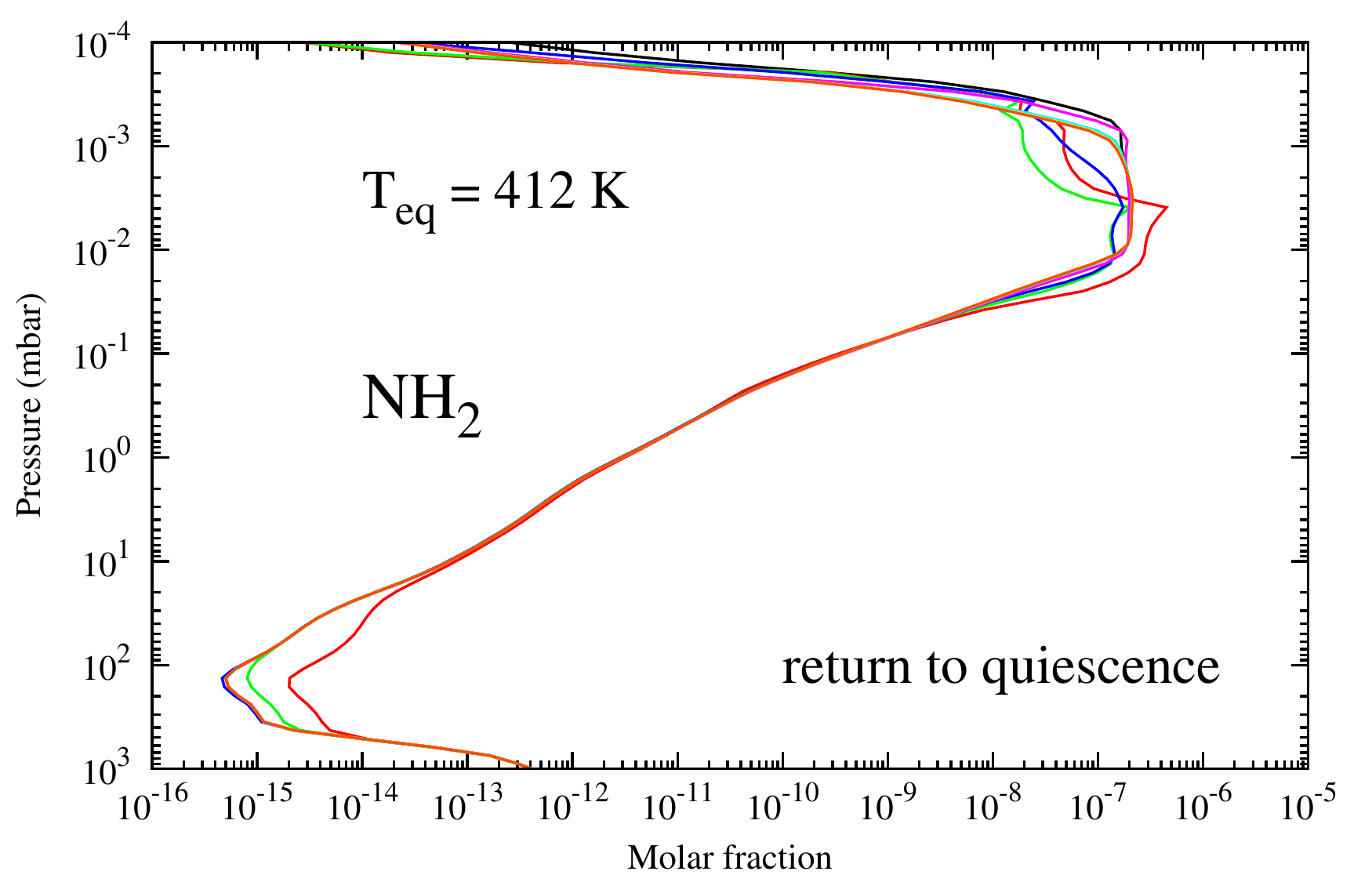}
\includegraphics[angle=0,width=\columnwidth]{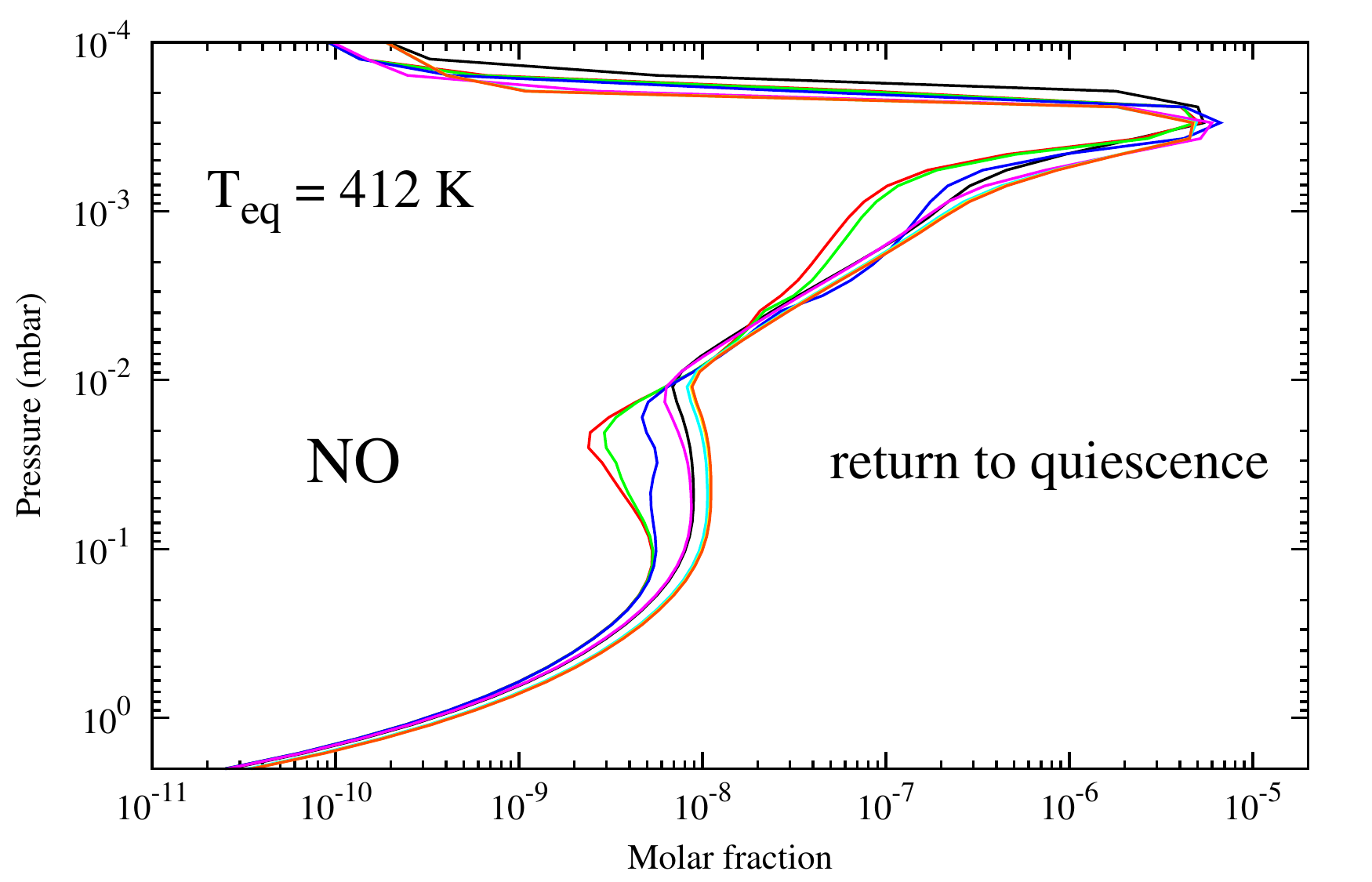}\\
\includegraphics[angle=0,width=\columnwidth]{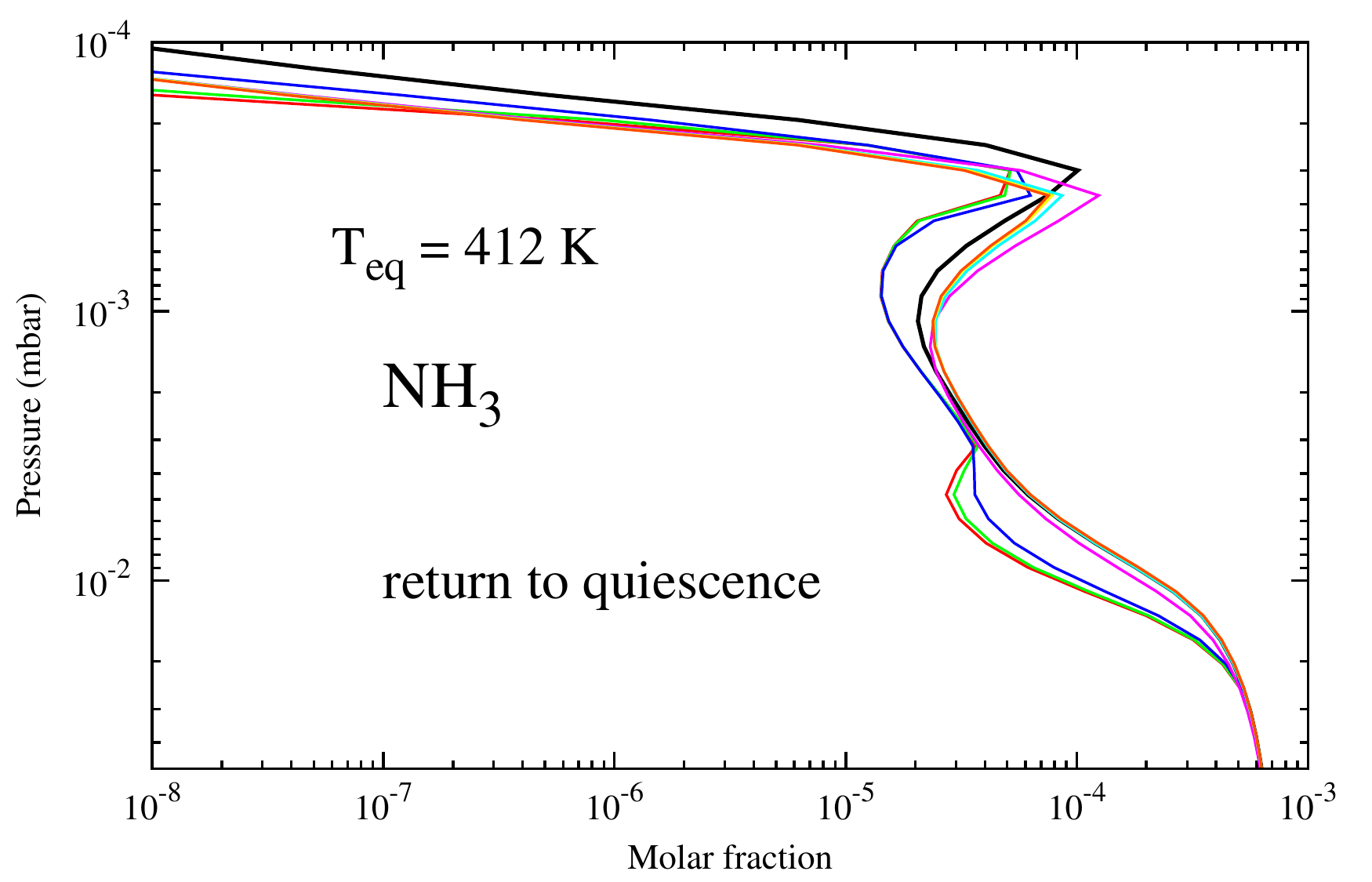}
\includegraphics[angle=0,width=\columnwidth]{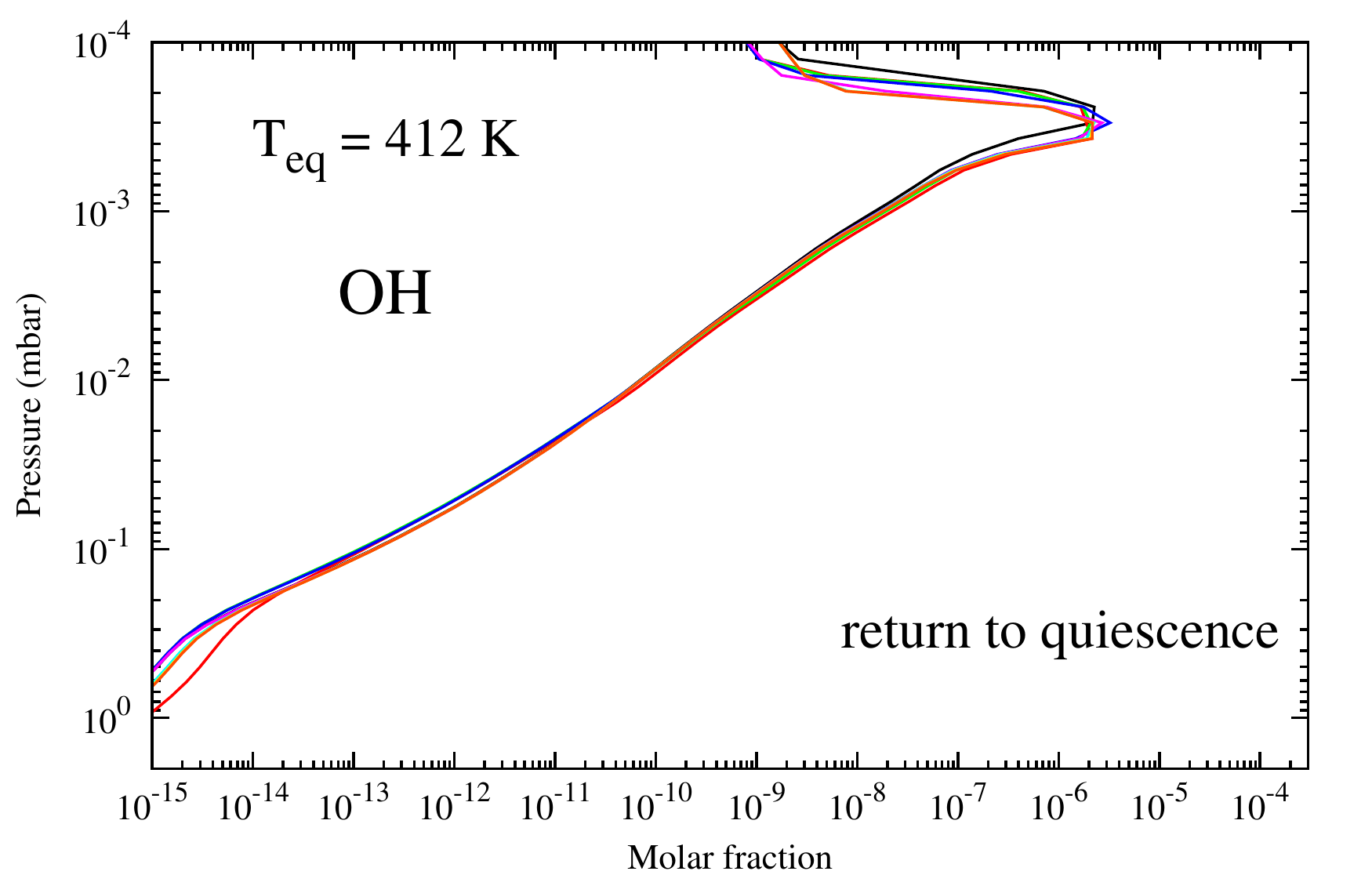}\\
\caption{Evolution of H, NH$_2$, NH$_3$, CO$_2$, NO, and OH mixing ratios after the flare event with the thermal profile corresponding to $T_{eq}$ = 412K. The legend for all figures is in the upper panels.} \label{fig:412K_return}
\end{figure*}

\begin{figure*}[!htb]
\centering
\includegraphics[angle=0,width=\columnwidth]{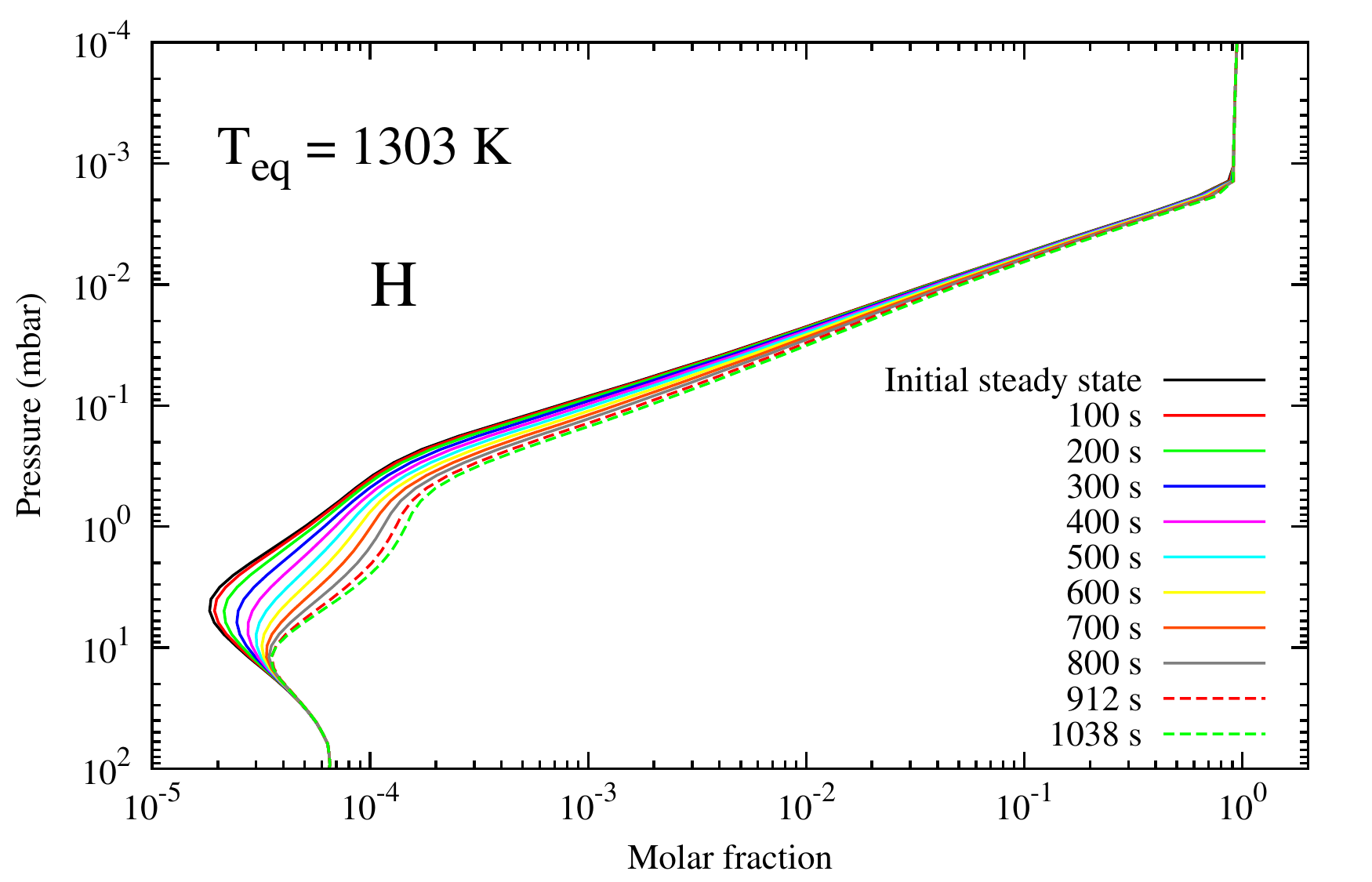}
\includegraphics[angle=0,width=\columnwidth]{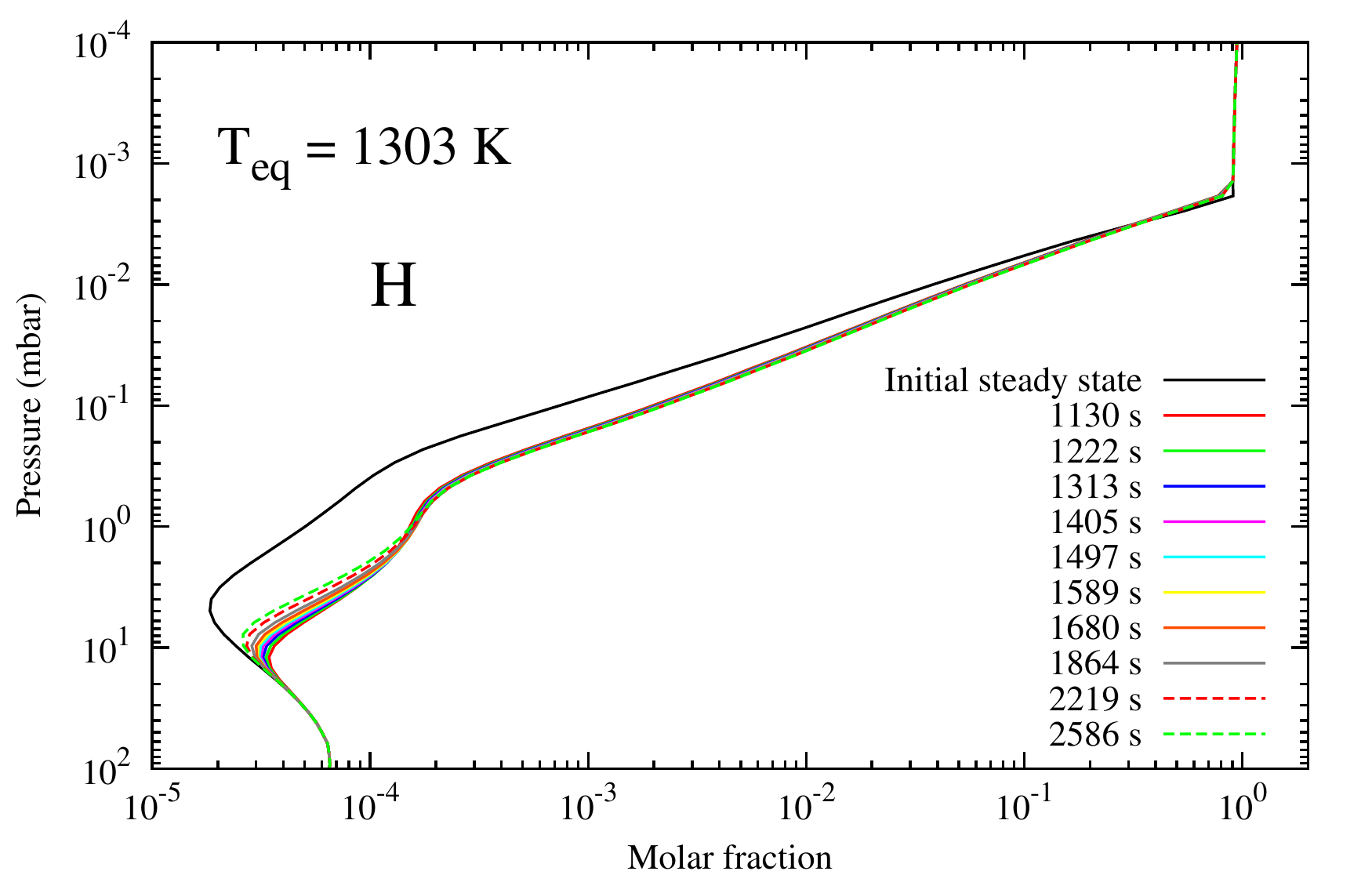}\\
\includegraphics[angle=0,width=\columnwidth]{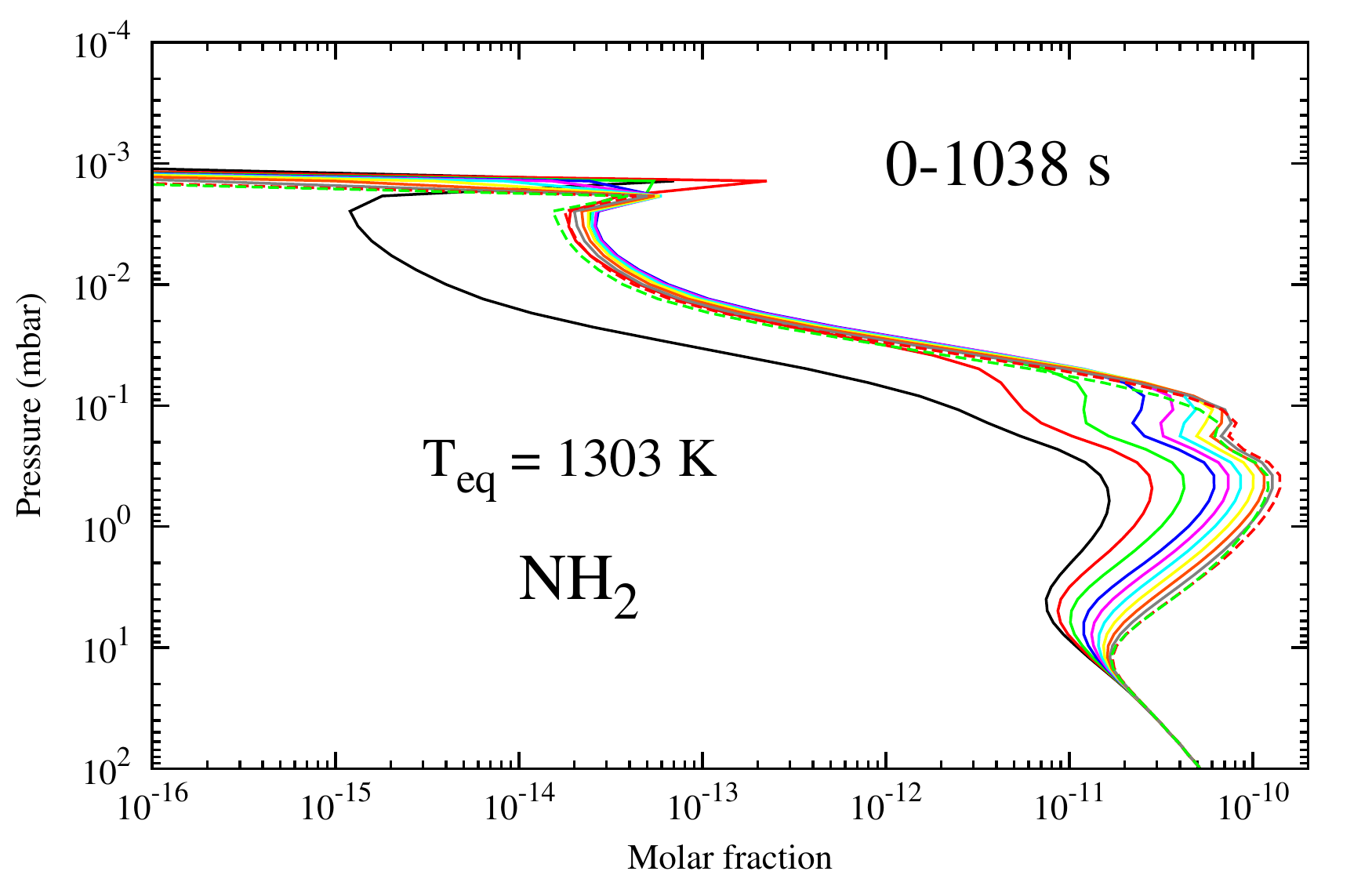}
\includegraphics[angle=0,width=\columnwidth]{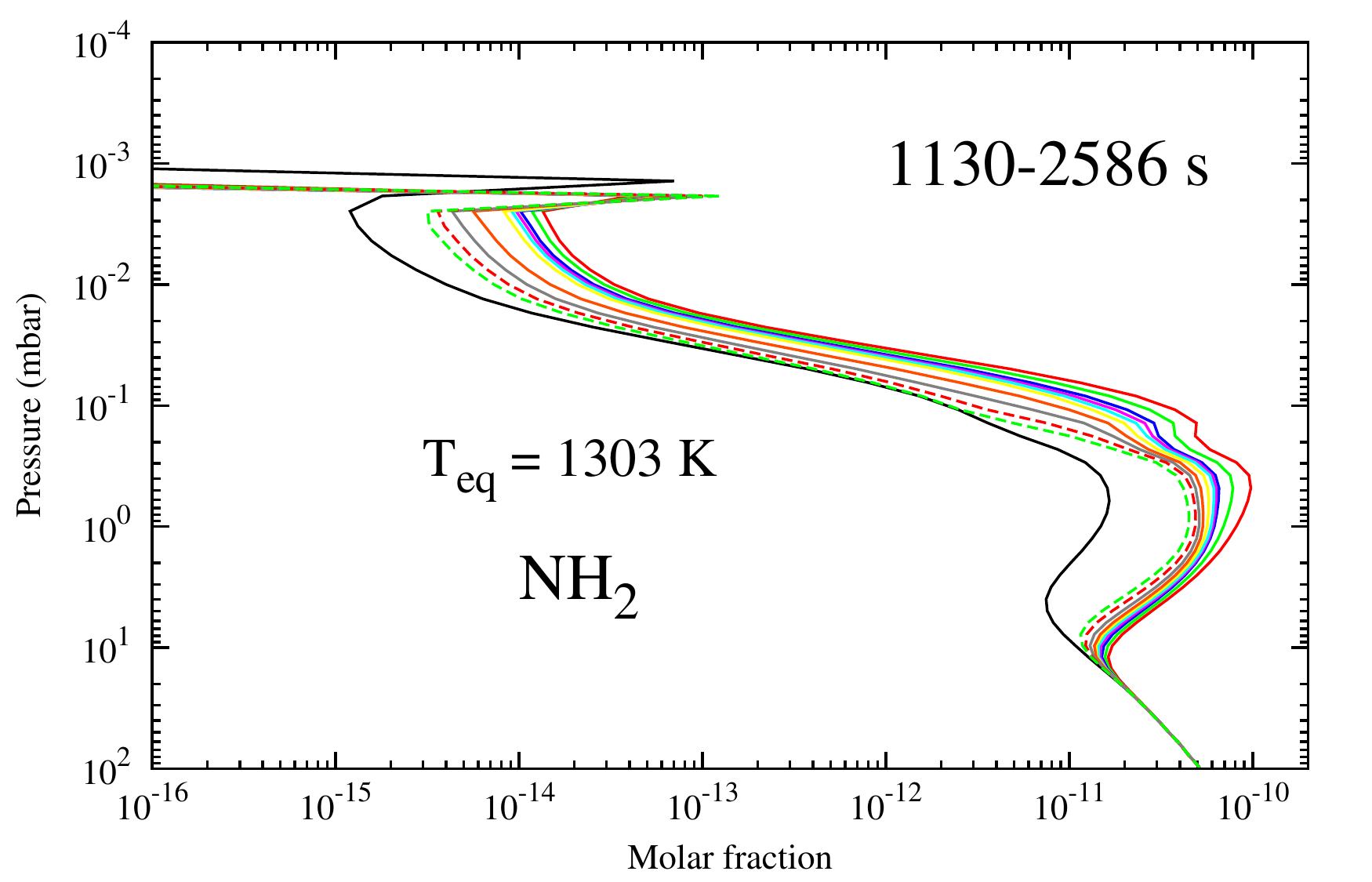}\\
\includegraphics[angle=0,width=\columnwidth]{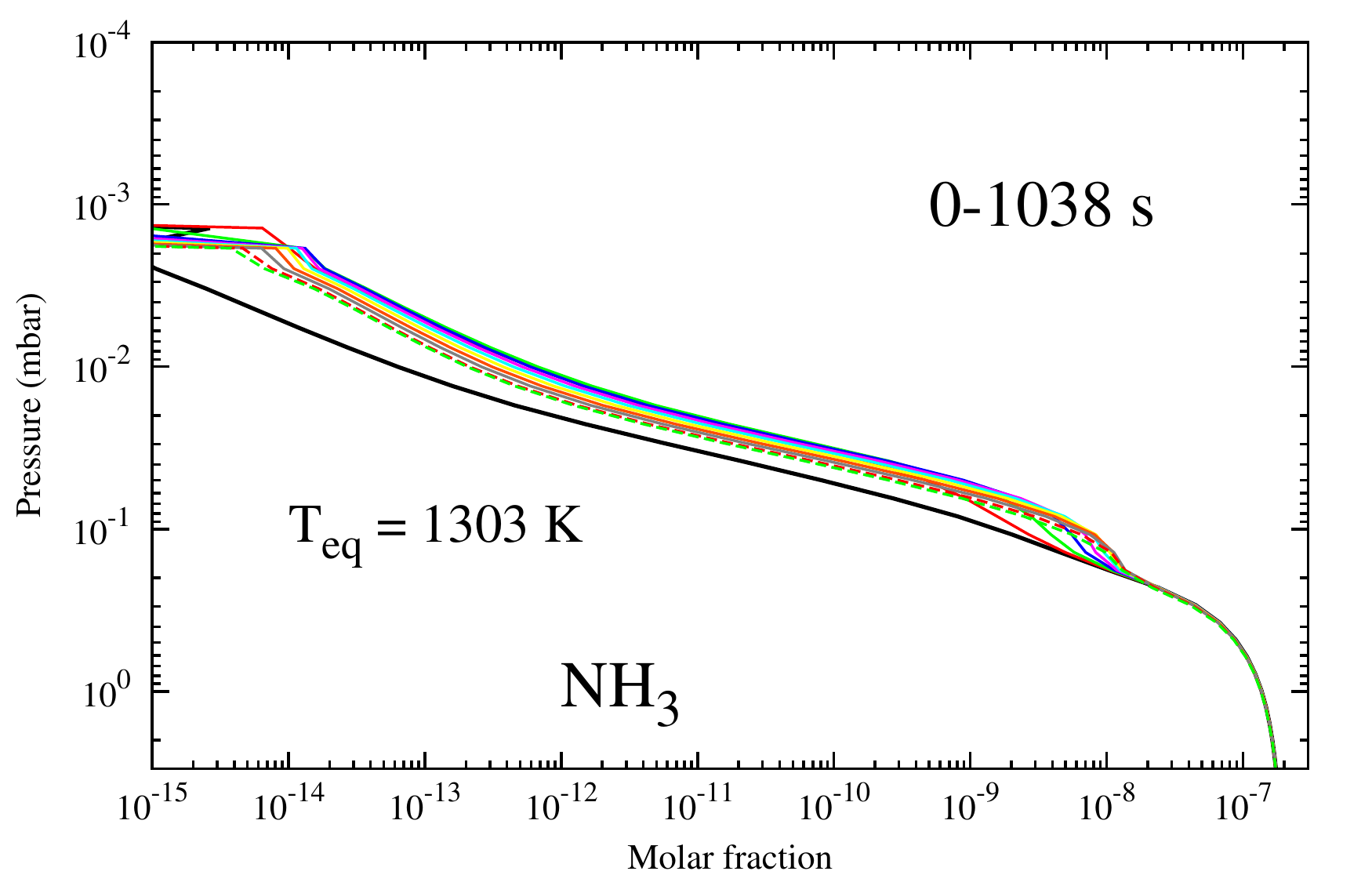}
\includegraphics[angle=0,width=\columnwidth]{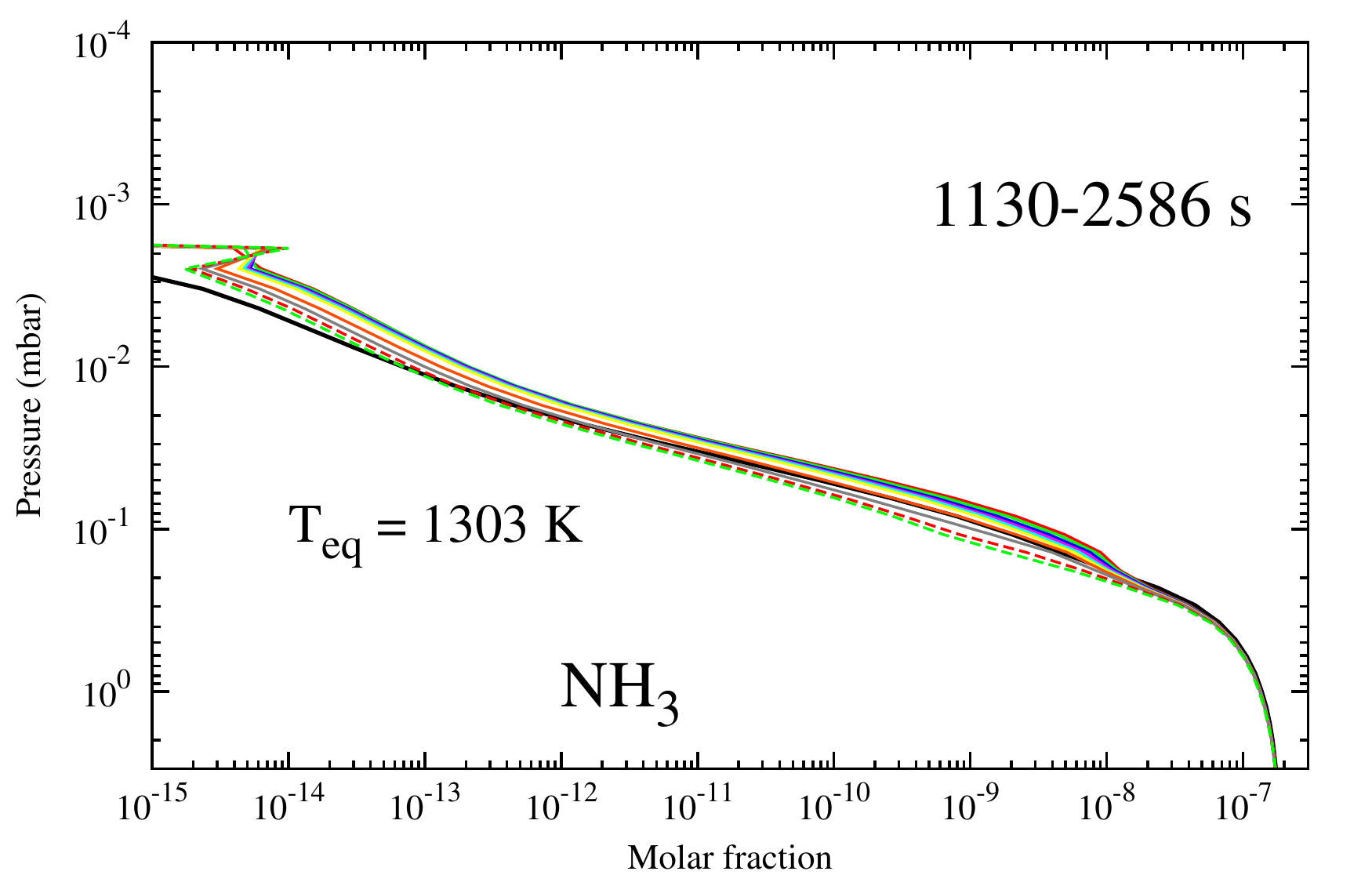}\\
\caption{Evolution of H, NH$_2$ and NH$_3$ mixing ratios during the flare event with the thermal profile corresponding to $T_{eq}$ = 1303K. The legend for all figures is in the upper panels.} \label{fig:1303K_NH3_H}
\end{figure*}

\begin{figure*}[!htb]
\centering
\includegraphics[angle=0,width=\columnwidth]{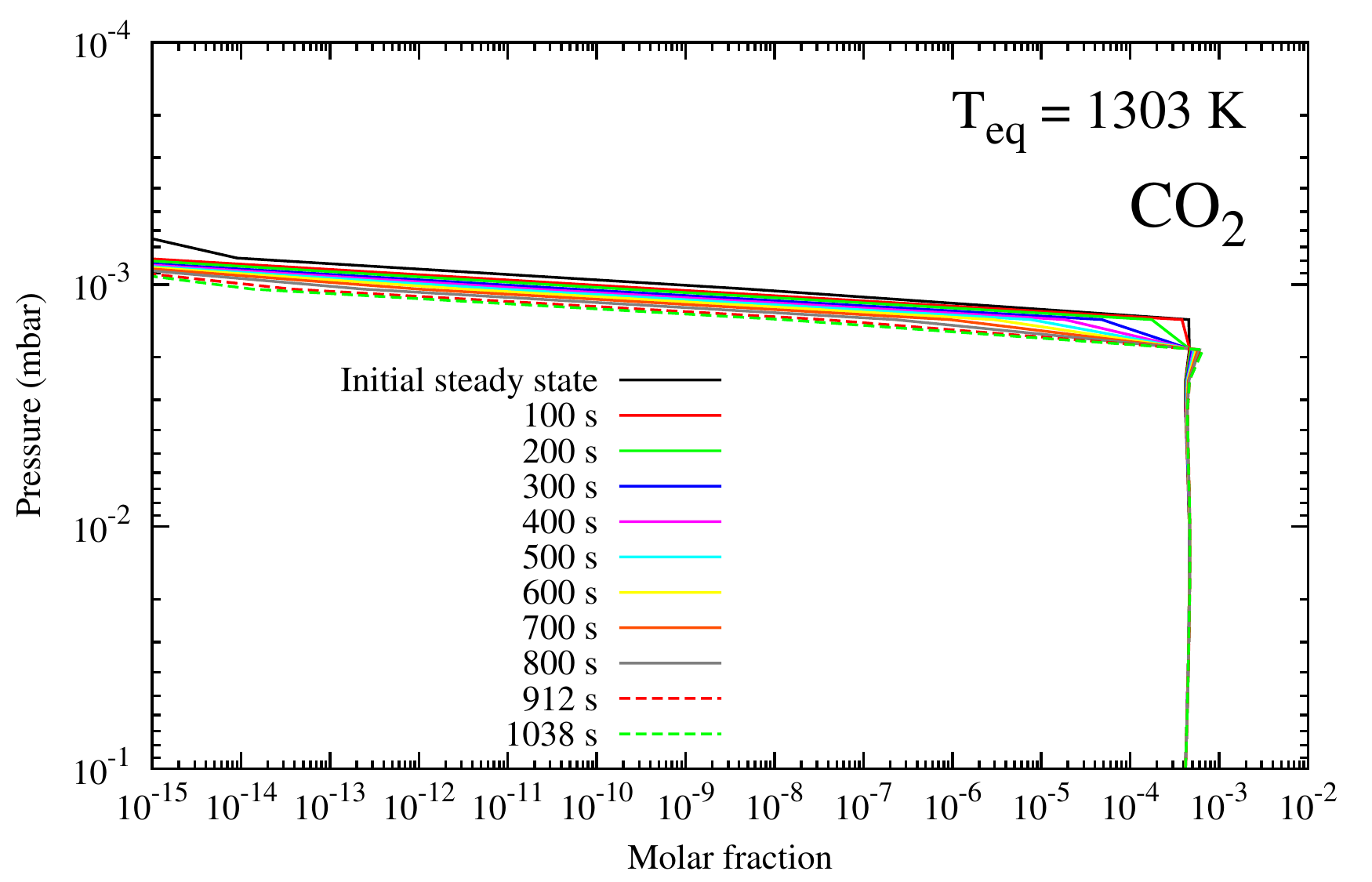}
\includegraphics[angle=0,width=\columnwidth]{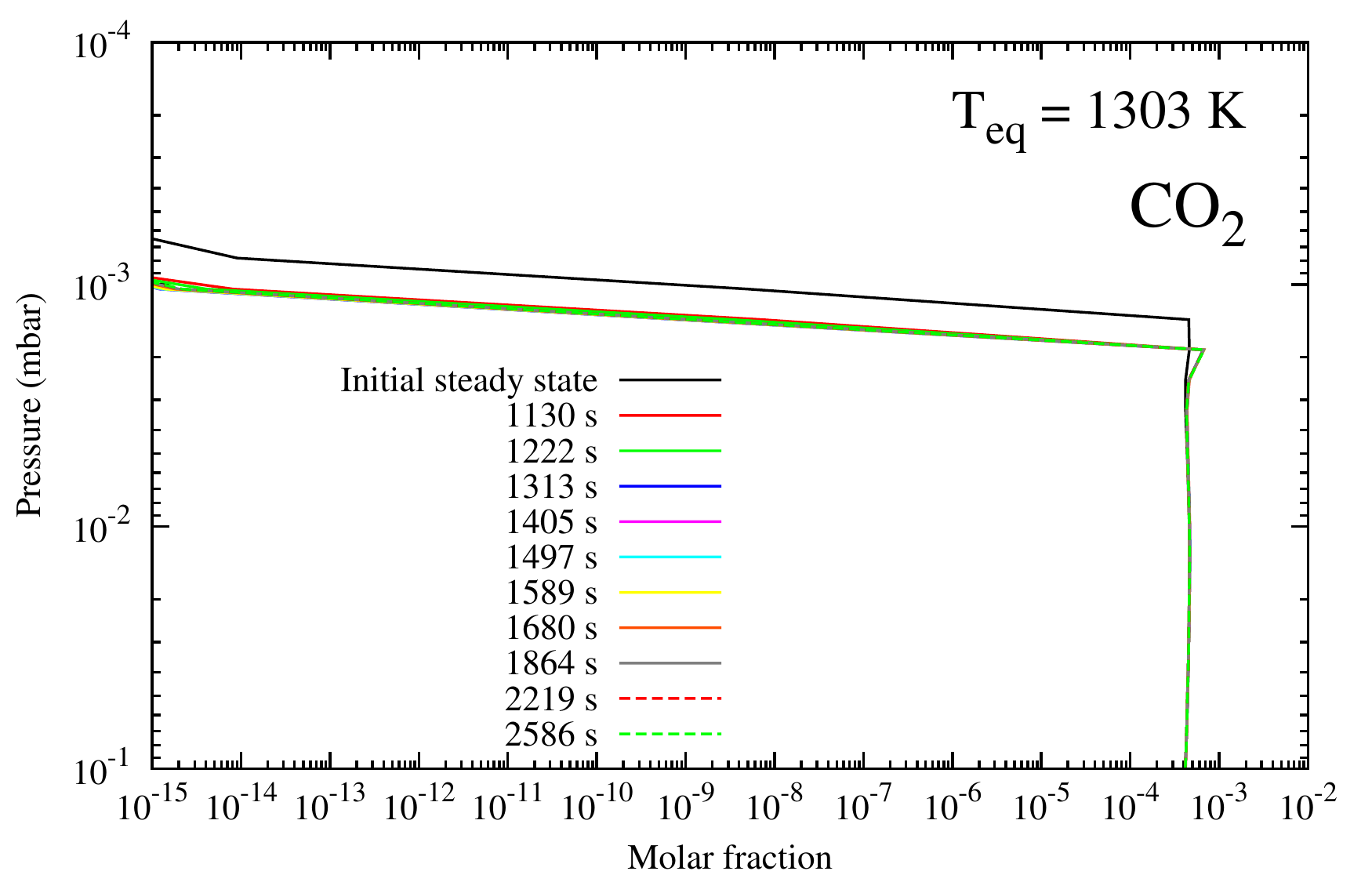}\\
\includegraphics[angle=0,width=\columnwidth]{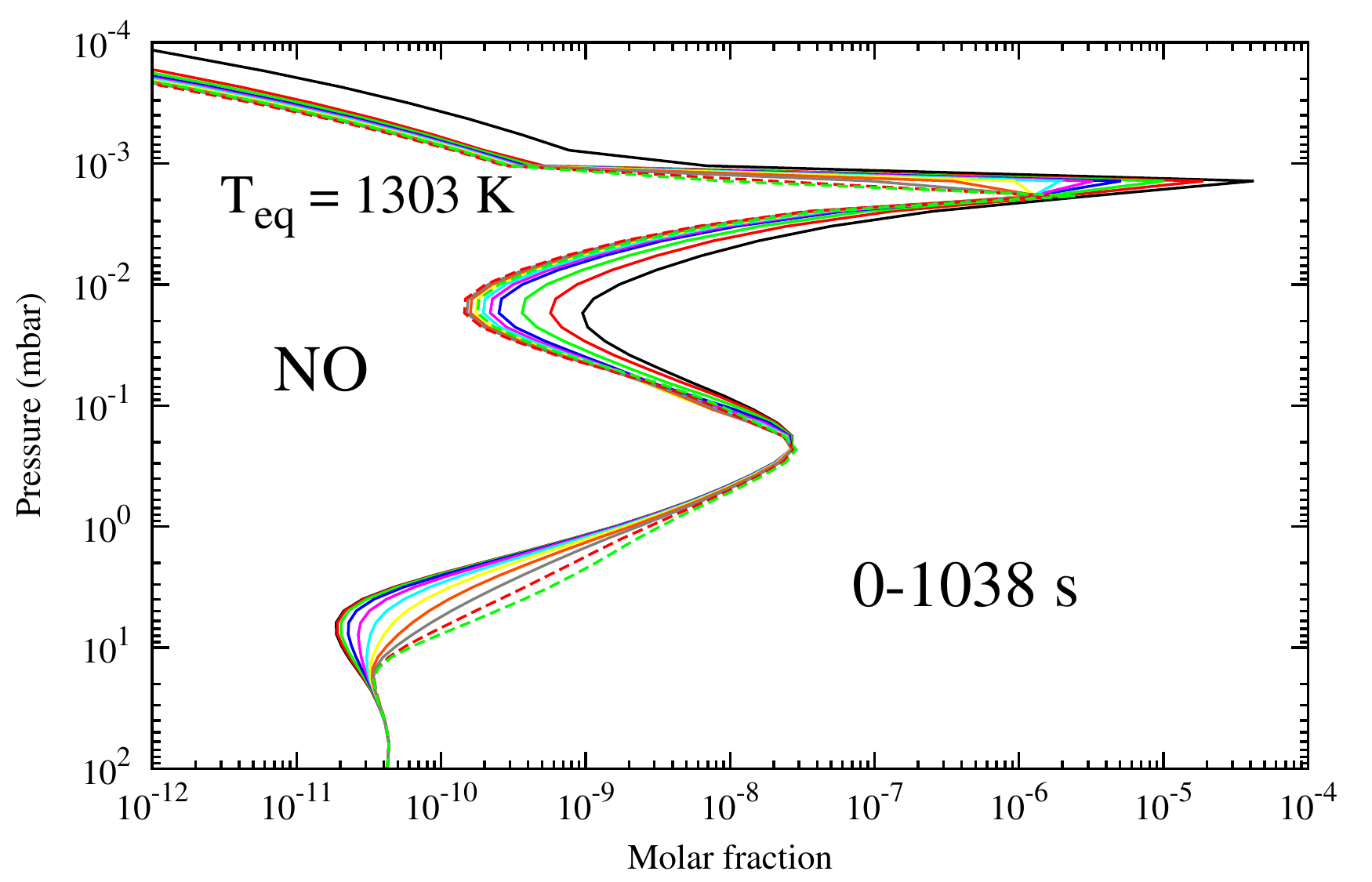}
\includegraphics[angle=0,width=\columnwidth]{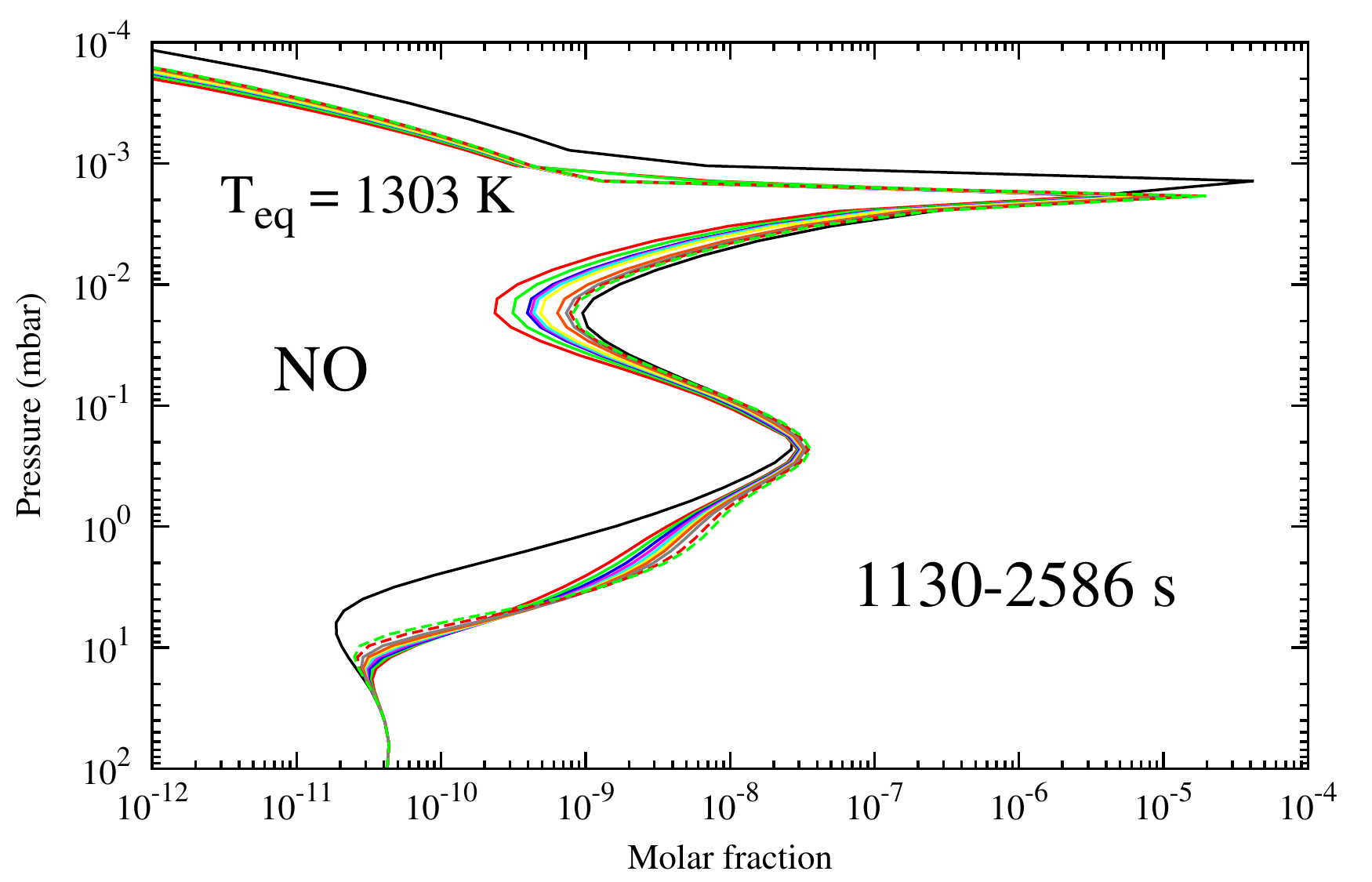}\\
\includegraphics[angle=0,width=\columnwidth]{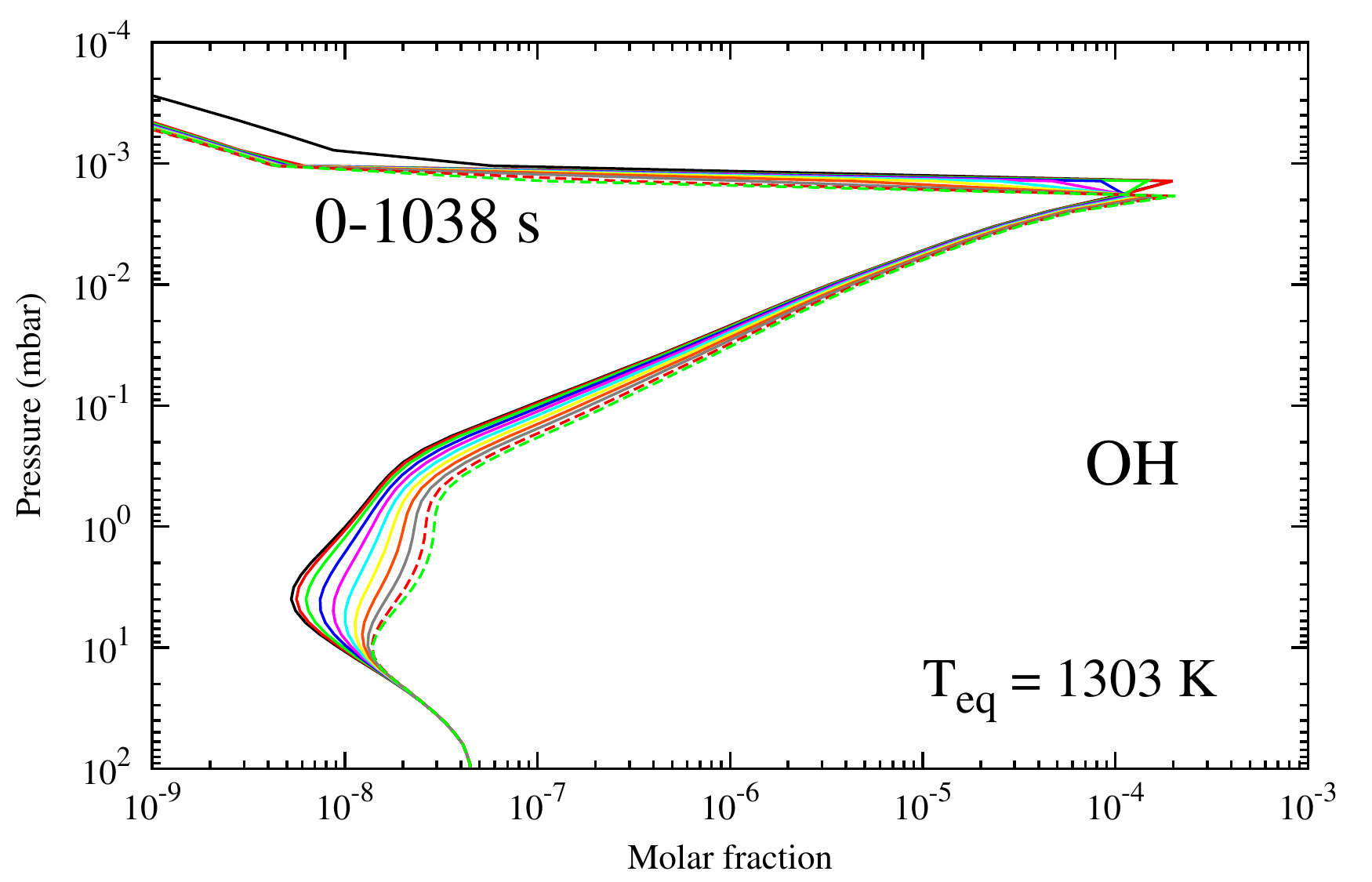}
\includegraphics[angle=0,width=\columnwidth]{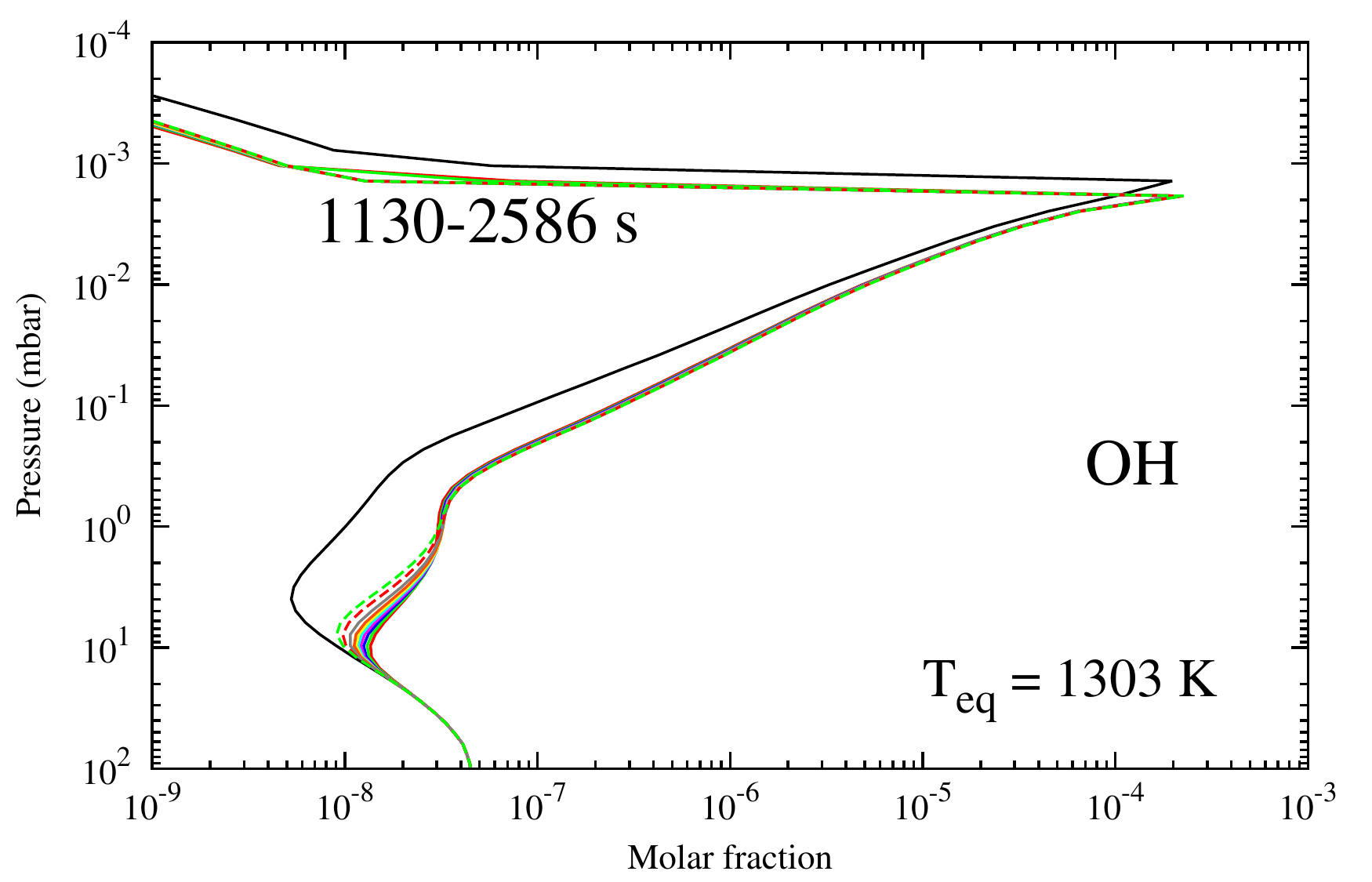}\\
\caption{Evolution of CO$_2$, NO and OH mixing ratios during the flare event with the thermal profile corresponding to $T_{eq}$ = 1303K. The legend for all figures is in the upper panels.} \label{fig:1303K_NO_OH}
\end{figure*}

Although many species are impacted by the increase of the UV irradiation due to the flare, we describe in the following the evolution of the mixing ratios of only six of them: hydrogen (H), the amidogen radical (NH$_2$), ammonia (NH$_3$), carbon dioxide (CO$_2$), nitric oxide (NO), and the hydroxyl radical (OH). These species are among the most abundant species and undergo deeper and more significant changes than the other chemical species in our model. Moreover, three of them are considered in our radiative transfer code to calculate synthetic spectra. Their abundances during the different phases of the flare are shown on Figs.~\ref{fig:412K_NH3_H}, \ref{fig:412K_NO_OH}, \ref{fig:1303K_NH3_H}, and \ref{fig:1303K_NO_OH}. For both thermal profiles, most species see their mixing ratios globally increase during the flare event, with a maximum abundance around 912 s (corresponding to the peak of the stellar flux). Exceptions are NH$_3$, CO$_2$, and NO, which can have a lower abundance during the stellar flare than at steady state. The other species considered in this calculation, but not plotted (H$_2$O, HCN, and CH$_4$), experienced changes only in the very upper atmosphere (P $<10^{-1}$ mbar and  P $<4\times10^{-4}$ mbar for the 1303 K and 412 K profiles, respectively). Their abundances globally decrease by $\sim 1 - 5$ orders of magnitude. The abundances after steady state, during the ``return to quiescence" phase are shown on Figs.~\ref{fig:412K_return} and \ref{fig:1303K_return}. Figure~\ref{fig:new-init-steady} shows a comparison between the initial and final steady states.

In several instances, the most important reactions controlling changes in species concentrations were identified. This was achieved using a module of our chemical pathway analysis software suite, which essentially keeps a tally of the cumulative reaction flux for each reaction in the scheme, such that the total and net reaction flux for any species over any time period can be identified. The net reaction fluxes were checked for equality against net changes in concentration.
 
\subsubsection{Thermal profile T$_{eq}$ = 412 K}

\paragraph{Hydrogen:}
The abundance of this species changes between $3\times10^{-4}$ and $7\times10^2$ mbar. At higher or lower pressures, the mixing ratio of hydrogen (y$_H$) remains identical to that at the steady state. Globally, during the impulsive phase (0 -- 912 s), the amount of H increases with time (by a factor 25 at 1 mbar). We note that between $3\times10^{-3}$ and $3\times10^{-2}$ mbar, the evolution of the mixing ratio does not evolve linearly with time, and one can notice some overlap and crossing between the different temporal profiles. This is caused by the peak of destruction, being in-between these two pressures, that moves downward to higher pressures (lower altitudes). The reactions responsible for the increase of y$_H$ depend on the pressure level. For pressures lower than $10^{-2}$ mbar, H$_2$O + h$\nu$ $\rightarrow$ H + OH and OH + H$_2$ $\rightarrow$ H + H$_2$O are the main production pathways. Around $10^{-2}$ mbar, the photodissociation NH$_3$ + h$\nu$ $\rightarrow$ NH$_2$ + H is the principal source of hydrogen, with a contribution of the two aforementioned reactions. Then, around 10 mbar, the main reactions producing hydrogen are NH$_2$ + H$_2$ $\rightarrow$ NH$_3$ + H and NH$_3$ + h$\nu$ $\rightarrow$ NH$_2$ + H, which form a limited catalytic cycle. Between 912~s and 2586 s, we see a decrease of the abundance of H, towards the initial steady state value. This is due to the decrease in efficiency of the production routes, which can no longer compete with the destruction pathways: N$_2$H$_2$ + H (+ M) $\rightarrow$ N$_2$H$_3$ (+ M) and N$_2$H$_3$ + H  $\rightarrow$ N$_2$H$_2$ + H$_2$ for P$\sim$ $10^{-2}$ mbar, and CH$_3$ + H (+ M) $\rightarrow$ CH$_4$ (+ M) and CH$_4$ + H  $\rightarrow$ CH$_3$ + H$_2$ for P$>$0.1 mbar.

\paragraph{Amidogen:}
This species experiences abundance variations from the top of the atmosphere down to $6\times10^2$ mbar. The mixing ratio of NH$_2$ increases during the impulsive phase and decreases after 912 s. Its time evolution is similar to that of hydrogen, since this species is mainly produced by one of the reactions responsible for the increase of H namely, NH$_3$ + h$\nu$ $\rightarrow$ NH$_2$ + H. We also note the important contribution of the reaction N$_2$H$_3$ + H $\rightarrow$ NH$_2$ + NH$_2$ in the production of amidogen. The reactions that destroy NH$_2$ are NH$_2$ + H (+ M) $\rightarrow$ NH$_3$ (+ M), NH$_2$ + NH$_2$ (+ M) $\rightarrow$ N$_2$H$_4$ (+ M), and NH$_2$ + H$_2$ $\rightarrow$ NH$_3$ + H. Their relative efficiencies depend on the pressure level and on the phase of the flare (their contributions become more important after 912 s). Interestingly, we note that at very low pressure ($\sim$ 4$\times$10$^{-4}$ mbar), the decrcease of NH$_2$ after 912 s is due to the reaction NH$_2$ + OH $\rightarrow$  NH$_2$OH that becomes dominant because of the increase of y$_{OH}$.

\paragraph{Ammonia:}
Unlike the other species considered here, ammonia globally decreases with time during the first 1038 seconds, down to $3\times10^{-2}$ mbar. At $5\times10^{-3}$ mbar, y$_{NH_3}$ decreases by a factor of 3. At lower pressures (P$\sim$1.5$\times$10$^{-4}$mbar), the abundance of ammonia decreases by a factor of $\sim$20. The loss of NH$_3$ is due mainly to the photodissociation, NH$_3$ + h$\nu$ $\rightarrow$ NH$_2$ + H. At some levels, we also note the contribution of other reactions: NH$_3$ + OH $\rightarrow$ NH$_2$ + H$_2$O at very low pressure ($\sim$ 4$\times$10$^{-4}$ mbar),  and NH$_3$ (+ M) $\rightarrow$ NH$_2$ + H (+ M) around 3$\times$10$^{-4}$ mbar.
From 1130 s, the evolution of NH$_3$ is more moderate. Around $4\times10^{-4}$ mbar, and for $8\times10^{-4}<P<9\times10^{-3}$ mbar, y$_{NH_3}$ increases together with time. This increase (which is most visible around 3$\times10^{-3}$ mbar) is due to the reaction NH$_2$ + H (+ M) $\rightarrow$ NH$_3$ (+ M). Apart from these pressures, the change in mixing ratio of ammonia is relatively small: we only observe a slight decrease between $3\times10^{-4}$ and $2\times10^{-4}$ mbar.


\paragraph{Carbon dioxide:}
CO$_2$ is affected by the stellar flare only in the very upper atmosphere down to $6\times10^{-4}$ mbar. 
In the first part of the flare, we observe two different evolutions for the abundance of this species. For pressures lower than $3\times10^{-4}$ mbar, it decreases with time (a factor  $\sim$15 at $\sim$1.5$\times$10$^{-4}$ mbar). Then, for pressures between $3\times10^{-4}$ and $6\times10^{-4}$ mbar, CO$_2$ sees its abundance slightly increasing (by a factor $\sim$1.3). During the second part of the flare, the abundance of CO$_2$ hardly changes. We observe a very small decrease of y$_{CO_2}$ for $P<3.5\times10^{-4}$ mbar. The loss of CO$_2$ is largely due to the photodissociation of CO$_2$. The branching to O($^1$D) is dominant throughout, but the relative yield of O($^3$P) increases at lower pressures ($\sim$ 4$\times$10$^{-4}$ mbar). The production of carbon dioxide is due to CO + OH $\rightarrow$  CO$_2$ + H. We will see in Sect. \ref{sect:spectra} that the variation of abundance of this species is quite small but sufficient to affect the synthetic spectra.

\paragraph{Nitric oxide:}
The abundance of this species changes from the top of the atmosphere down to 0.5 mbar. We observe different regimes depending on the pressure. During the first 1038 s, for pressures lower than $2.5\times10^{-4}$ mbar, y$_{NO}$ decreases slightly with time from 0 to 1038~s. The decrease is mainly due to the photodissociation NO + h$\nu$ $\rightarrow$ N($^4$S) + O($^3$P), which is somewhat tempered by the formation reactions N($^4$S) + OH $\rightarrow$ NO + H. We also observe a decrease of the abundance of NO for $4\times10^{-4} <P<3\times10^{-3}$ mbar (up to a factor 3), and for $2\times10^{-2}<P<0.5 $ mbar (up to a factor 3 also) which is due to the reaction NH$_2$ + NO $\rightarrow$ N$_2$ + H$_2$O. In between these two regimes, y$_{NO}$ increases with time because it is produced by HNO + H $\rightarrow$ NO + H$_2$. In the second part of the flare, the abundance of NO varies less. There is a small decrease of its abundance between $3\times10^{-3}$ and 0.5 mbar, and also around 2$\times10^{-4}$ mbar.

\paragraph{Hydroxyl radical:}
The abundance of OH varies from the top of the atmosphere down to $\sim$500 mbar. From 0 to 912 s, we globally observe an increase of the abundance of OH (by a factor 100 at $10^{-3}$ mbar), except for pressures lower than $3\times10^{-4}$mbar. From 912 to 2586 s, we observe a decrease of the abundance of OH towards the initial steady state. This species evolves similarly to hydrogen because the abundance of OH is highly connected to H through the dominant forming reaction H$_2$O + h$\nu$ $\rightarrow$ H + OH. The reaction that mainly destroys OH is OH + H$_2$ $\rightarrow$ H + H$_2$O.
 
\subsubsection{Thermal profile T$_{eq}$ = 1303 K}

\paragraph{Hydrogen:}
The amount of hydrogen varies from 30 mbar up to $10^{-3}$ mbar. y$_H$ increases significantly up to 1038 s. The maximum enhancement occurs at 3 mbar (a factor $\sim$4.5). The reactions responsible for the increase of hydrogen are different depending on the level pressure. We found that in the very upper atmosphere (down to $10^{-1}$ mbar), the increase of y$_H$ is due to the photodissociation of water (H$_2$O + h$\nu$ $\rightarrow$  H + OH) and to the reaction OH + H$_2$ $\rightarrow$ H + H$_2$O. At higher pressures ($\sim$1 mbar), the production of hydrogen is due mainly to NH$_2$ + H$_2$ $\rightarrow$ NH$_3$ + H and NH$_3$ + h$\nu$ $\rightarrow$ NH$_2$ + H, with a small contribution of CO + OH $\rightarrow$ CO$_2$ + H. Around 3 mbar (where we observe the maximum change in H abundance), the reactions that produce the most hydrogen are OH + H$_2$ $\rightarrow$ H + H$_2$O, OH + H$_2$ $\rightarrow$ OH + H, and CN + H$_2$ $\rightarrow$ HCN + H.\\
From 1130 to 2586 s, the variations of the abundances are small. We observe a transition between two regimes around 0.5 mbar. For smaller pressures, y$_H$ keeps increasing slightly (by a factor less than $\sim$1.25 between 1130 s and 2586 s), but for higher pressures, the amount of hydrogen decreases with time (by a factor $\sim$1.5 at 10 mbar).

\paragraph{Amidogen:}
As for H, the abundance of NH$_2$ is affected down to 30 mbar. We observe three regimes of variations. For pressures greater than $\sim$0.2 mbar, the abundance increases with time (up to 912 s). 
Between $1\times10^{-3}$ and $2\times10^{-3}$ mbar, there is a production peak of NH$_2$. Here and above, after a sudden increase of y$_{NH_2}$ (by a factor $\sim$3 at $1\times10^{-2}$mbar) from 0 to 100 s, the amount of amidogen decreases with time. Finally, between $2\times10^{-3}$  and 0.2 mbar, the temporal evolution of the abundance of NH$_2$ is more variable but remains greater than the initial steady state. From 1038 s, y$_{NH_2}$ diminishes with time at all pressures.
We analyzed the chemical pathways involved in the production and the destruction of NH$_2$ and found that over the entire duration of the flare the production of amidogen was due to NH$_3$ + h$\nu$ $\rightarrow$ NH$_2$ + H, so the production rate of NH$_2$ follows closely that of ammonia. The dominant removal process of NH$_2$ is by reaction with H$_2$: NH$_2$ + H$_2$ $\rightarrow$ NH$_3$ + H. However, we note that around 2 mbar, the main reactions involved in the formation/destruction of amidogen are different than at lower pressures: NH$_2$ + H $\rightarrow$ NH + H$_2$ for the destruction; HNCO + H $\rightarrow$ NH$_2$ + CO and OH + HCN $\rightarrow$ NH$_2$ + CO for the production.

\paragraph{Ammonia:}
This species sees its abundance affected by the flare from the top of the atmosphere down to $\sim$0.2 mbar.
As observed for NH$_2$, y$_{NH_3}$ undergoes an initial increase (by a factor 15 at $2\times10^{-3}$ mbar) , which is due to the fact that the production of ammonia via the reaction NH$_2$ + H$_2$ $\rightarrow$ NH$_3$ + H dominates over the process of destruction through photolysis. At longer times, the amount of ammonia decreases, ending up lower than the initial steady-state at pressures greater than $10^{-2}$ mbar for $t>$ 1864 s. At this stage, NH$_3$ removal (dominated by the photodissociation NH$_3$ + h$\nu$ $\rightarrow$ NH$_2$ + H) is dominant over ammonia production.

\paragraph{Carbon dioxide:}
The abundance of CO$_2$ changes down to $3\times10^{-3}$ mbar. One can observe a decrease of the mixing ratio during the flare of four orders of magnitude at $9\times10^{-4}$ mbar between the initial steady state and 1038 s. As for the warm atmosphere case, this decrease is due to the photodissociation of CO$_2$: CO$_2$ +  h$\nu$ $\rightarrow$ CO + O($^1$D). In the second part of the flare, there is almost no change in the abundance of CO$_2$, due to the balance between the production and destruction processes.

\paragraph{Nitric oxide:}
NO is affected by the stellar flare down to 30 mbar. Its evolution depends on the pressure level. For the first part of the flare (0-1038 s), from 30 to 0.2 mbar, y$_{NO}$ increases as a function of time until the inflection point of the abundance profile (located at $\sim$ 0.2 mbar). At these lower altitudes, where NO increases, NO is produced mainly by thermal dissociation of HNO (HNO + H $\rightarrow$ NO + H$_2$) and reaction with nitrogen atom (N($^4$S) + OH $\rightarrow$ NO + H). The maximum enhancement of the mixing ratio (a factor 10) is found at 5 mbar. Above this point, the abundance profile of NO decreases with time. At $1.2\times10^{-3}$ mbar (pressure corresponding to the production peak of NO), the abundance of NO decreases by a factor of 4000 between 0 and 1038 s.We also notice that this peak is shifted downward with time. At these higher altitudes, the NO decrease is due mainly to photodissociation (NO + h$\nu$ $\rightarrow$ N($^4$S) + O($^3$P)) and, at the highest altitudes, also to reactions with carbon atoms (C + NO $\rightarrow$ CO + N($^4$S) and C + NO $\rightarrow$ CN +  O($^3$P)), produced by photodissociation of CO. From 1130 s, we observe a transition in the abundance at 4 mbar. Below this pressure, y$_{NO}$ decreases with time, and for higher pressures, y$_{NO}$ increases with time. 

\paragraph{Hydroxyl radical:}
This species sees its abundances change down to $\sim$20 mbar. From 0 to 1038 s, y$_{OH}$ increases with time from 20 mbar up to $2\times10^{-3}$ mbar, the pressure at which there is a production peak. We observe an increase by a factor of four at 3 mbar between the abundances at 0 and at 1038 s. 

At these lower altitudes ($\sim$2 mbar) the reversible reactions O + H$_2$ $\leftrightarrow$ OH + H and H + H$_2$O $\leftrightarrow$ OH + H$_2$ are finely balanced. Interestingly, production of OH via photolysis of water (H$_2$O + h$\nu$ $\rightarrow$ H + OH) tips the balance in favour of overall OH production, even at these high pressures. At lower pressures, in addition of the photodissociation of water, OH produced by the reaction O($^3$P) + H$_2$ $\rightarrow$ OH + H also affects the balance. Finally, at high altitudes OH decreases due to the dominance of OH photodissociation and its reaction with carbon and oxygen atoms, namely OH + C $\rightarrow$ CO + H and OH + O($^3$P) $\rightarrow$ O$_2$ + H. From 1130 s, we observe a decrease of the abundance with time for pressures between 20 and 1 mbar. For pressures between 1 and $2\times10^{-3}$ mbar, y$_{OH}$ increases slightly. Between $2\times10^{-3}$ and $10^{-3}$ mbar, the abundance of OH slightly decreases with time. Finally, for pressures lower than $10^{-3}$ mbar, y$_{OH}$ increases with time.

\begin{figure*}[!htb]
\centering
\includegraphics[angle=0,width=\columnwidth]{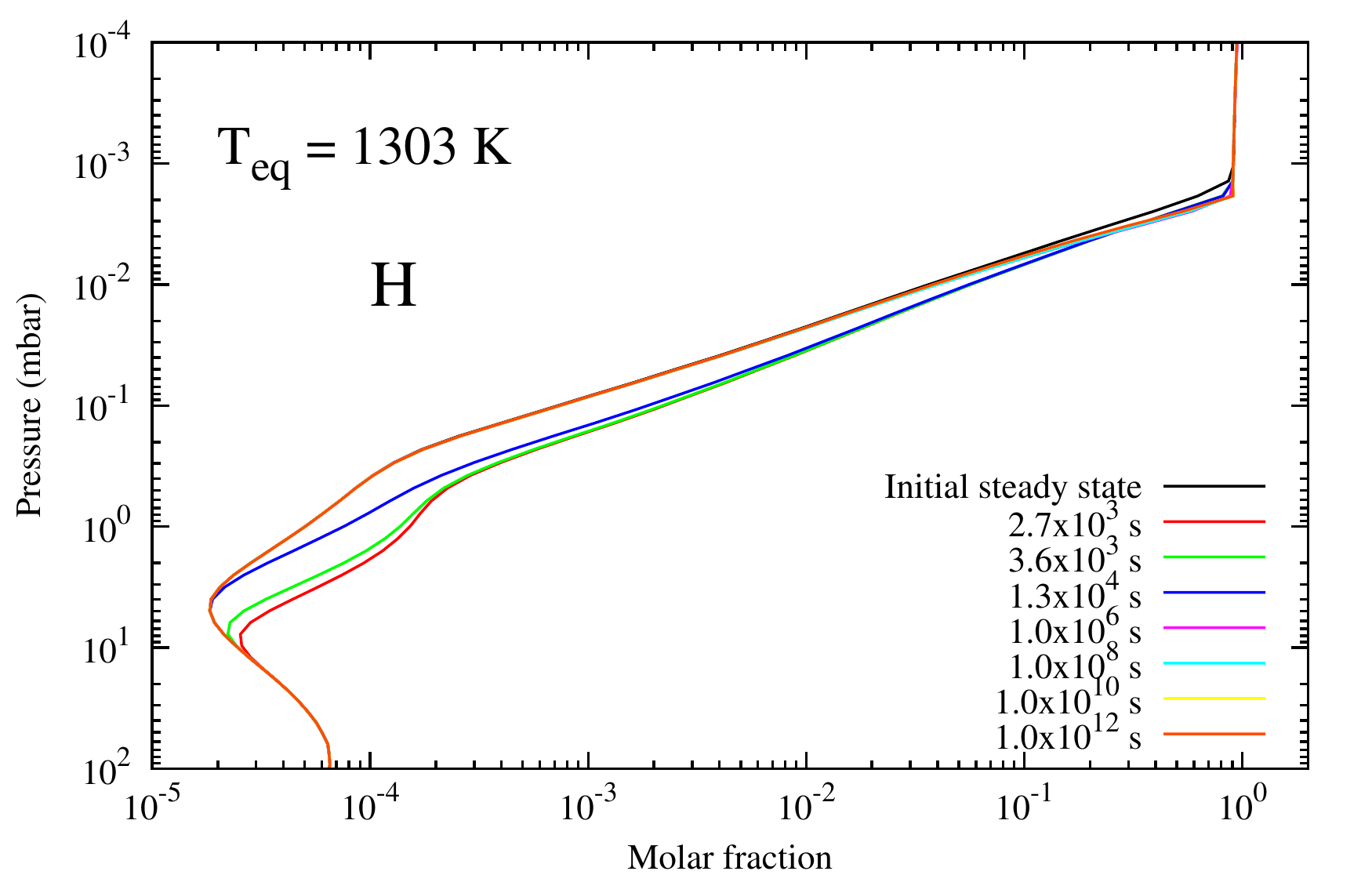}
\includegraphics[angle=0,width=\columnwidth]{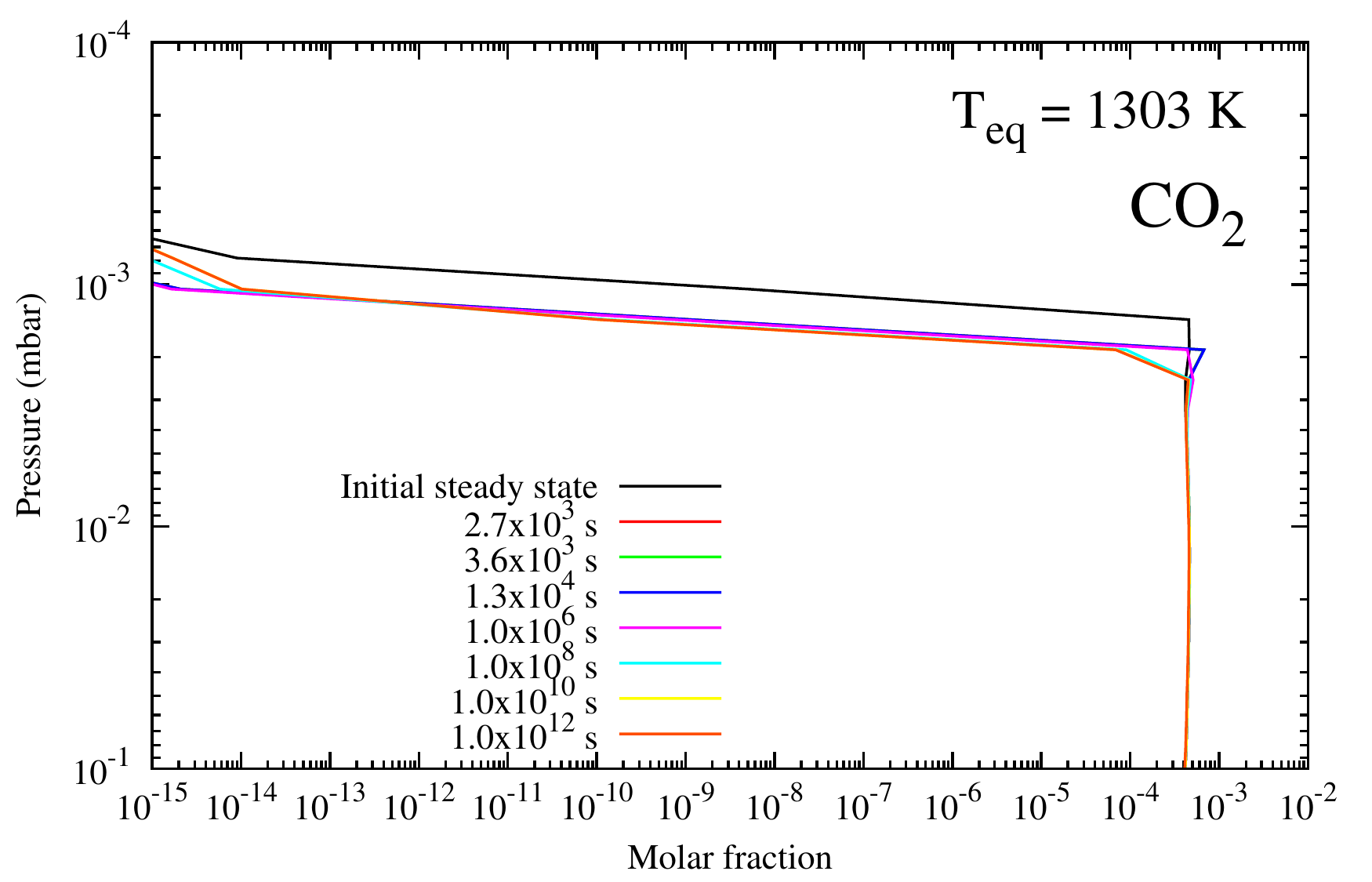}\\
\includegraphics[angle=0,width=\columnwidth]{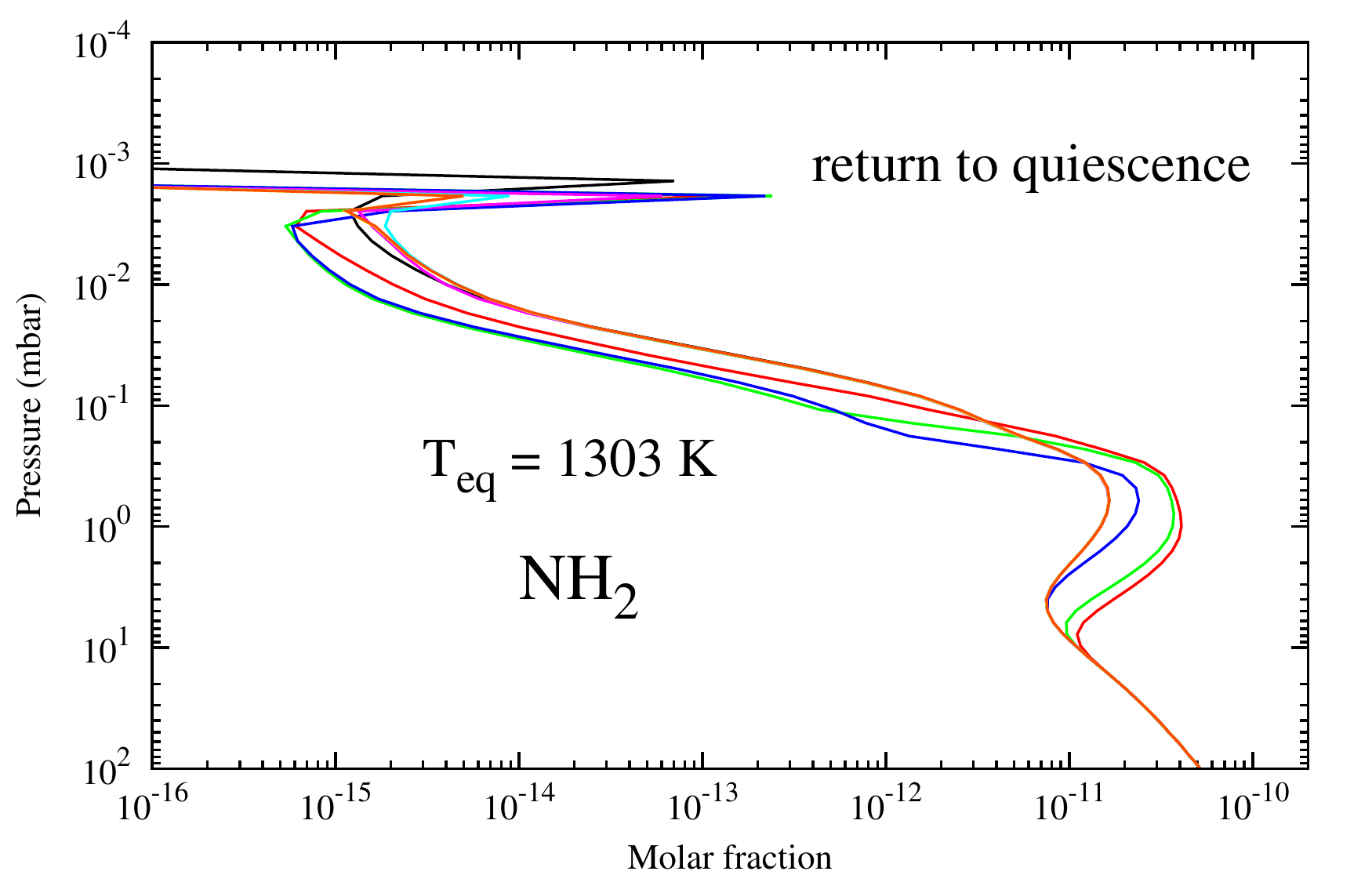}
\includegraphics[angle=0,width=\columnwidth]{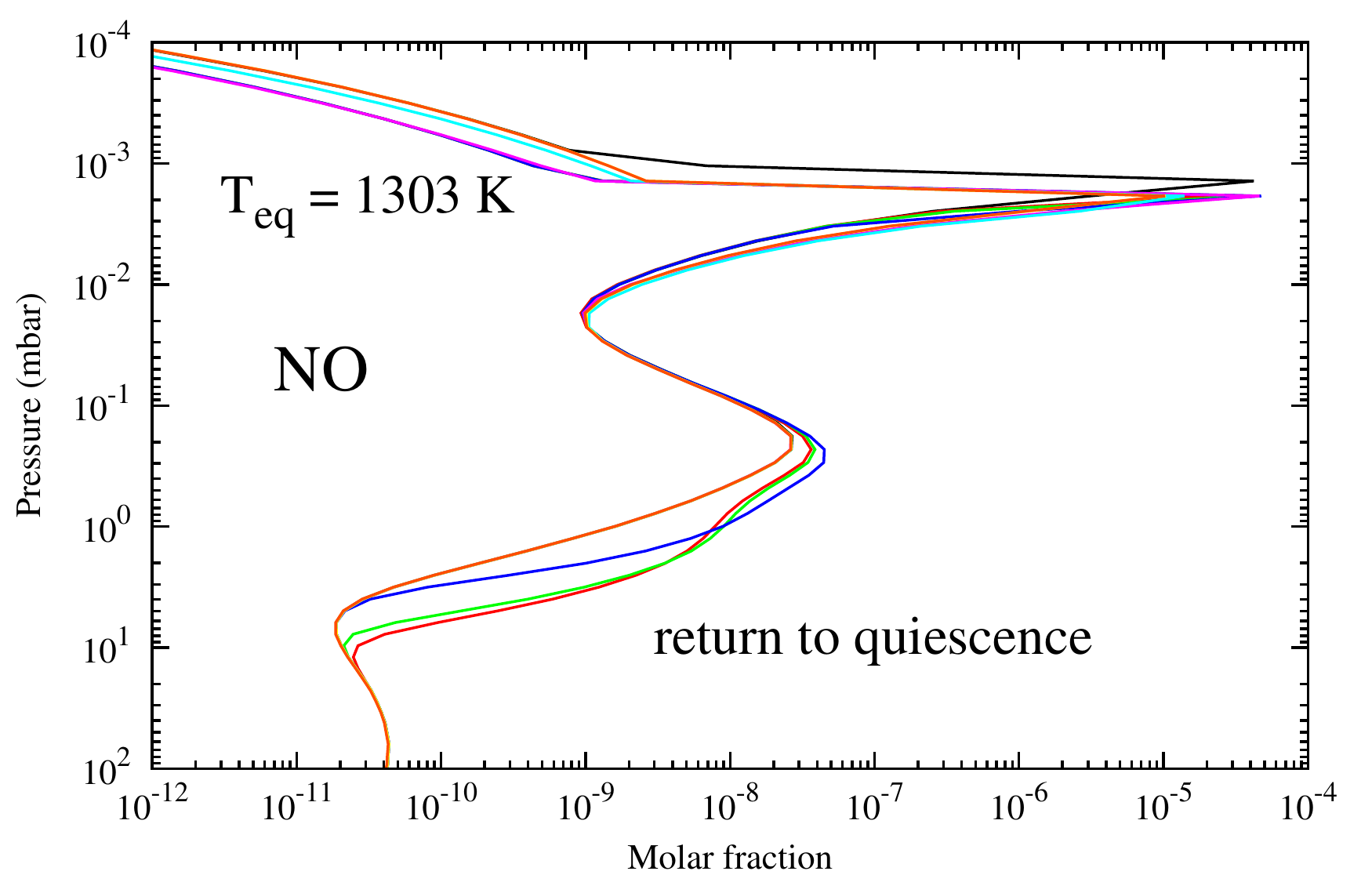}\\
\includegraphics[angle=0,width=\columnwidth]{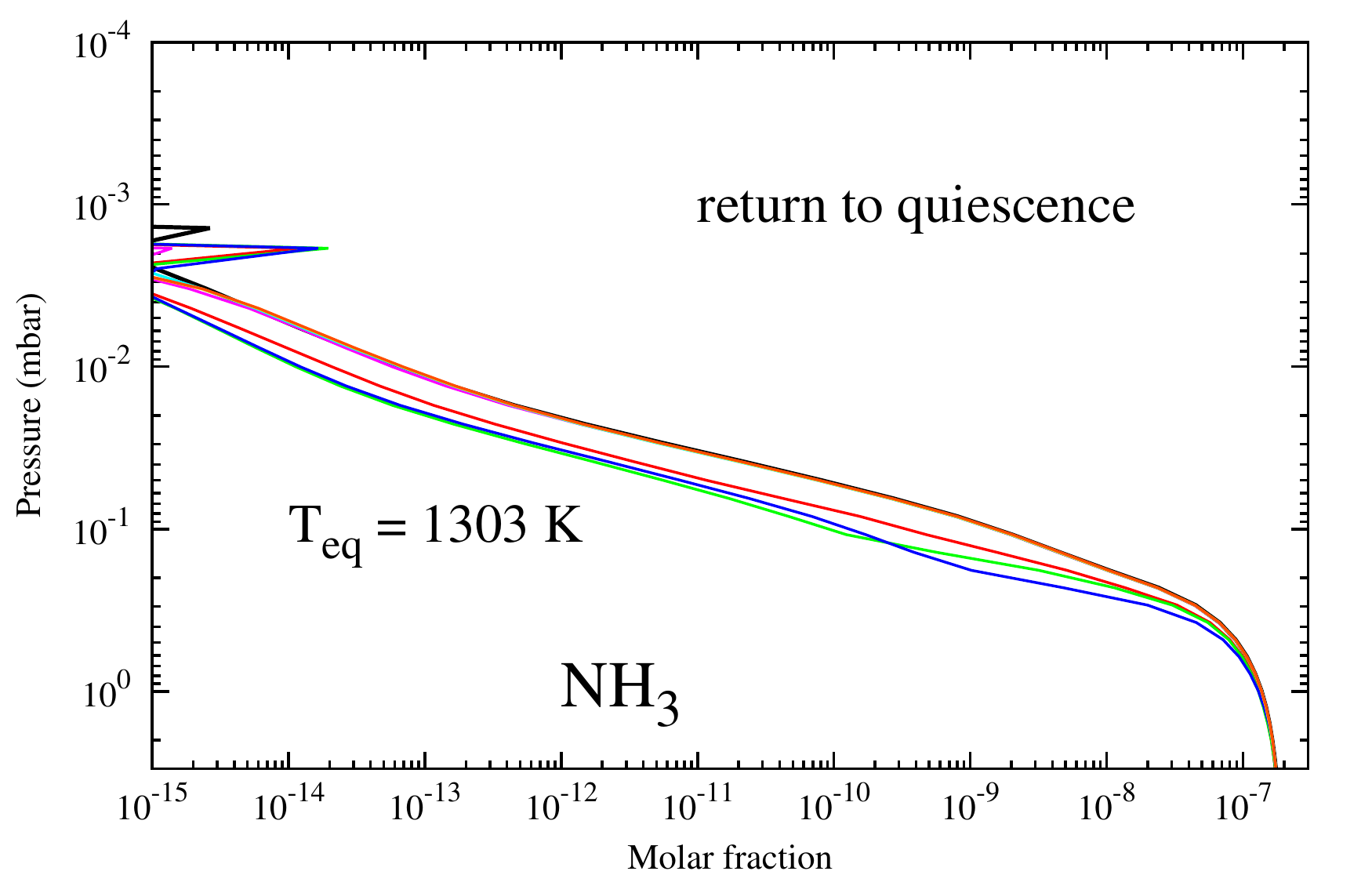}
\includegraphics[angle=0,width=\columnwidth]{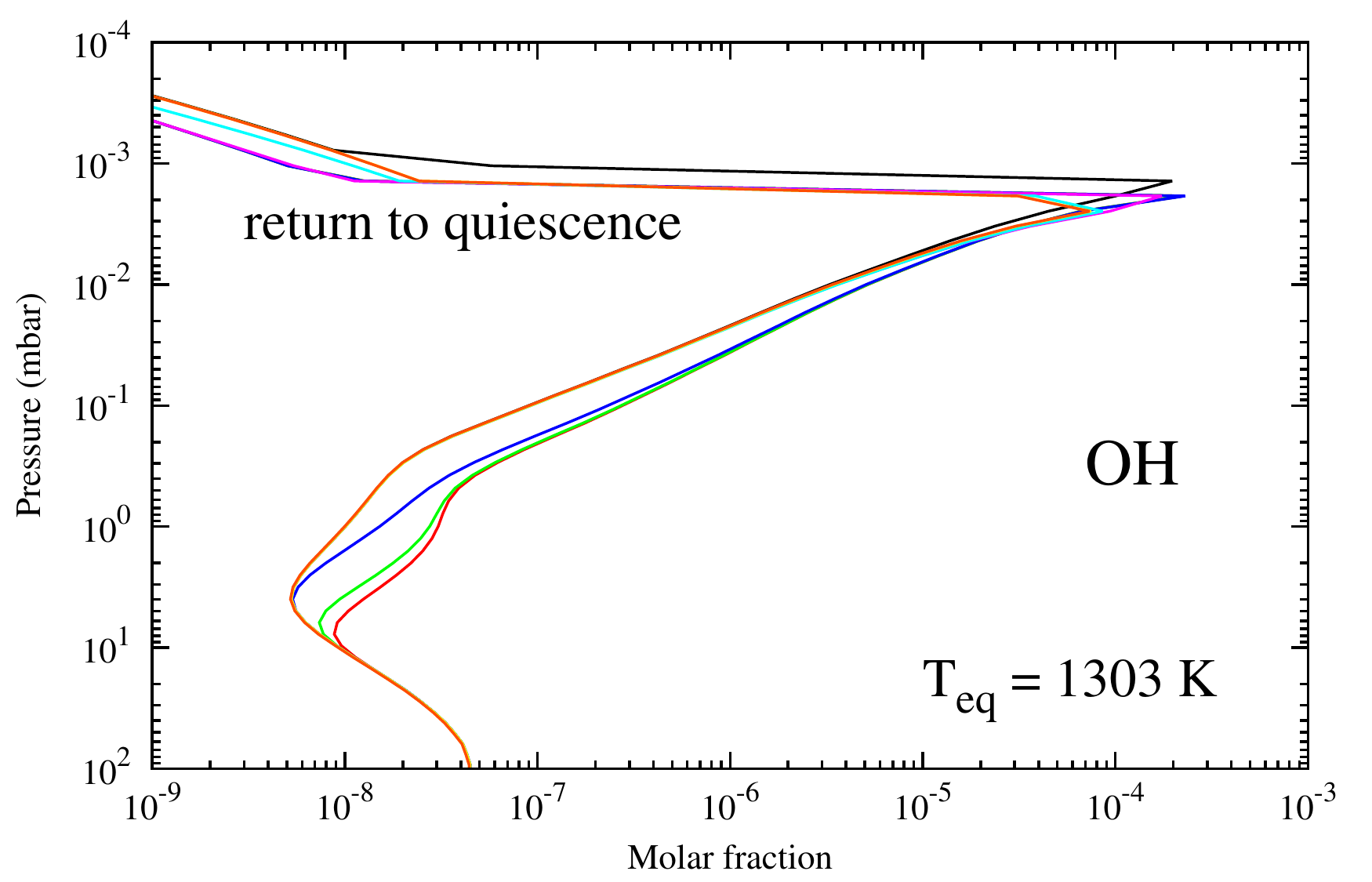}\\
\caption{Evolution of H, NH$_2$, NH$_3$, CO$_2$, NO, and OH mixing ratios after the flare event with the thermal profile corresponding to $T_{eq}$ = 1303K. The legend for all figures is in the upper panels.} \label{fig:1303K_return}
\end{figure*}

\subsection{Long-term effect of the flare}
\begin{figure*}[!htb]
\centering
\includegraphics[angle=0,width=\columnwidth]{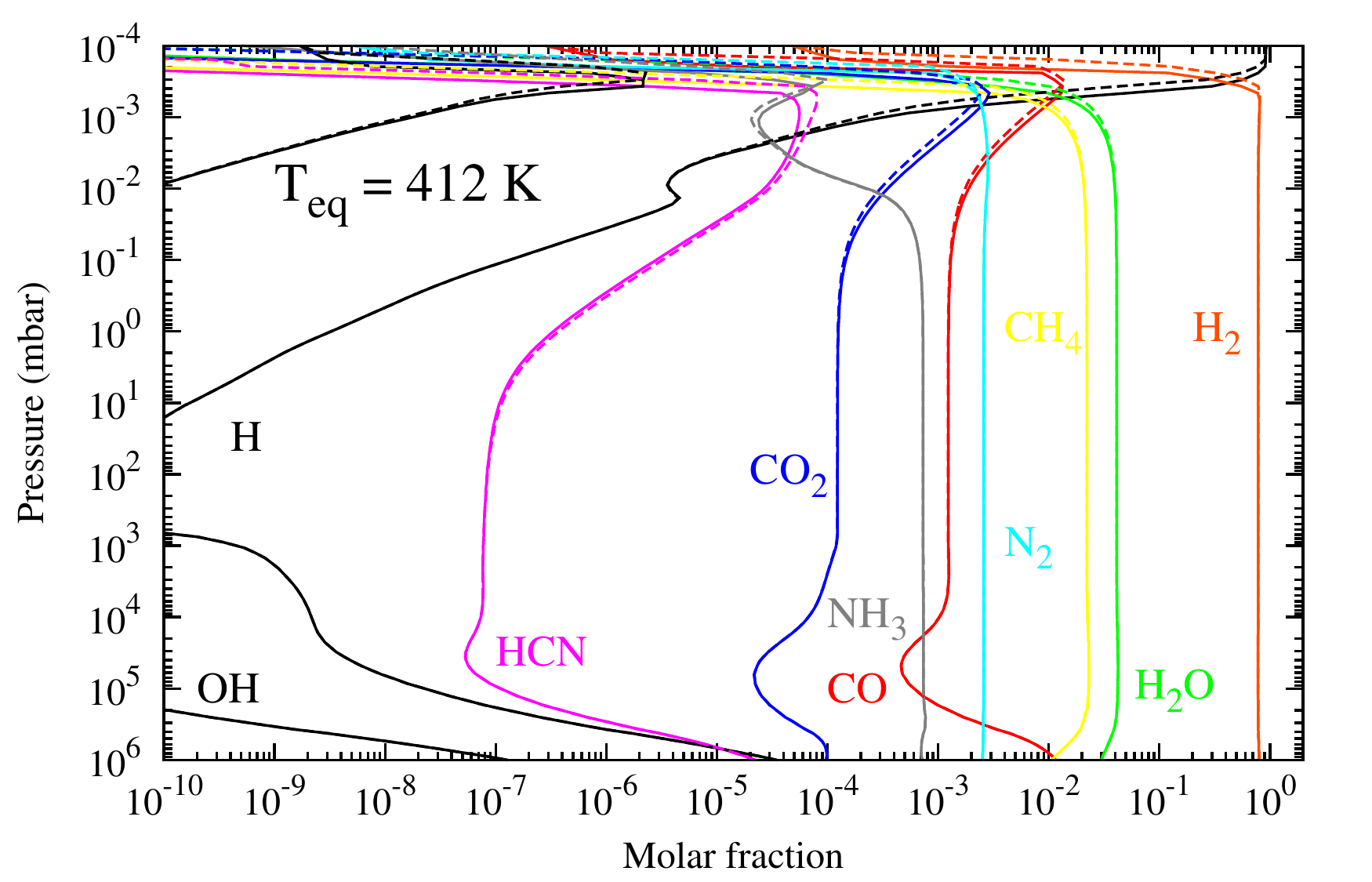}
\includegraphics[angle=0,width=\columnwidth]{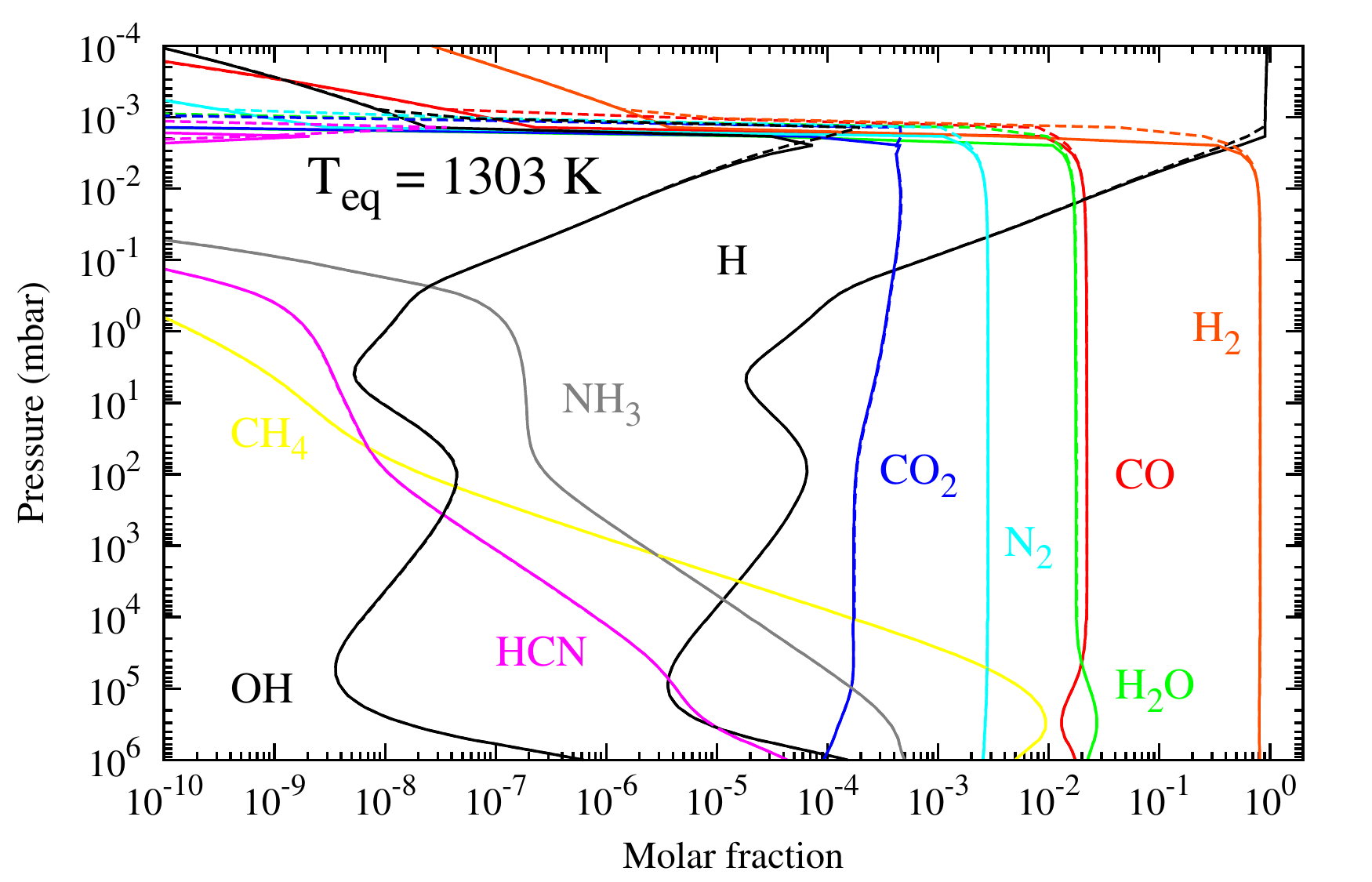}
\caption{Atmospheric composition at the initial steady-state (dashed line) and at the final steady-state  (solid line) with the thermal profile corresponding to $T_{eq}$ = 412 K (left) $T_{eq}$ = 1303K (right). They correspond to the compositions obtained $10^{12}$ s after the flare event.} 
\label{fig:new-init-steady}
\end{figure*}

After 2586 s, the stellar flux returns to quiescence. We study how long it takes for all species to return to steady-state. For the two atmosphere cases considered here, the final steady-states are reached after $10^{12}$ s (between $10^{9}$ and $10^{12}$ s the variations are very minor). However, we see in Figs.~\ref{fig:412K_return}, \ref{fig:1303K_return}, and \ref{fig:new-init-steady} that some species exhibit differences between the initial and the post-flare steady-states in the upper atmospheres (and down to 2 mbar for NO in the case of the warm planet). These variations are more important for the T$_{eq}$=1303 K atmosphere. In the hot atmosphere, except for H, which hardly changes concentration, the major species under consideration in this study are less abundant after the flare event than before. This is not the case for the warm atmosphere, where we observe abundances changes in either direction, depending on species and pressure level. We suggest that these new steady-states are due to the permanent modification of the photolysis rates and the coupling between the different layers. The intense flare effectively bleaches the upper atmosphere, thus exposing lower levels also to more intense radiation. The resulting induced chemistry partly maintains this increased atmospheric transparency even when photon fluxes return to background levels after the flare. We made numerous tests (i.e. cut off the photolysis processes after the flare event, apply the flare event from thermochemical equilibrium, change the initial conditions) that confirm our hypothesis. In a previous model, in which the mixing ratio of dihydrogen was 2000 times higher than what is shown here (i.e. 10$\%$ of y$_H$), the initial and post-flare steady-states were identical. This shows that the differences expected between the two steady-states depend also on the chemical composition of the upper atmosphere. 

\subsection{Recurrence of flares}\label{sect:recurrence}
A flare is not a unique event in the lifetime of an active star. \cite{Pettersen1984}, who studied the flare activity of AD Leo during observations of this star between 1971 and 1984, saw no obvious periodicity in the occurrence of the flares, but did determine that the duration between two flares ranged from 1 to 137 minutes. Such a random distribution of flare events has also been observed on other active stars  \citep[e.g.,][]{Lacy1976}.

\begin{figure*}[!htb]
\centering
\includegraphics[angle=0,width=\columnwidth]{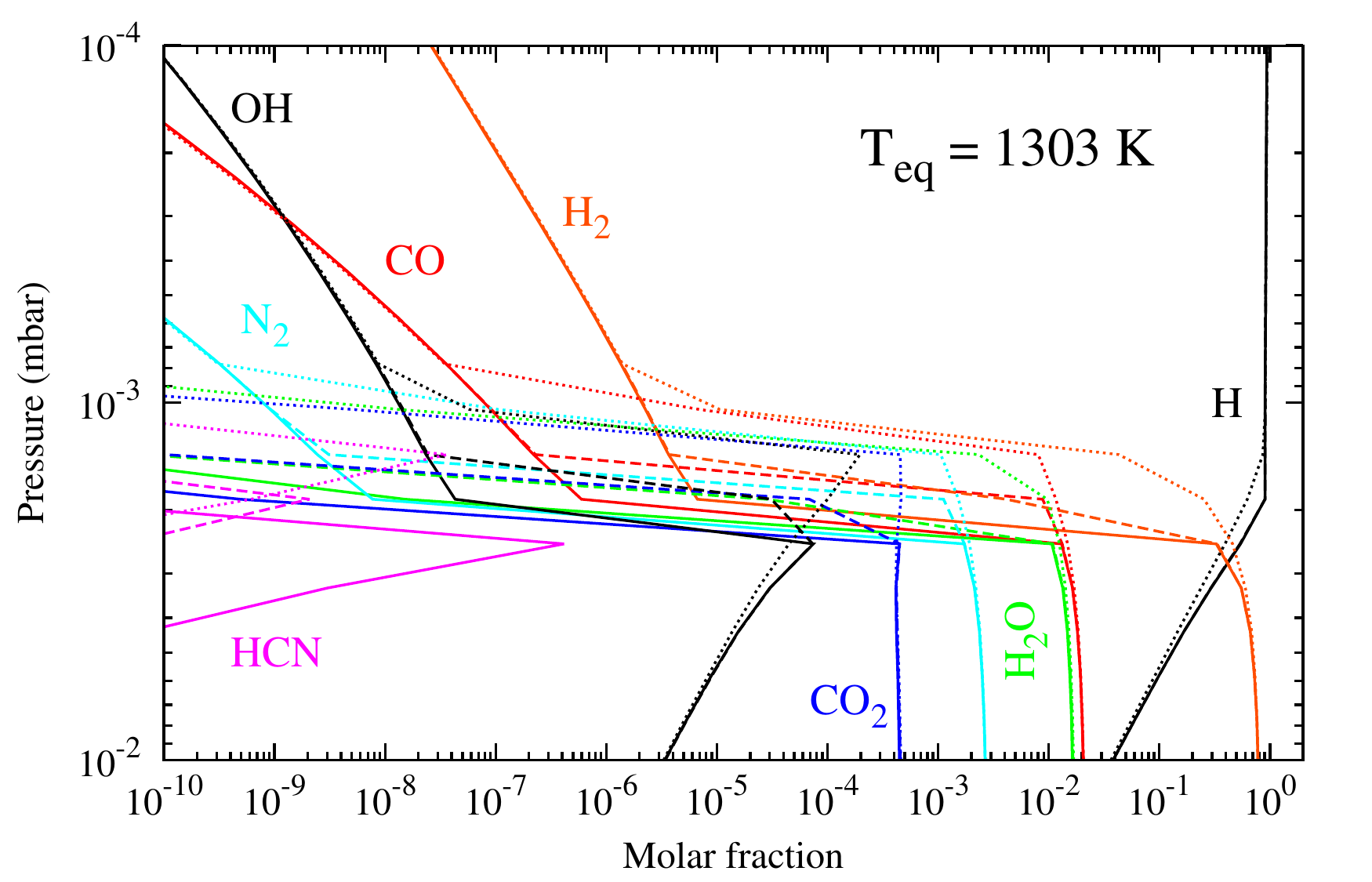}
\caption{Atmospheric composition at the initial steady-state (dotted line), at the first final steady-state (after one flare - dashed line), and at the second final steady-state (after a second flare - solid line) with the thermal profile corresponding to $T_{eq}$ = 1303K. They correspond to the compositions obtained $10^{12}$ s after each flare event.} 
\label{fig:frompost-steady}
\end{figure*}

\begin{figure*}[!htb]
\centering
\includegraphics[angle=0,width=0.9\columnwidth]{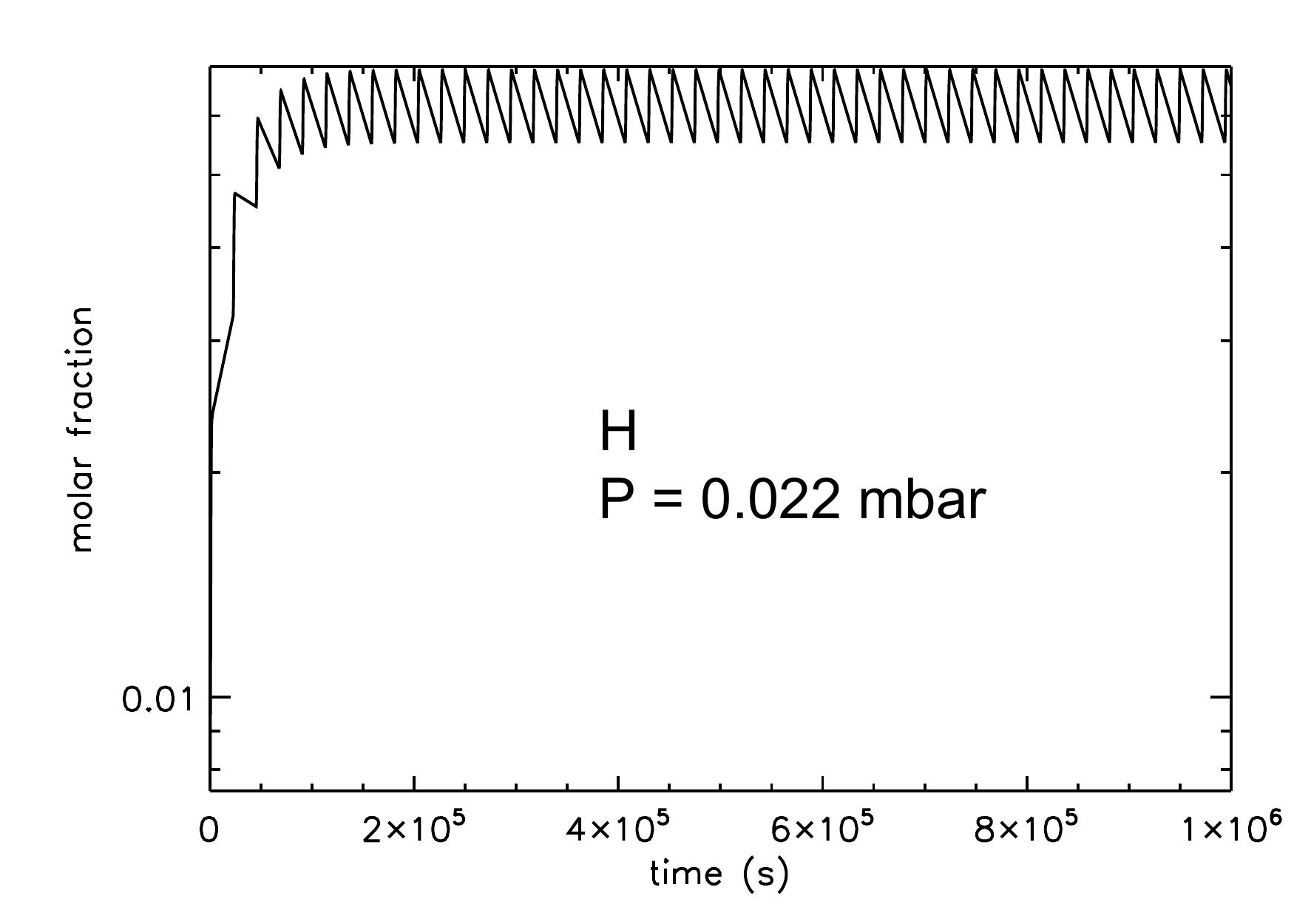}
\includegraphics[angle=0,width=0.9\columnwidth]{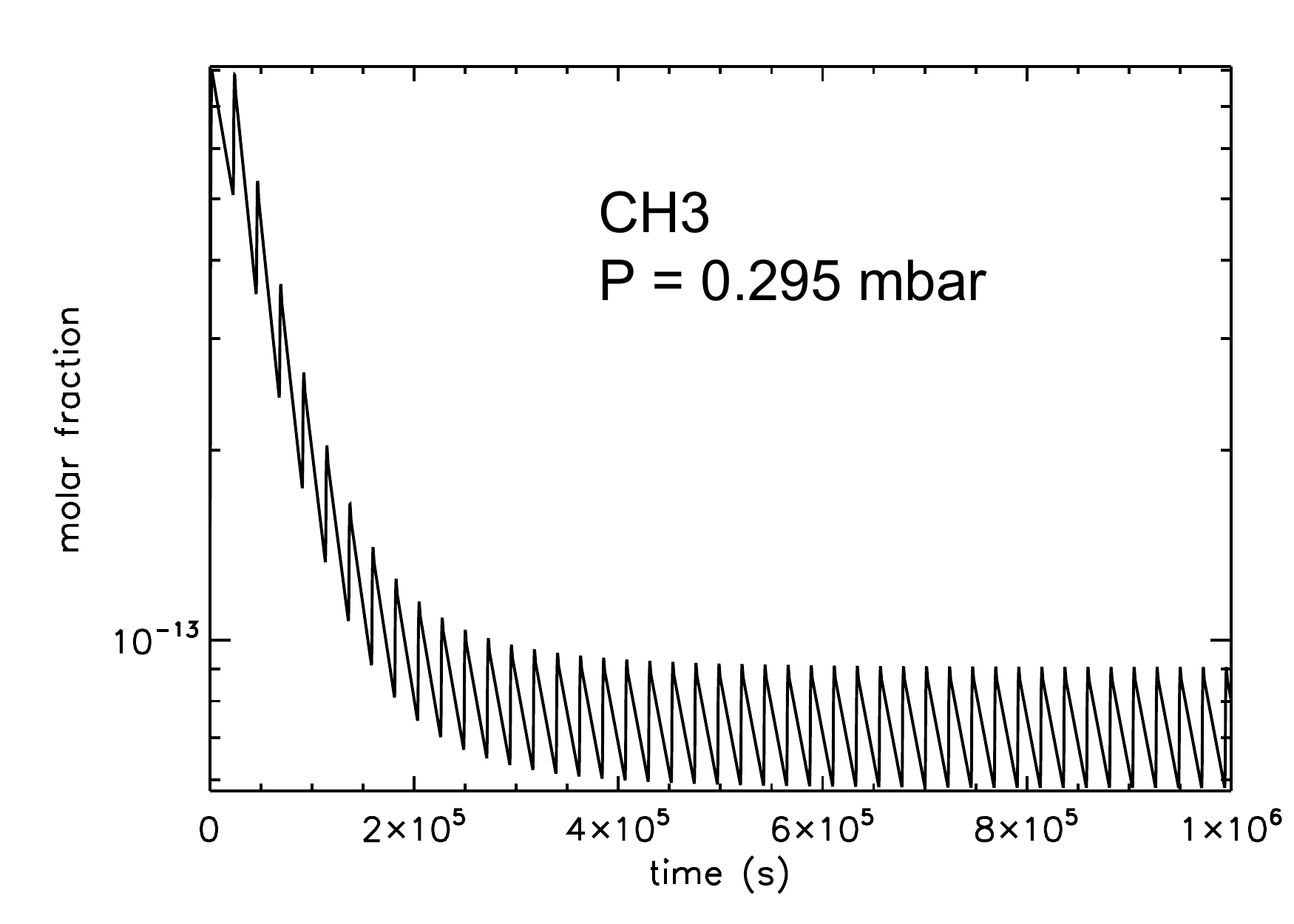}
\includegraphics[angle=0,width=0.9\columnwidth]{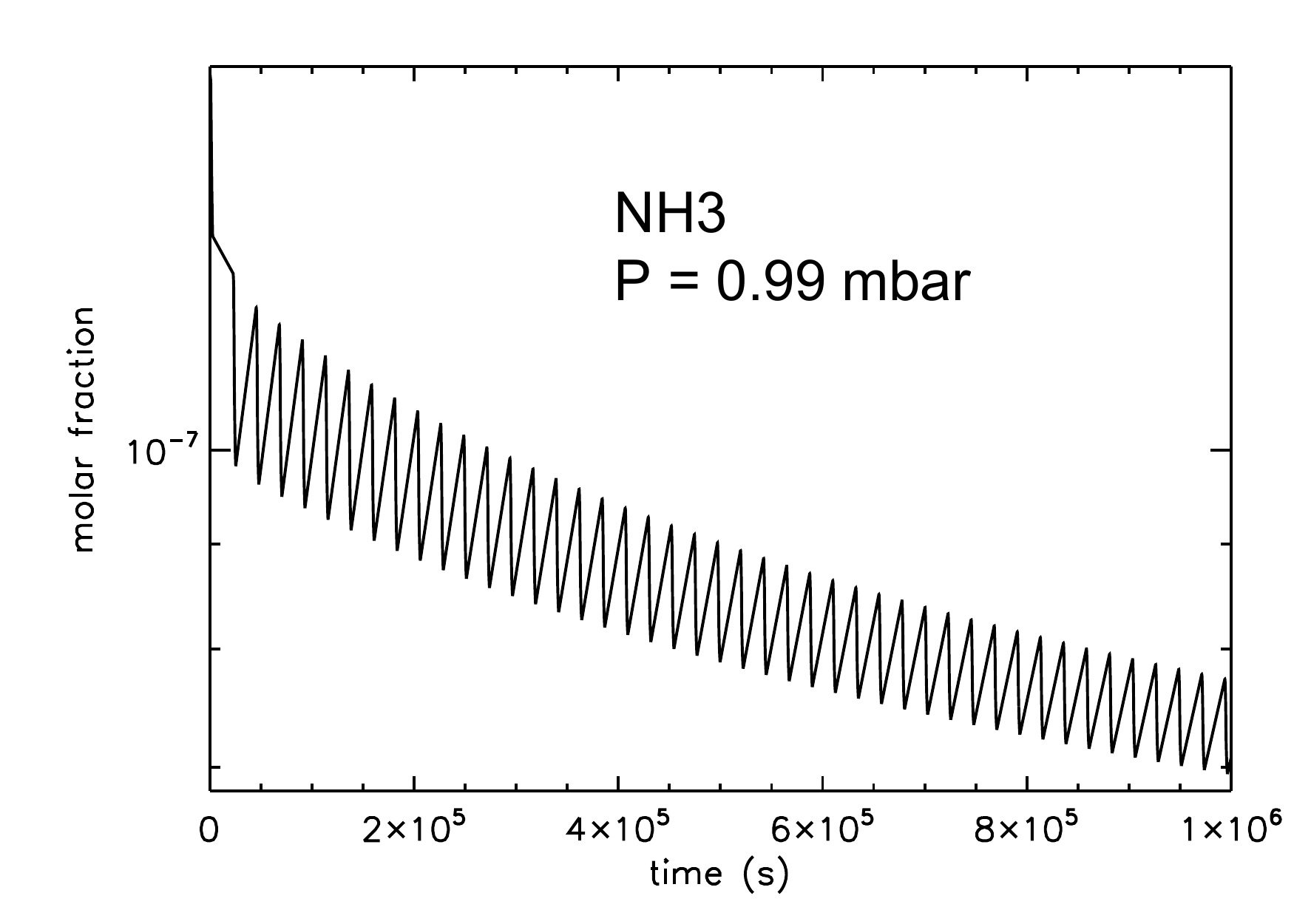}\includegraphics[angle=0,width=0.9\columnwidth]{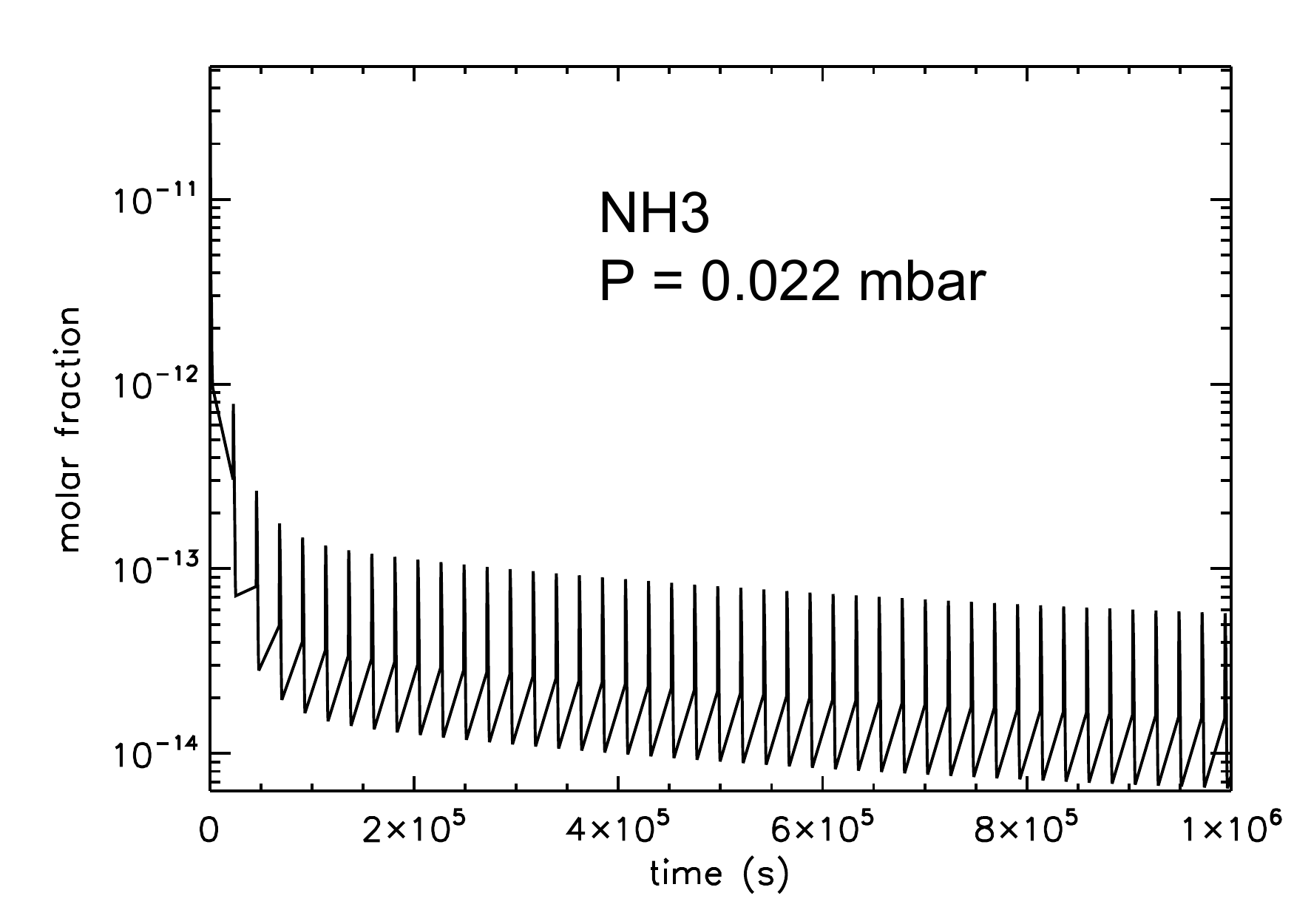}
\caption{Temporal evolution of the molar fractions of H (top left), CH$_3$ (top right), and NH$_3$ (bottom) at different pressures level with the thermal profile corresponding to $T_{eq}$ = 1303, assuming that the host star underwent 47 flares.} 
\label{fig:rec}
\end{figure*}

To account for the recurrence of flares, we made two tests:\\
1. We re-ran our time-dependent model from the post-flare steady-state, up to a new steady-state. This case is not realistic as it implies that flares occur every $10^{12}$s, but can be considered as an extreme case for low-activity stars. Here again, we observed variations between the initial and final steady-states, but to a lesser extent than for our nominal case (Fig.~\ref{fig:frompost-steady}). We note that hydrogen returns to the initial steady-state (i.e. the previous post-flare steady-state) after this second flare event.\\
2. We ran our time-dependent model from the initial steady-state, imposing a period between flares of $2\times10^{4}$s (i.e. $\sim$ 5 h), a more realistic time scale and close to the maximum duration between flares observed by \cite{Pettersen1984} on AD Leo. We chose this upper value, which is more likely to be representative of the occurence of flares observed by \cite{tofflemire2012, hunt2012} and of the one in lower spectral type (M0-M3) stars due to their lower magnetic activity \citep{kowalski2009, Davenport2012}. We simulate a repetition of 47 flare cycles (representing $\sim$10$^{6}$s s or $\sim$11 years). We found that the abundance of species continuously evolves at each flare event, but it seems that the maximum (and minimum) abundances reached at each cycle tend towards a limiting value (Fig.~\ref{fig:rec}) The number of cycles necessary to regulate the variations of abundances depends on both species and pressure level. Hydrogen reaches the same maximum abundance of 7$\times$10$^{-2}$ at 0.022 mbar after only 6 cycles, whereas methyl radical (CH$_3$) needs about 20 cycles to reach its maximum abundance of 9$\times$10$^{-14}$ at 0.295 mbar. For ammonia, more than 47 cycles are necessary to reach a limiting value.

\subsection{Synthetic spectra}\label{sect:spectra}

\begin{figure*}[!htb]
\centering
\includegraphics[angle=0,width=\columnwidth]{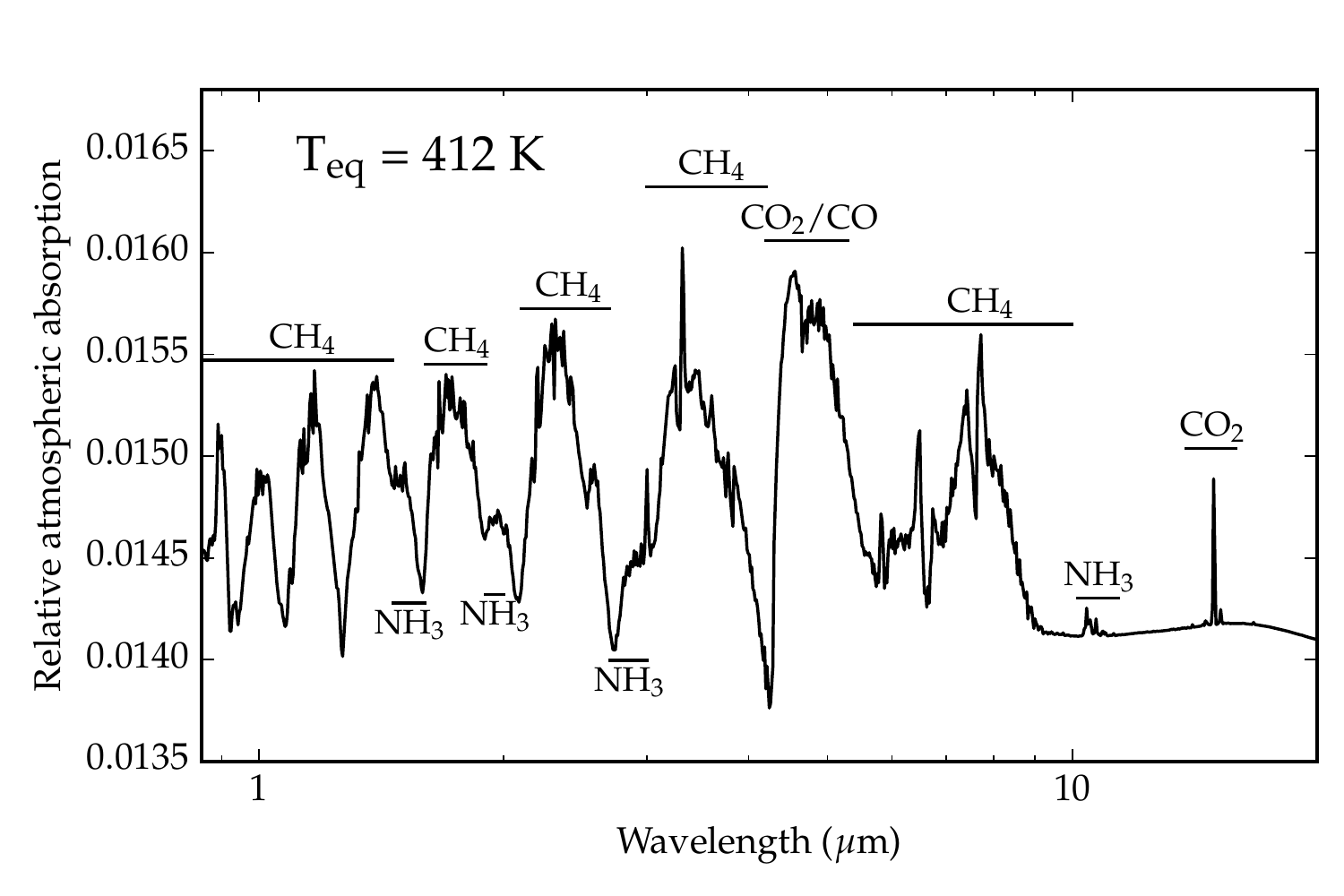}
\includegraphics[angle=0,width=\columnwidth]{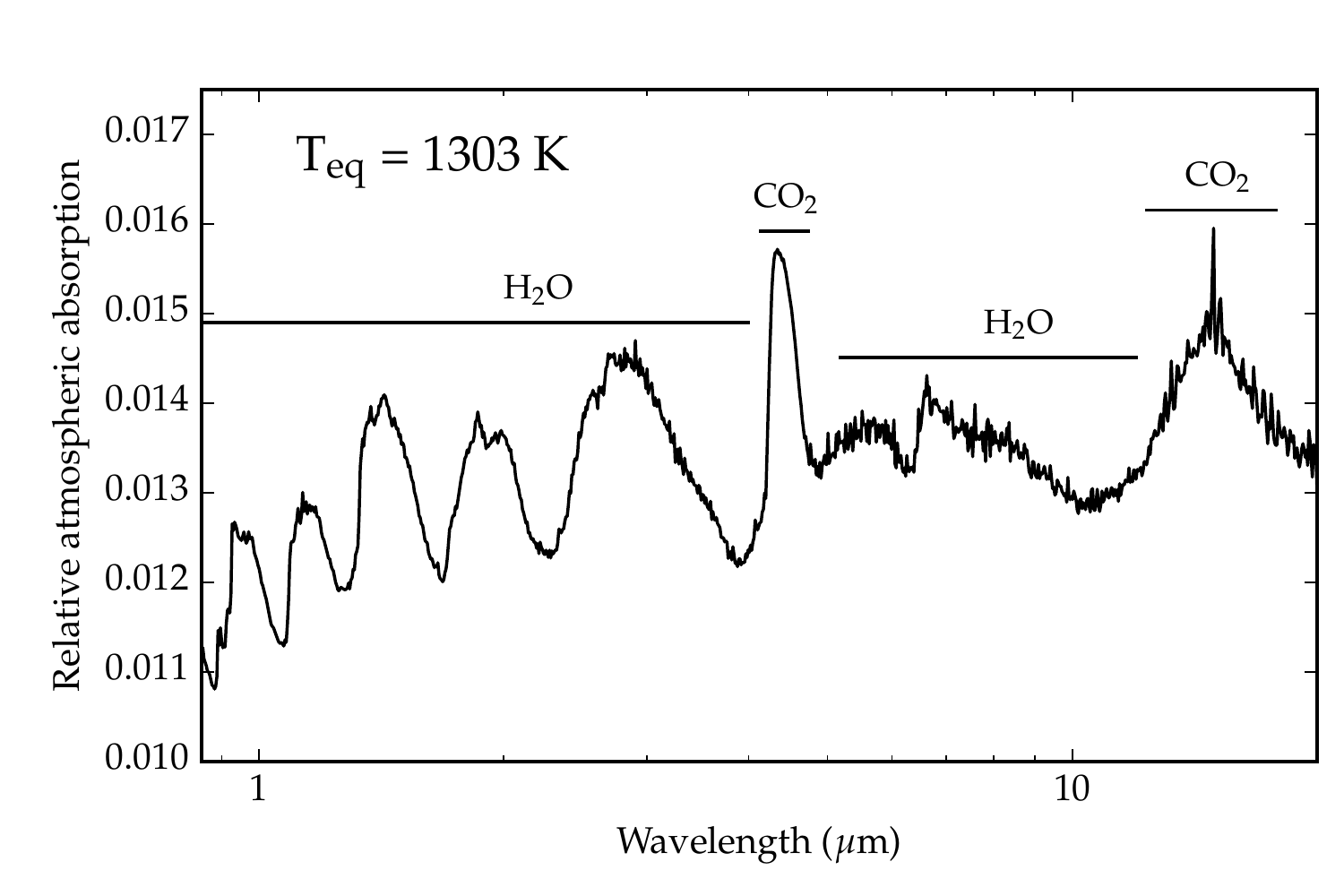}
\caption{Synthetic transmission spectra for the initial steady state for  $T_\mathrm{eq} = 412$\,K (left)  and  $T_\mathrm{eq} = 1303$\,K (right) cases. The resolution is 100 across the entire spectrum.}
\label{fig:spectra}
\end{figure*}

\begin{figure*}[p]
\centering
\includegraphics[angle=0,width=\columnwidth]{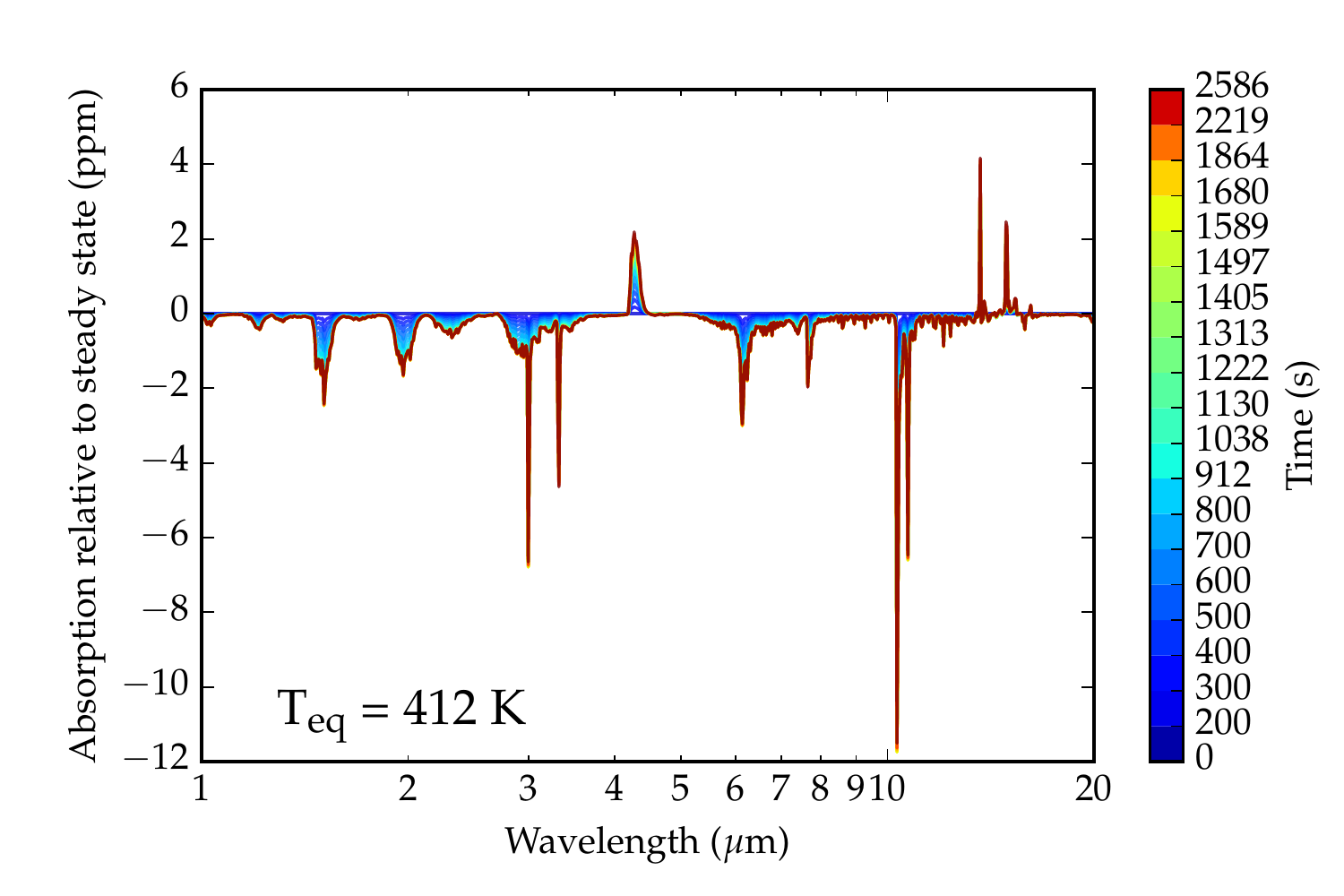}
\includegraphics[angle=0,width=\columnwidth]{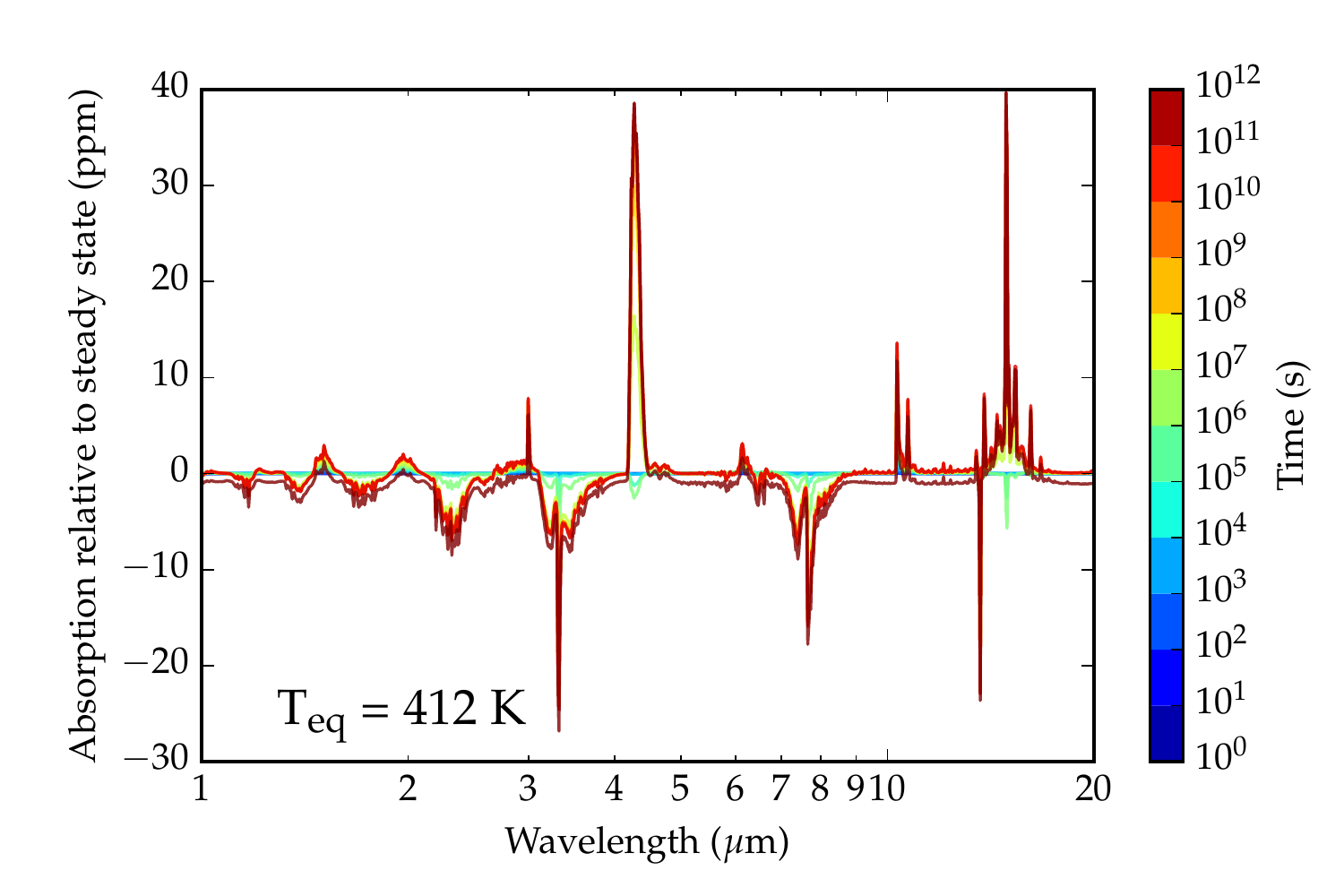}
\includegraphics[angle=0,width=\columnwidth]{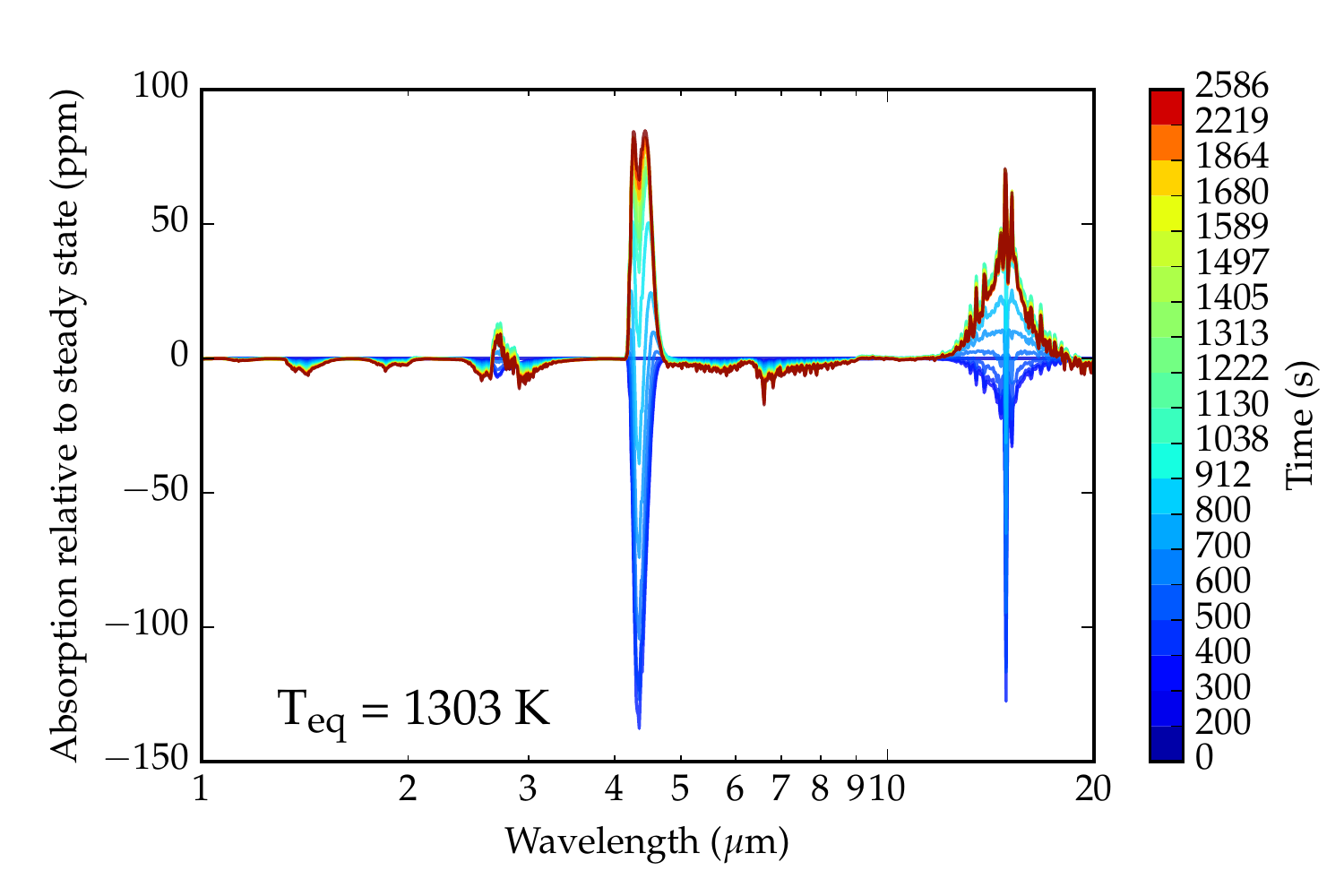}
\includegraphics[angle=0,width=\columnwidth]{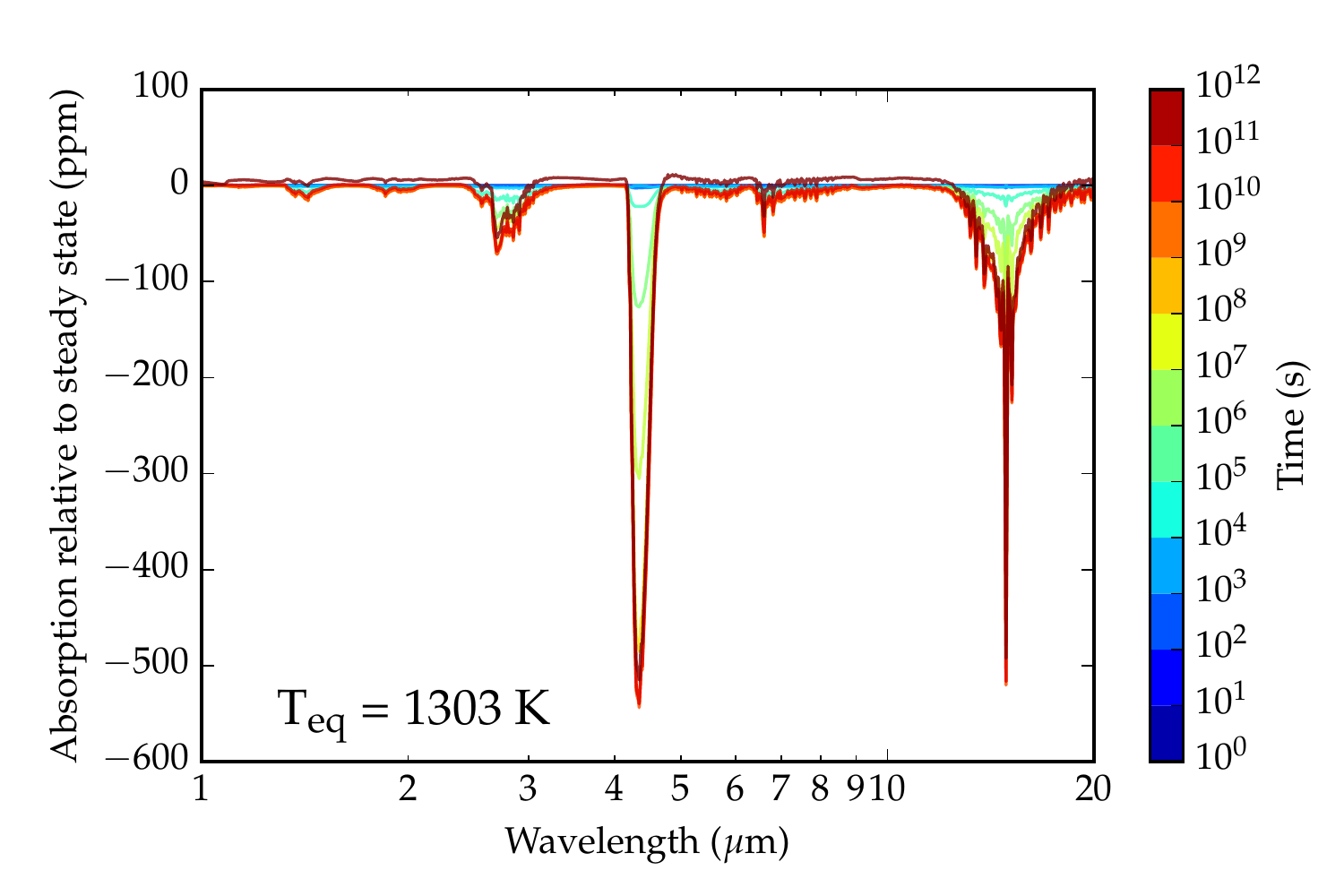}
\caption{Difference in relative absorption between the initial steady state and the instantaneous transmission spectra obtained during the different phases of the flare, for the  $T_\mathrm{eq} = 412$\,K (top)  and  $T_\mathrm{eq} = 1303$\,K (bottom) cases. The left plots refer to the impulsive and gradual phases, while the right plots to the return to quiescence phase. The color legends of the right plots are represented with a logarithm scale. The resolution is 100 across the entire spectrum.}
\label{fig:spectradiff}
\end{figure*}
\begin{figure*}[p]
\centering
\includegraphics[angle=0,width=\columnwidth]{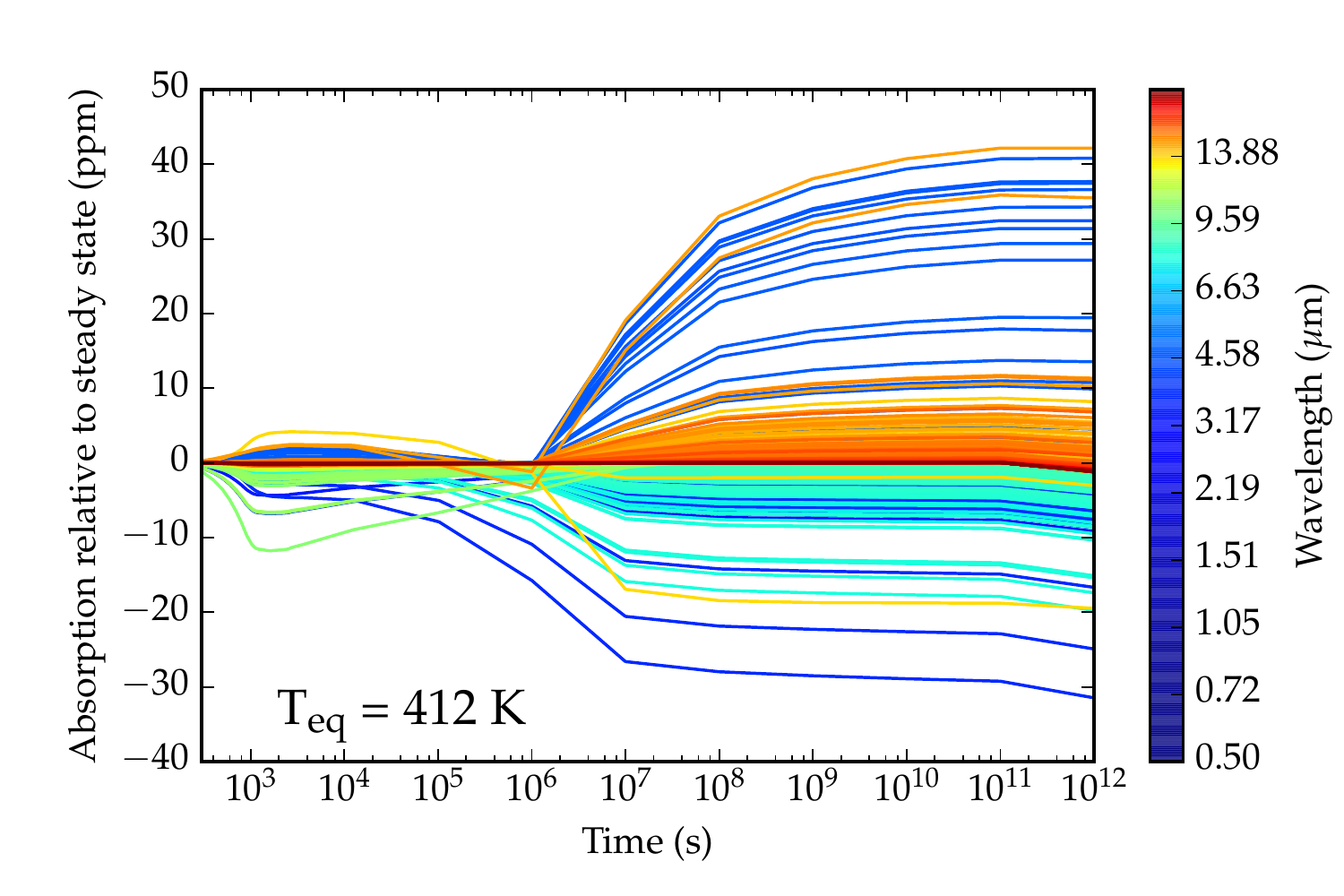} 
\includegraphics[angle=0,width=\columnwidth]{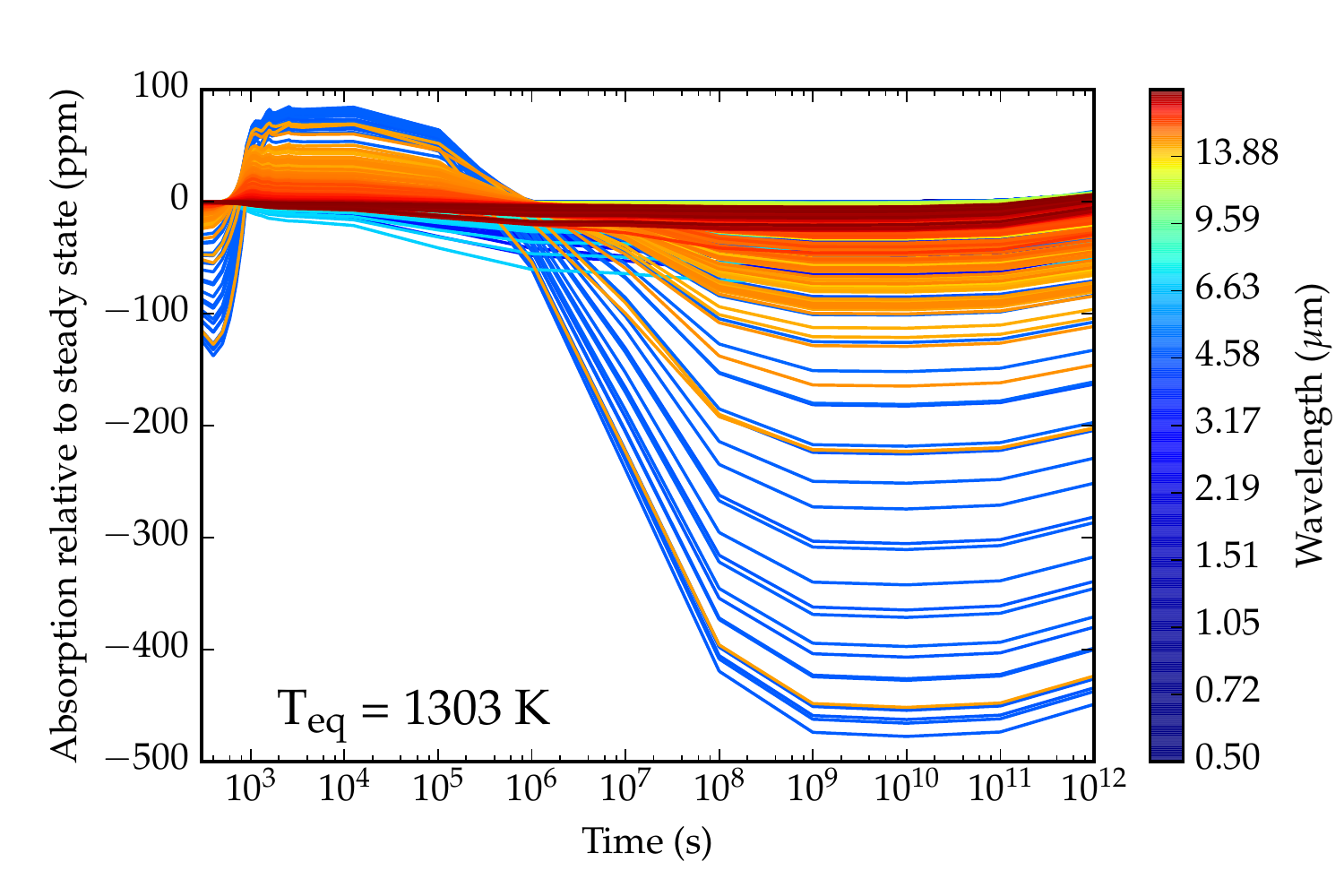}
\caption{Systematic shifts as a function of time ($x$-axis) and wavelength (colorbar) with respect to the steady state, for the $T_\mathrm{eq} = 412$\,K (left) $T_\mathrm{eq} = 1303$\,K (right) cases. Each line represent a wavelength bin corresponding to a constant resolution of 50. This is the shift, relative to the initial steady state, as a function of wavelength and time that would be seen in the transit depth. It is equivalent to Figure \ref{fig:spectradiff}, but with the $x$-axis and colorbar inverted. }
\label{fig:spdiff2}
\end{figure*}

We computed synthetic transmission spectra during the different phases of the flare and compared them to that of the initial steady state. Figure \ref{fig:spectra} shows the instantaneous transmission spectra for the initial steady state for the planets with $T_\mathrm{eq} = 412$\,K and $T_\mathrm{eq} = 1303$\,K. The planetary radius is defined at 1 bar. The spectra are binned at a  constant resolution of $R = 300$. Figure \ref{fig:spectradiff} shows the differences between the initial steady state and the transmission spectra obtained during the different phases of the flare. It can be seen that while small changes occur in the warm atmosphere, significant changes occur in the hotter planet's atmosphere. The strongest changes are seen in the CO$_2$ features at 4.6 and 14 $\mu$m in the hotter planet case, with total amplitude variations of, respectively, 220 and 200 ppm in the impulsive phase, and up to 500 ppm in the return to quiescence phase. Other changes with amplitude smaller than about 50 ppm are seen in other part of the spectra, especially in the return to quiescence phase. Weaker variations of only 40 ppm are seen in the 412 K planet, especially in the CO/CO$_2$ features at 4.6 $\mu$m and 14 $\mu$m. Other very small changes of about 20 ppm are seen in other parts of the spectra.

Two effects contribute to the observed changes. Firstly, changes in the molecular abundances vary the strength of the absorption features specific to the different molecules. Secondly, these changes modify  the mean molecular weight ($\mu$) for the different atmospheric layers, amplifying or reducing the amplitude of the absorption features throughout the entire spectrum. This is because the scale height, i.e. the altitude at which the atmospheric pressure decreases by $1/e$, is inversely proportional to the mean molecular weight ($H = k T/\mu g$; where: $k$ is the Boltzmann constant, $T$ is the atmospheric temperature, $g$ is the gravity acceleration and $\mu$ the mean molecular weight). Relatively small variations in mean molecular weight are seen in the warm planet case as the flare evolves. For the hotter planet case, we found that $\mu$ decreases (and hence $H$ increases) at high altitudes (i.e. low pressures) during the flare, indicating that the density of the upper atmosphere decreases. However, we found that the decrease in abundances in the atmosphere outweighs the effect of an increased scale height on the transmission spectra. 

We note that while the mean molecular weight for each atmospheric layer is calculated taking into account the entire set of 105 molecules contained in the chemical model, we only compute the spectral opacities for a set of 7 molecules which are the most abundant. The changes observed in the transmission spectra are therefore limited to these molecules. Other absorbers that might contribute to additional observing features will however show much weaker -- if not unobservable -- features.

\subsubsection{Effects on current and future observations}

Transmission spectra of exoplanetary atmospheres are obtained by observing simultaneously the transit of the planet at different wavelengths. By tracing the transit depth as a function of wavelength it is possible to reconstruct the transmission spectrum of the planet, as different molecules absorb differently in different spectral regions, making the planet appear smaller or larger at different wavelengths. A varying atmospheric composition during the transit event will introduce a time and wavelength-dependent shift on the transit light curve. 

Figure \ref{fig:spdiff2} shows the expected change in transit depth as a function of wavelength and time with respect to the transit depths measured during the initial steady state. It must be noted that the timescale of the flare is of the same order of magnitude as the typical transit duration ($\sim10^4$ s). Therefore, if a transit is observed during the flare, the transit depth would change during the transit, and shifts of up to 500 ppm are expected for the hotter planet. Clearly, these changes will be wavelength dependent, as shown by Figure \ref{fig:spectra}. Changes of only $\approx 40$ ppm would be seen in the warm planet.

The bulk of recent observations of exoplanetary atmospheres have been obtained from space using the \emph{Hubble Space Telescope} Wide Field Camera 3 (\emph{HST} WFC3), covering the $1.1-1.7$ $\mu$m range. Observations of hot Jupiters with WFC3 claim an uncertainty in transit depth per wavelength bin of 40 -- 100 ppm  \citep{Fraine2014,Deming2013,Knutson2014, tsiaras2016}, for resolutions of about 50. For smaller super-Earths such as GJ1214b, an uncertainty of 60 ppm was achieved \citep{Kreidberg2014}, but only by combining observations of 12 transits, while for 55 Cancri e a relative uncertainty of 22 ppm was achieved  for a single transit observation \citep{tsiaras2016b}.

Other space-based instruments that have been used to obtain exoplanetary spectra include IRAC, IRS and MIPS  onboard the \emph{Spitzer Space Telescope}  \citep[e.g.,][]{Tinetti2007,Grillmair2008,Richardson2006}, and NICMOS, STIS and COS onboard \emph{HST} \citep[e.g.,][]{Swain2008,Sing2008,Linsky2010}. Several ground-based facilities have also been used \citep[e.g.,][]{Bean2010,Snellen2010,Swain2010,Sedaghati2015,Mandell2011}. Future missions, such as JWST, will achieve a similar or greater level of precision \citep{Barstow2015}. 

The predicted changes in the spectra during a flare are therefore comparable to the sensitivity of the current and future instruments, especially for the 1303 K planet case, where relative changes of up to 500 ppm are seen. We note however that the current instrumentation cover spectral ranges that are either outside the spectral features with the strongest changes (HST/WFC and STIS), or have only photometric channels with relatively low sensitivity (Spitzer/IRAC).
Only with future instruments such as JWST, which will cover a spectral range that includes the CO/CO$_2$ features that show the strongest variations, it will therefore be possible to observe changes of spectral features of planetary atmospheres due to stellar activity in the form of flares.

\subsubsection{Recurrence of flares}

\begin{figure}
\centering
\includegraphics[angle=0,width=\columnwidth]{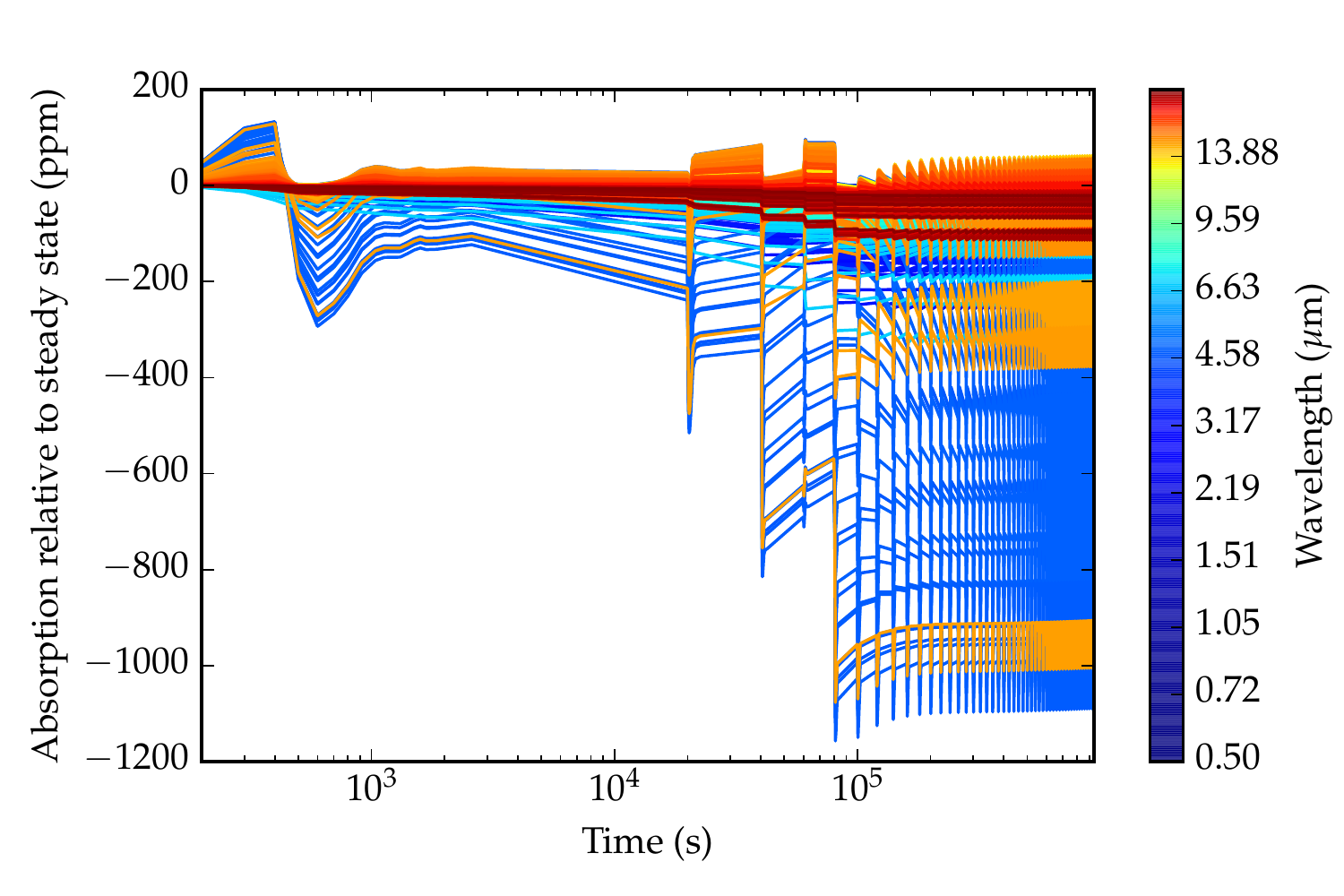} 
\caption{Systematic shifts as a function of time ($x$-axis) and wavelength (colorbar) with respect to the steady state, for the case in which the flare comes back every $\approx 5 h$ (See Section 3.3). Caption as in Figure \ref{fig:spdiff2}. }
\label{fig:spdiff3}
\end{figure}

Figure \ref{fig:spdiff3} shows the shifts as a function of wavelength and time that would occur over the period of ten days assuming that the flare recurs every five hours. It can be seen that, compared to a single flare event, the changes in the transmission spectrum are even more dramatic. We see variations of up to 1200 ppm after one day, which would be  detectable with future instruments onboard JWST. This shows even more clearly how strong stellar activity influences the atmospheric chemistry and the resulting spectra.

\section{Discussions}\label{sec:disc}

In this study, we focused on the effect of stellar flares on the chemical composition of the planet's atmosphere. One major simplification in this paper is that we did not change the thermal profile during the flare event, nor during the recovery phase. In all rigour, the temperature of the atmosphere should be modified by both the increase of the irradiation (that could possibly also take place in the IR wavelength range) and the change of the atmospheric composition. However, if the change of temperature is about the same order as the one found in \cite{segura2010effect} (i.e. less than 8 K), the chemical composition and thus the findings of this paper will not be significantly changed. However, because of the shorter radiative timescale of warmer planets \citep{showman2008}, the change of temperature might be higher. A more complete study will be necessary to draw firm conclusions and for that purpose, we are currently developing a 3D self-consistent model including dynamics and chemistry. Thus, a more detailed study will come in a forthcoming paper. \\

Another aspect of stellar flares which is not discussed here, is the effect of energetic particles. Indeed, a flare event can be accompanied by an ejection of energetic particles \citep{segura2010effect}. In Earth-like planets, a high flux of energetic particles indirectly destroys ozone, because it favours the formation of nitrogen oxides, but in warm hydrogen-rich atmospheres, like the ones studied in this study, ozone is not present. Most studies dealing with the thermospheres of exoplanets \citep[e.g.,][]{yelle2004, garcia2007, koskinen2007} suggest that the charged particles will affect the atmospheric part above the one we studied here (P$\lesssim 10^{-3}$/$10^{-4}$ mbar). However, the work of \cite{koskinen2010} suggests that the ionization peak in the atmosphere of planets orbiting young host stars with a strong X-ray emission could be located down to 1 mbar. Thus, it would be interesting to quantify the effect of the possible increase of energetic particles expected during a stellar flare using one of the rare complex ion-neutral chemical scheme developed for warm exoplanet atmospheres \citep{lavvas2014, Rimmer2015}.\\

Concerning the spectra, we assumed a cloud-free atmosphere, in accordance to the thermal profiles and the atmospheric model. However, if clouds were considered, their effect would be to flatten the spectra, especially at small wavelengths. But a cloud deck would not hide the strongest features at long wavelengths (such as the ones caused by CO$_2$). Given that the spectral changes are seen at these strong features, the actual differences would stay relatively constant.\\
Finally, although the investigation of this effect is beyond the scope of this paper, we also note that the spectrum of the star varies as the flare evolves, introducing another systematic and time dependent change on the transit light curves. However, stronger variations will only be seen in the bluest bands ($<$ 450 nm) compared to the red and in infrared, since emission in the blue is due to material heated by magnetic reconnection up to tens of thousands of degrees \citep{Davenport2012}. 

\section{Conclusions}\label{sec:concl}

We studied the effect of a stellar flare on the chemical composition of sub-Neptune/super-Earth -like planets orbiting around active M stars. As a case-study, we chose AD Leo and used data recorded during a stellar flare event this star underwent in 1985. Two hypothetic planets have been modelled with different thermal profiles ($T_{eq} = 412$\,K and $T_{eq} = 1303$\,K). Using the 1D photochemical model of \cite{ven2012}, we followed the evolution of the atmospheric compositions during the flare. In both cases, these compositions change with time. Prominent species having relative abundances that are most affected are H, OH, CO$_2$, NH$_2$, NO and NH$_3$. We found that a single flare event could alter irreversibly the chemical composition of warm/hot planetary atmospheres, with final steady-state (reached 10$^{12}$ s after the end of the flare, i.e. $\sim$30 000 years) being significantly different than the initial steady-state in both cases. Even if the frequency at which flares occur is random, it has been observed on AD Leo that the typical time between two flares is about few hours \citep{Pettersen1984,hunt2012}, so far less than the time needed to reach the post-flare steady-state. To discuss this fact, we simulated a series of flares occurring every $\sim$five hours and found that the chemical abundances of species oscillate at each flare around a mean value that evolves with time towards a limiting value. The number of flare events required to reach a limiting value depends not only on the species but also on pressure (thus also local temperature). This result suggests that planets around very active stars (with frequent flares) could never be at a steady-state but are constantly and permanently altered by flare events.

Using the radiative transfer model of \cite{Waldmann2015}, we computed synthetic transmission spectra for the two planet cases during the different phases of the flare (including the return to quiescence phase) and found variations in both cases. We see relative variations of up to 500 ppm in the CO$_2$ bands at respectively 4.6 and 14 $\mu$m in the hotter planet case, and changes of up to 40 ppm in the CO$_2$ /CO at 4.2 $\mu$m feature in the the warm planet case. In particular, we note that the transmission spectra of the initial and final steady states are different, especially for the closer-in planet. Stronger variations of up to 1200 ppm are seen in the hotter planet case assuming that the flare comes back every five hours over a period of ten days. These variations, although difficult to detect with current instrumentation, could be easily detected using JWST.\\

\begin{acknowledgements}

We thank Sarah Rugheimer for the quality of her referee work. Our paper has been highly improved with her useful comments. We also thank Ant\'{i}gona Segura and Lucianne Walkowicz for providing the stellar fluxes of AD Leo. We thank Ludmila Carone and Ingo Waldmann, for their insightful comments, Vivien Parmentier, Kevin Heng  for useful discussions about thermal profiles and radiative transfer and Jonathan Fortney for providing the thermal profiles files. O.V. acknowledges support from the KU Leuven IDO project IDO/10/2013 and from the FWO Postdoctoral Fellowship programme. This work was supported by STFC (ST/K502406/1) and the ERC projects ExoLights (617119) and ExoMol (267219).
\end{acknowledgements}

\bibliographystyle{apj}
\bibliography{VENOT_Flares}

\begin{thebibliography}{}
\expandafter\ifx\csname natexlab\endcsname\relax\def\natexlab#1{#1}\fi

\bibitem[{{Ag{\'u}ndez} {et~al.}(2014){Ag{\'u}ndez}, {Venot}, {Selsis}, \&
  {Iro}}]{agu2014}
{Ag{\'u}ndez}, M., {Venot}, O., {Selsis}, F., \& {Iro}, N. 2014, \apj, 781, 68

\bibitem[{Barstow {et~al.}(2015)Barstow, Aigrain, Irwin, Kendrew, \&
  Fletcher}]{Barstow2015}
Barstow, J.~K., Aigrain, S., Irwin, P. G.~J., Kendrew, S., \& Fletcher, L.~N.
  2015, Monthly Notices of the Royal Astronomical Society, 448, 2546

\bibitem[{Bean {et~al.}(2010)Bean, Miller-Ricci~Kempton, \& Homeier}]{Bean2010}
Bean, J.~L., Miller-Ricci~Kempton, E., \& Homeier, D. 2010, Nature, 468, 669

\bibitem[{Borysow(2002)}]{Borysow2002}
Borysow, A. 2002, A{\&}A, 390, 779

\bibitem[{Borysow {et~al.}(2001)Borysow, Jorgensen, \& Fu}]{Borysow2001}
Borysow, A., Jorgensen, U.~G., \& Fu, Y. 2001, Journal of Quantitative
  Spectroscopy {\&} Radiative Transfer, 68, 235

\bibitem[{{Cavali{\'e}} {et~al.}(2014){Cavali{\'e}}, {Moreno}, {Lellouch},
  {Hartogh}, {Venot}, {Orton}, {Jarchow}, {Encrenaz}, {Selsis}, {Hersant}, \&
  {Fletcher}}]{Cavalie2014}
{Cavali{\'e}}, T., {Moreno}, R., {Lellouch}, E., {et~al.} 2014, \aap, 562, A33

\bibitem[{Charnay {et~al.}(2015)Charnay, Meadows, Misra, Leconte, \&
  Arney}]{charnay2014}
Charnay, B., Meadows, V., Misra, A., Leconte, J., \& Arney, G. 2015, The
  Astrophysical Journal Letters, 813, L1

\bibitem[{Charpinet {et~al.}(2011)Charpinet, Fontaine, Brassard, Green,
  Van~Grootel, Randall, Silvotti, Baran, {\O}stensen, Kawaler,
  {et~al.}}]{charpinet2011}
Charpinet, S., Fontaine, G., Brassard, P., {et~al.} 2011, Nature, 480, 496

\bibitem[{Davenport {et~al.}(2012)Davenport, Becker, Kowalski, Hawley, Schmidt,
  Hilton, Sesar, \& Cutri}]{Davenport2012}
Davenport, J. R.~A., Becker, A.~C., Kowalski, A.~F., {et~al.} 2012, ApJ, 748,
  58

\bibitem[{Deming {et~al.}(2013)Deming, Wilkins, McCullough, Burrows, Fortney,
  Agol, Dobbs-Dixon, Madhusudhan, Crouzet, Desert, Gilliland, Haynes, Knutson,
  Line, Magic, Mandell, Ranjan, Charbonneau, Clampin, Seager, \&
  Showman}]{Deming2013}
Deming, D., Wilkins, A., McCullough, P., {et~al.} 2013, ApJ, 774, 95

\bibitem[{{Dobrijevic} {et~al.}(2010){Dobrijevic}, {Cavali{\'e}},
  {H{\'e}brard}, {Billebaud}, {Hersant}, \& {Selsis}}]{dob2010}
{Dobrijevic}, M., {Cavali{\'e}}, T., {H{\'e}brard}, E., {et~al.} 2010, \planss,
  58, 1555

\bibitem[{{Dobrijevic} {et~al.}(2014){Dobrijevic}, {H{\'e}brard}, {Loison}, \&
  {Hickson}}]{dob2014}
{Dobrijevic}, M., {H{\'e}brard}, E., {Loison}, J.~C., \& {Hickson}, K.~M. 2014,
  \icarus, 228, 324

\bibitem[{{Favata} {et~al.}(2000){Favata}, {Micela}, \& {Reale}}]{Favata2000}
{Favata}, F., {Micela}, G., \& {Reale}, F. 2000, \aap, 354, 1021

\bibitem[{{Fortney} {et~al.}(2008){Fortney}, {Lodders}, {Marley}, \&
  {Freedman}}]{Fortney2008}
{Fortney}, J.~J., {Lodders}, K., {Marley}, M.~S., \& {Freedman}, R.~S. 2008,
  \apj, 678, 1419

\bibitem[{{Fortney} {et~al.}(2005){Fortney}, {Marley}, {Lodders}, {Saumon}, \&
  {Freedman}}]{Fortney2005}
{Fortney}, J.~J., {Marley}, M.~S., {Lodders}, K., {Saumon}, D., \& {Freedman},
  R. 2005, \apjl, 627, L69

\bibitem[{{Fortney} {et~al.}(2013){Fortney}, {Mordasini}, {Nettelmann},
  {Kempton}, {Greene}, \& {Zahnle}}]{Fortney2013}
{Fortney}, J.~J., {Mordasini}, C., {Nettelmann}, N., {et~al.} 2013, \apj, 775,
  80

\bibitem[{Fraine {et~al.}(2014)Fraine, Deming, Benneke, Knutson, Jord{\'a}n,
  Espinoza, Madhusudhan, Wilkins, \& Todorov}]{Fraine2014}
Fraine, J., Deming, D., Benneke, B., {et~al.} 2014, Nature, 513, 526

\bibitem[{{Garc{\'{\i}}a Mu{\~n}oz}(2007)}]{garcia2007}
{Garc{\'{\i}}a Mu{\~n}oz}, A. 2007, \planss, 55, 1426

\bibitem[{Gear(1971)}]{gear1971numerical}
Gear, C. 1971, Numerical initial value problems in ordinary differential
  equations, Prentice-Hall series in automatic computation (Prentice-Hall)

\bibitem[{Grillmair {et~al.}(2008)Grillmair, Burrows, Charbonneau, Stauffer,
  Meadows, von Braun, \& Levine}]{Grillmair2008}
Grillmair, C.~J., Burrows, A., Charbonneau, D., {et~al.} 2008, Nature, 456, 767

\bibitem[{{Gueymard}(2004)}]{gueymard2004}
{Gueymard}, C. 2004, Solar Energy, 76, 423

\bibitem[{Hawley \& Pettersen(1991)}]{hawley1991}
Hawley, S.~L., \& Pettersen, B.~R. 1991, The Astrophysical Journal, 378, 725

\bibitem[{Hindmarsh(1983)}]{hindmarsh1983odepack}
Hindmarsh, A. 1983, IMACS Transactions on Scientific Computation, 1, 55

\bibitem[{Hollis {et~al.}(2013)Hollis, Tessenyi, \& Tinetti}]{Hollis2013}
Hollis, M. D.~J., Tessenyi, M., \& Tinetti, G. 2013, Computer Physics
  Communications, 184, 2351

\bibitem[{{Hunt-Walker} {et~al.}(2012){Hunt-Walker}, {Hilton}, {Kowalski},
  {Hawley}, \& {Matthews}}]{hunt2012}
{Hunt-Walker}, N.~M., {Hilton}, E.~J., {Kowalski}, A.~F., {Hawley}, S.~L., \&
  {Matthews}, J.~M. 2012, \pasp, 124, 545

\bibitem[{{Judge} {et~al.}(2003){Judge}, {Solomon}, \& {Ayres}}]{Judge2003}
{Judge}, P.~G., {Solomon}, S.~C., \& {Ayres}, T.~R. 2003, \apj, 593, 534

\bibitem[{Knutson {et~al.}(2014)Knutson, Benneke, Deming, \&
  Homeier}]{Knutson2014}
Knutson, H.~A., Benneke, B., Deming, D., \& Homeier, D. 2014, Nature, 505, 66

\bibitem[{{Koskinen} {et~al.}(2007){Koskinen}, {Aylward}, {Smith}, \&
  {Miller}}]{koskinen2007}
{Koskinen}, T.~T., {Aylward}, A.~D., {Smith}, C.~G.~A., \& {Miller}, S. 2007,
  \apj, 661, 515

\bibitem[{{Koskinen} {et~al.}(2010){Koskinen}, {Cho}, {Achilleos}, \&
  {Aylward}}]{koskinen2010}
{Koskinen}, T.~T., {Cho}, J.~Y.-K., {Achilleos}, N., \& {Aylward}, A.~D. 2010,
  \apj, 722, 178

\bibitem[{{Kowalski} {et~al.}(2009){Kowalski}, {Hawley}, {Hilton}, {Becker},
  {West}, {Bochanski}, \& {Sesar}}]{kowalski2009}
{Kowalski}, A.~F., {Hawley}, S.~L., {Hilton}, E.~J., {et~al.} 2009, \aj, 138,
  633

\bibitem[{Kreidberg {et~al.}(2014)Kreidberg, Bean, Desert, Benneke, Deming,
  Stevenson, Seager, Berta-Thompson, Seifahrt, \& Homeier}]{Kreidberg2014}
Kreidberg, L., Bean, J.~L., Desert, J.-M., {et~al.} 2014, Nature, 505, 69

\bibitem[{{Lacy} {et~al.}(1976){Lacy}, {Moffett}, \& {Evans}}]{Lacy1976}
{Lacy}, C.~H., {Moffett}, T.~J., \& {Evans}, D.~S. 1976, \apjs, 30, 85

\bibitem[{{Lavvas} {et~al.}(2014){Lavvas}, {Koskinen}, \& {Yelle}}]{lavvas2014}
{Lavvas}, P., {Koskinen}, T., \& {Yelle}, R.~V. 2014, \apj, 796, 15

\bibitem[{Lewis {et~al.}(2010)Lewis, Showman, Fortney, Marley, Freedman, \&
  Lodders}]{lew2010}
Lewis, N.~K., Showman, A.~P., Fortney, J.~J., {et~al.} 2010, \apj, 720, 344

\bibitem[{Line {et~al.}(2011)Line, Vasisht, Chen, Angerhausen, \&
  Yung}]{line2011thermochemical}
Line, M., Vasisht, G., Chen, P., Angerhausen, D., \& Yung, Y. 2011, \apjl, 738,
  32

\bibitem[{Linsky {et~al.}(2010)Linsky, Yang, France, Froning, Green, Stocke, \&
  Osterman}]{Linsky2010}
Linsky, J.~L., Yang, H., France, K., {et~al.} 2010, ApJ, 717, 1291

\bibitem[{Mandell {et~al.}(2011)Mandell, Drake~Deming, Blake, Knutson, Mumma,
  Villanueva, \& Salyk}]{Mandell2011}
Mandell, A.~M., Drake~Deming, L., Blake, G.~A., {et~al.} 2011, ApJ, 728, 18

\bibitem[{{Morley} {et~al.}(2013){Morley}, {Fortney}, {Kempton}, {Marley},
  {Visscher}, \& {Zahnle}}]{Morley2013}
{Morley}, C.~V., {Fortney}, J.~J., {Kempton}, E.~M.-R., {et~al.} 2013, \apj,
  775, 33

\bibitem[{Moses {et~al.}(2011)Moses, Visscher, Fortney, Showman, Lewis,
  Griffith, Klippenstein, Shabram, Friedson, Marley,
  {et~al.}}]{moses2011disequilibrium}
Moses, J.~I., Visscher, C., Fortney, J.~J., {et~al.} 2011, \apj, 737, 15

\bibitem[{{Mousis} {et~al.}(2014){Mousis}, {Fletcher}, {Lebreton}, {Wurz},
  {Cavali{\'e}}, {Coustenis}, {Courtin}, {Gautier}, {Helled}, {Irwin}, {Morse},
  {Nettelmann}, {Marty}, {Rousselot}, {Venot}, {Atkinson}, {Waite}, {Reh},
  {Simon}, {Atreya}, {Andr{\'e}}, {Blanc}, {Daglis}, {Fischer}, {Geppert},
  {Guillot}, {Hedman}, {Hueso}, {Lellouch}, {Lunine}, {Murray}, {O`Donoghue},
  {Rengel}, {S{\'a}nchez-Lavega}, {Schmider}, {Spiga}, {Spilker}, {Petit},
  {Tiscareno}, {Ali-Dib}, {Altwegg}, {Bolton}, {Bouquet}, {Briois}, {Fouchet},
  {Guerlet}, {Kostiuk}, {Lebleu}, {Moreno}, {Orton}, \& {Poncy}}]{Mousis2014}
{Mousis}, O., {Fletcher}, L.~N., {Lebreton}, J.-P., {et~al.} 2014, \planss,
  104, 29

\bibitem[{Muirhead {et~al.}(2012)Muirhead, Johnson, Apps, Carter, Morton,
  Fabrycky, Pineda, Bottom, Rojas-Ayala, Schlawin, {et~al.}}]{muirhead2012}
Muirhead, P.~S., Johnson, J.~A., Apps, K., {et~al.} 2012, The Astrophysical
  Journal, 747, 144

\bibitem[{Nejad(2005)}]{nejad2005}
Nejad, L.~A. 2005, Astrophysics and Space Science, 299, 1

\bibitem[{Pettersen \& Coleman(1981)}]{pettersen1981}
Pettersen, B.~R., \& Coleman, L.~A. 1981, The Astrophysical Journal, 251, 571

\bibitem[{{Pettersen} {et~al.}(1984){Pettersen}, {Coleman}, \&
  {Evans}}]{Pettersen1984}
{Pettersen}, B.~R., {Coleman}, L.~A., \& {Evans}, D.~S. 1984, \apjs, 54, 375

\bibitem[{Radhakrishnan \& Hindmarsh(1993)}]{radhakrishnan1993description}
Radhakrishnan, K., \& Hindmarsh, A. 1993, Lawrence Livermore National
  Laboratory Report

\bibitem[{Richardson {et~al.}(2006)Richardson, Harrington, Seager, \&
  Deming}]{Richardson2006}
Richardson, L.~J., Harrington, J., Seager, S., \& Deming, D. 2006, ApJ, 649,
  1043

\bibitem[{{Rimmer} \& {Helling}(2015)}]{Rimmer2015}
{Rimmer}, P.~B., \& {Helling}, C. 2015, ArXiv e-prints, arXiv:1510.07052

\bibitem[{Rojas-Ayala {et~al.}(2012)Rojas-Ayala, Covey, Muirhead, \&
  Lloyd}]{rojas2012}
Rojas-Ayala, B., Covey, K.~R., Muirhead, P.~S., \& Lloyd, J.~P. 2012, The
  Astrophysical Journal, 748, 93

\bibitem[{Rothman {et~al.}(2009)Rothman, Gordon, Barbe, Benner, Bernath, Birk,
  Boudon, Brown, Campargue, Champion, Chance, Coudert, Dana, Devi, Fally,
  {Flaud, J.-M}, Gamache, Goldman, Jacquemart, Kleiner, Lacome, Lafferty,
  {Mandin, J.-Y}, Massie, Mikhailenko, Miller, Moazzen-Ahmadi, Naumenko,
  Nikitin, Orphal, Perevalov, Perrin, Predoi-Cross, Rinsland, Rotger,
  {\v{S}}ime{\v c}kov{\'a}, Smith, Sung, Tashkun, Tennyson, Toth, Vandaele, \&
  Vander~Auwera}]{Rothman2009}
Rothman, L.~S., Gordon, I.~E., Barbe, A., {et~al.} 2009, Journal of
  Quantitative Spectroscopy {\&} Radiative Transfer, 110, 533

\bibitem[{Rothman {et~al.}(2010)Rothman, Gordon, Barber, Dothe, Gamache,
  Goldman, Perevalov, Tashkun, \& Tennyson}]{Rothman2010}
Rothman, L.~S., Gordon, I.~E., Barber, R.~J., {et~al.} 2010, Journal of
  Quantitative Spectroscopy {\&} Radiative Transfer, 111, 2139

\bibitem[{Rothman {et~al.}(2013)Rothman, Gordon, Babikov, Barbe, Chris~Benner,
  Bernath, Birk, Bizzocchi, Boudon, Brown, Campargue, Chance, Cohen, Coudert,
  Devi, Drouin, Fayt, {Flaud, J.-M}, Gamache, Harrison, Hartmann, Hill, Hodges,
  Jacquemart, Jolly, Lamouroux, Le~Roy, Li, Long, Lyulin, Mackie, Massie,
  Mikhailenko, M{\"u}ller, Naumenko, Nikitin, Orphal, Perevalov, Perrin,
  Polovtseva, Richard, Smith, Starikova, Sung, Tashkun, Tennyson, Toon,
  Tyuterev, \& Wagner}]{Rothman2013}
Rothman, L.~S., Gordon, I.~E., Babikov, Y., {et~al.} 2013, Journal of
  Quantitative Spectroscopy {\&} Radiative Transfer, 130, 4

\bibitem[{{Rugheimer} {et~al.}(2015){Rugheimer}, {Kaltenegger}, {Segura},
  {Linsky}, \& {Mohanty}}]{Rugheimer2015}
{Rugheimer}, S., {Kaltenegger}, L., {Segura}, A., {Linsky}, J., \& {Mohanty},
  S. 2015, \apj, 809, 57

\bibitem[{{Sanchis-Ojeda} {et~al.}(2015){Sanchis-Ojeda}, {Rappaport},
  {Pall{\'e}}, {Delrez}, {DeVore}, {Gandolfi}, {Fukui}, {Ribas}, {Stassun},
  {Albrecht}, {Dai}, {Gaidos}, {Gillon}, {Hirano}, {Holman}, {Howard},
  {Isaacson}, {Jehin}, {Kuzuhara}, {Mann}, {Marcy}, {Miles-P{\'a}ez},
  {Monta{\~n}{\'e}s-Rodr{\'{\i}}guez}, {Murgas}, {Narita}, {Nowak}, {Onitsuka},
  {Paegert}, {Van Eylen}, {Winn}, \& {Yu}}]{Sanchis-Ojeda2015}
{Sanchis-Ojeda}, R., {Rappaport}, S., {Pall{\'e}}, E., {et~al.} 2015, ArXiv
  e-prints, arXiv:1504.04379

\bibitem[{Sedaghati {et~al.}(2015)Sedaghati, Boffin, Csizmadia, Gibson, Kabath,
  Mallonn, \& Van~den Ancker}]{Sedaghati2015}
Sedaghati, E., Boffin, H. M.~J., Csizmadia, S., {et~al.} 2015, A{\&}A, 576, L11

\bibitem[{Segura {et~al.}(2005)Segura, Kasting, Meadows, Cohen, Scalo, Crisp,
  Butler, \& Tinetti}]{segura2005biosignatures}
Segura, A., Kasting, J.~F., Meadows, V., {et~al.} 2005, Astrobiology, 5, 706

\bibitem[{Segura {et~al.}(2010)Segura, Walkowicz, Meadows, Kasting, \&
  Hawley}]{segura2010effect}
Segura, A., Walkowicz, L.~M., Meadows, V., Kasting, J., \& Hawley, S. 2010,
  Astrobiology, 10, 751

\bibitem[{Shkolnik {et~al.}(2009)Shkolnik, Liu, \& Reid}]{shkolnik2009}
Shkolnik, E., Liu, M.~C., \& Reid, I.~N. 2009, The Astrophysical Journal, 699,
  649

\bibitem[{{Showman} {et~al.}(2008){Showman}, {Cooper}, {Fortney}, \&
  {Marley}}]{showman2008}
{Showman}, A.~P., {Cooper}, C.~S., {Fortney}, J.~J., \& {Marley}, M.~S. 2008,
  \apj, 682, 559

\bibitem[{Sing {et~al.}(2008)Sing, Vidal-Madjar, D{\'e}sert, Lecavelier~des
  Etangs, \& Ballester}]{Sing2008}
Sing, D.~K., Vidal-Madjar, A., D{\'e}sert, J.~M., Lecavelier~des Etangs, A., \&
  Ballester, G. 2008, ApJ, 686, 658

\bibitem[{Snellen {et~al.}(2010)Snellen, de~Kok, de~Mooij, \&
  Albrecht}]{Snellen2010}
Snellen, I. A.~G., de~Kok, R.~J., de~Mooij, E. J.~W., \& Albrecht, S. 2010,
  Nature, 465, 1049

\bibitem[{{Stauffer} {et~al.}(1994){Stauffer}, {Caillault}, {Gagne}, {Prosser},
  \& {Hartmann}}]{stauffer1994}
{Stauffer}, J.~R., {Caillault}, J.-P., {Gagne}, M., {Prosser}, C.~F., \&
  {Hartmann}, L.~W. 1994, \apjs, 91, 625

\bibitem[{Swain {et~al.}(2008)Swain, Vasisht, \& Tinetti}]{Swain2008}
Swain, M., Vasisht, G., \& Tinetti, G. 2008, Nature, 452, 329

\bibitem[{Swain {et~al.}(2010)Swain, Deroo, Griffith, Tinetti, Thatte, Vasisht,
  Chen, Bouwman, Crossfield, Angerhausen, Afonso, \& Henning}]{Swain2010}
Swain, M., Deroo, P., Griffith, C.~A., {et~al.} 2010, Nature, 463, 637

\bibitem[{Tennyson \& Yurchenko(2012)}]{Tennyson2012}
Tennyson, J., \& Yurchenko, S.~N. 2012, Monthly Notices of the Royal
  Astronomical Society, 425, 21

\bibitem[{Tinetti {et~al.}(2007)Tinetti, Vidal-Madjar, Liang, Beaulieu, Yung,
  Carey, Barber, Tennyson, Ribas, Allard, Ballester, Sing, \&
  Selsis}]{Tinetti2007}
Tinetti, G., Vidal-Madjar, A., Liang, M.-C., {et~al.} 2007, Nature, 448, 169

\bibitem[{Tofflemire {et~al.}(2012)Tofflemire, Wisniewski, Kowalski, Schmidt,
  Kundurthy, Hilton, Holtzman, \& Hawley}]{tofflemire2012}
Tofflemire, B.~M., Wisniewski, J.~P., Kowalski, A.~F., {et~al.} 2012, The
  Astronomical Journal, 143, 12

\bibitem[{{Toon} {et~al.}(1989){Toon}, {McKay}, {Ackerman}, \&
  {Santhanam}}]{toon89}
{Toon}, O.~B., {McKay}, C.~P., {Ackerman}, T.~P., \& {Santhanam}, K. 1989,
  \jgr, 94, 16287

\bibitem[{{Tsiaras} {et~al.}(2015){Tsiaras}, {Waldmann}, {Rocchetto}, {Varley},
  {Morello}, \& {Tinetti}}]{tsiaras2016}
{Tsiaras}, A., {Waldmann}, I.~P., {Rocchetto}, M., {et~al.} 2015, ArXiv
  e-prints, arXiv:1511.07796

\bibitem[{{Tsiaras} {et~al.}(2016){Tsiaras}, {Rocchetto}, {Waldmann}, {Venot},
  {Varley}, {Morello}, {Damiano}, {Tinetti}, {Barton}, {Yurchenko}, \&
  {Tennyson}}]{tsiaras2016b}
{Tsiaras}, A., {Rocchetto}, M., {Waldmann}, I.~P., {et~al.} 2016, \apj, 820, 99

\bibitem[{{Venot} {et~al.}(2014){Venot}, {Ag{\'u}ndez}, {Selsis}, {Tessenyi},
  \& {Iro}}]{ven2014}
{Venot}, O., {Ag{\'u}ndez}, M., {Selsis}, F., {Tessenyi}, M., \& {Iro}, N.
  2014, \aap, 562, A51

\bibitem[{{Venot} {et~al.}(in prep){Venot}, B{\'e}nilan, Fray, Gazeau,
  {et~al.}}]{venot_NH3}
{Venot}, O., B{\'e}nilan, Y., Fray, N., Gazeau, M.-C., {et~al.} in prep

\bibitem[{{Venot} {et~al.}(2015){Venot}, {H{\'e}brard}, {Ag{\'u}ndez}, {Decin},
  \& {Bounaceur}}]{ven2015}
{Venot}, O., {H{\'e}brard}, E., {Ag{\'u}ndez}, M., {Decin}, L., \& {Bounaceur},
  R. 2015, \aap, 577, A33

\bibitem[{{Venot} {et~al.}(2012){Venot}, H\'ebrard, Ag\'undez, Dobrijevic,
  Selsis, Hersant, Iro, \& Bounaceur}]{ven2012}
{Venot}, O., H\'ebrard, E., Ag\'undez, M., {et~al.} 2012, \aap, 546, A43

\bibitem[{{Venot} {et~al.}(2013){Venot}, Fray, B{\'e}nilan, Gazeau,
  H{\'e}brard, Larcher, Schwell, Dobrijevic, Selsis, {et~al.}}]{venot2013high}
{Venot}, O., Fray, N., B{\'e}nilan, Y., {et~al.} 2013, \aap, 551

\bibitem[{{Vilhu} \& {Walter}(1987)}]{vilhu1987}
{Vilhu}, O., \& {Walter}, F.~M. 1987, \apj, 321, 958

\bibitem[{Waldmann {et~al.}(2015)Waldmann, Tinetti, Rocchetto, Barton,
  Yurchenko, \& Tennyson}]{Waldmann2015}
Waldmann, I., Tinetti, G., Rocchetto, M., {et~al.} 2015, ApJ, 802, 107

\bibitem[{{Yelle}(2004)}]{yelle2004}
{Yelle}, R.~V. 2004, \icarus, 170, 167

\end{thebibliography}

\end{document}